\documentclass[english,aps,manuscript,aip,jcp,english,dvipsnames,preprint]{revtex4-1}

\usepackage[T1]{fontenc}
\usepackage[latin9]{inputenc}
\usepackage{color}
\usepackage{amsmath}
\usepackage{amssymb}
\usepackage{graphicx}
\usepackage{esint}
\usepackage{subscript}

\makeatletter

\providecommand{\tabularnewline}{\\}

 \@ifundefined{textcolor}{}
 {%
   \definecolor{BLACK}{gray}{0}
   \definecolor{WHITE}{gray}{1}
   \definecolor{RED}{rgb}{1,0,0}
   \definecolor{GREEN}{rgb}{0,1,0}
   \definecolor{BLUE}{rgb}{0,0,1}
   \definecolor{CYAN}{cmyk}{1,0,0,0}
   \definecolor{MAGENTA}{cmyk}{0,1,0,0}
   \definecolor{YELLOW}{cmyk}{0,0,1,0}
 }

\def\part#1{\left(#1\right)}

\usepackage{amsthm}
\usepackage{graphics}
\usepackage{times}
\usepackage{bm}
\usepackage{hyperref}
\usepackage{color}
\usepackage{braket}

\RequirePackage[dvipsnames]{xcolor} 
\definecolor{RoyalBlue}{cmyk}{1, 0.80, 0, 0}

\hypersetup{
colorlinks=true,
breaklinks=true, 
hyperfootnotes=true,
urlcolor=RoyalBlue, 
citecolor=RoyalBlue, 
linkcolor=RoyalBlue, 
}

\makeatother

\usepackage{babel}
\begin{document}

\title{``Divide and Conquer'' Semiclassical Molecular Dynamics: A practical
method for Spectroscopic calculations of High Dimensional Molecular
Systems}

\author{Giovanni \surname{Di Liberto}}

\affiliation{Dipartimento di Chimica, Università degli Studi di Milano, via C.
Golgi 19, 20133 Milano, Italy}

\author{Riccardo \surname{Conte}}

\affiliation{Dipartimento di Chimica, Università degli Studi di Milano, via C.
Golgi 19, 20133 Milano, Italy}

\author{Michele \surname{Ceotto}}

\affiliation{Dipartimento di Chimica, Università degli Studi di Milano, via C.
Golgi 19, 20133 Milano, Italy}
\email{michele.ceotto@unimi.it}

\begin{abstract}
We extensively describe our recently established ``divide-and-conquer''
semiclassical method {[}M. Ceotto, G. Di Liberto and R. Conte, \textit{Phys.
Rev. Lett.} \textbf{119}, 010401 (2017){]} and propose a new implementation
of it to increase the accuracy of results. The technique permits to
perform spectroscopic calculations of high dimensional systems by
dividing the full-dimensional problem into a set of smaller dimensional
ones. The partition procedure, originally based on a dynamical analysis
of the Hessian matrix, is here more rigorously achieved through a
hierarchical subspace-separation criterion based on Liouville's theorem.
Comparisons of calculated vibrational frequencies to exact quantum
ones for a set of molecules including benzene show that the new implementation
performs better than the original one and that, on average, the loss
in accuracy with respect to full-dimensional semiclassical calculations
is reduced to only 10 wavenumbers. Furthermore, by investigating the
challenging Zundel cation, we also demonstrate that the ``divide-and-conquer''
approach allows to deal with complex strongly anharmonic molecular
systems. Overall the method very much helps the assignment and physical
interpretation of experimental IR spectra by providing accurate vibrational
fundamentals and overtones decomposed into reduced dimensionality
spectra.
\end{abstract}
\maketitle

\section{Introduction\label{sec:Introduction}}

The simulation of vibrational spectra of high-dimensional systems
is an important open issue in quantum mechanics. The challenge is
to beat the curse of dimensionality that plagues any quantum method
in both electronic and nuclear spectroscopy simulations. In fact,
the exact treatment of quantum problems often implies the set-up of
a grid. As a consequence, the computational cost scales exponentially
with dimensionality, and only simulations involving a few atoms can
be exactly performed.\citep{Bowman_Meyer_Polyatomic_2008,Avila_Carrington_C2H4_2011,avila_carrington_exact12D_2011,avila_carrington_punedbases_2012,Thomas_Carrington_SevenAtoms_2015}
Alternatively, perturbative quantum methods have also been successfully
applied to many systems, but they are intrinsically limited to a single
reference geometry. \citep{Barone_AutomatedVPT2_2005,Puzzarini_Barone_Open-shell_2010,Puzzarini_Barone_Uracil_2011,Biczysko_Barone_abinitioIRgly_2012,Bludsky_Hobza_Anharmonicgly_2000,bloino_Biczysko_vibronic_2016}
High dimensional systems, such as peptides, are instead usually simulated
through \textit{ad-hoc} scaled harmonic approaches or by means of
classical mechanics, either using force fields\citep{petersen_Voth_pocket_2005,vanommeslaeghe_Mackerell_charmm_2010,wang_Case_gaffFF_2004}
or employing ab initio molecular dynamics (AIMD)\citep{marx_mathias_IRfluxional_2011,gaigeot_gaigeot_floppypeptides_2010,thomas_Kirchner_vibroAIMD_2013,Pollak_gomez_spectroscopy_1992,Tuckerman_iftimie_AIMD_2005,hase_pratihar_directdynamics_2017,schlegel_Frisch_AIMD_2001}
approaches in which the nuclear forces are calculated using electronic
structure codes. In classical simulations the curse of dimensionality
is significantly tamed with respect to quantum mechanical counterparts.
However, a purely classical dynamics simulation is unable to describe
tunneling effects, zero point energies, overtones and other important
spectroscopic quantum features.

Semiclassical dynamics employs classical trajectories to reproduce
quantum mechanical effects. In semiclassical methods, spectra are
calculated in a time-dependent way, i.e. by Fourier transforming the
survival amplitude or the autocorrelation function of some observables
(such as the dipole moment).\citep{Heller_SCspectroscopy_1981} Semiclassical
methods based on the coherent states Herman-Kluk propagator\citep{Herman_Kluk_SCnonspreading_1984,Antipov_Nandini_Mixedqcl_2015,Walton_Manolopoulos_FrozenGaussianCO2_1995,Elran_Kay_ImprovingHK_1999,Kay_Multidim_1994,Kay_Numerical_1994,Kay_Integralexpression_1994}
and the initial value representation (SCIVR)\citep{Miller_Atom-Diatom_1970,Miller_S-Matrix_1970,Nandini_Church_Mixedqcl_2015}
are robust, have been proven to reproduce quantum effects quite quantitatively,\citep{Zhang_Pollak_Deeptunneling_2004,Miller_Addingquantumtoclassical_2001,Kay_Atomsandmolecules_2005,Heller_SCspectroscopy_1981,Shalashilin_Child_CCS_2004,bonnet1_rayez_quasiclscattering_1997,bonnet_rayez_Gaussweighting_2004,crespos_Bonnet_H2Pdscattering_2017,Garashchuk_gu_quantumdynamics_2016,Garashchuk_Prezhdo_Bohmian_2011,Conte_Pollak_ThawedGaussian_2010,Conte_Pollak_ContinuumLimit_2012,Kondorskiy_Nanbu_Nonadiabatic_2015,Nakamura_Ohta_SCDevelopment_2016,Koda_SCIVRWigner_2015,Koda_Mixedsemiclassical_2016,Cina_Chapman_2007_SCsmallmolecules,Cina_Chapman_2011_smallmolecules,Cina_cheng_2014_variationalquantumclass,Grossmann_Xavier_SCderivation_1998,Harabati_Grossmann_LongtimeSCIVR_2004,Grossmann_SConPES_1999,Bonella_Coker_Linearizedpathintegral_2005,Bonella_Kapral_quantum-classical_2010,Gottwald_Ivanov_2017}
and have been shown to have an accuracy in spectra calculations often
within 1\% of exact results.\citep{Miller_PNAScomplexsystems_2005,Kay_Atomsandmolecules_2005}
Recently, the multiple-coherent (MC)-SCIVR technique has been developed.
It allows to perform on-the-fly semiclassical molecular dynamics simulations
given a few input trajectories.\citep{Ceotto_AspuruGuzik_Curseofdimensionality_2011,Ceotto_AspuruGuzik_Multiplecoherent_2009,Ceotto_AspuruGuzik_PCCPFirstprinciples_2009,Ceotto_Tantardini_Copper100_2010,Conte_Ceotto_NH3_2013,Gabas_Ceotto_Glycine_2017,Ceotto_Hase_AcceleratedSC_2013,Zhuang_Ceotto_Hessianapprox_2012}
The approach is amenable to ab initio direct molecular dynamics, thus
avoiding the effort to construct an accurate analytical potential
energy surface which may be quite demanding especially for large systems,\citep{Braams_Bowman_PermutInvariant_2009,Jiang_Guo_NeuralNetworks_2014,Conte_Bowman_Manybody_2015,Homayoon_Bowman_H2-H2O_2015,Paukku_Truhlar_N4_2013,Conte_Bowman_CollisionsCH4-H2O_2015,Varga_Truhlar_PESN2O2_2016,Houston_Bowman_RoamingH2CO_2016,Conte_Bowman_GaussianBinning_2013}
and permits to faithfully reproduce quantum effects like quantum resonances,\citep{Ceotto_AspuruGuzik_PCCPFirstprinciples_2009}
intra-molecular and long-range dipole splittings, and the quantum
resonant ammonia umbrella inversion.\citep{Conte_Ceotto_NH3_2013}
Nevertheless, all SCIVR methods run out of steam when straightforwardly
applied to problems involving large-sized systems. 

Understanding the reasons of such a limitation is the first step to
do for dealing with the curse of dimensionality and possibly overcoming
it. The semiclassical wavepacket for a system of $N$ degrees of freedom
consists in the direct product of N monodimensional (Gaussian) wavefunctions
$\left|\Psi\left(t\right)\right\rangle =\left|\psi_{1}\left(t\right)\right\rangle ...\left|\psi_{N}\left(t\right)\right\rangle $.
When the time-dependent overlap $\left\langle \Psi\left(0\right)|\Psi\left(t\right)\right\rangle $
is Fourier transformed to generate the spectrum, the simulation time
should have been long enough to provide a significant overlap. In
other words, if the trajectory does not periodically return to the
surroudings of the phase space region where it started, a noisy signal
will be collected. If, instead, the multidimensional classical trajectory
is such that $\left(\mathbf{p}\left(t\right),\mathbf{q}\left(t\right)\right)$
approaches several times $\left(\mathbf{p}\left(0\right),\mathbf{q}\left(0\right)\right)$,
then the overlap $\left\langle \Psi\left(0\right)|\Psi\left(t\right)\right\rangle $
is sizable and the signal associated to the vibrational features will
prevail on the noise. The curse of dimensionality occurs because each
monodimensional coherent state overlap $\left\langle \psi_{i}\left(0\right)|\psi_{i}\left(t\right)\right\rangle $
should be significant for \emph{all} dimensions at the same time.
Even for uncoupled oscillators with non-commensurable frequencies
the concomitant overlapping event is rarer and rarer as the dimensionality
increases, and the simulation time has to be much prolonged.\citep{Ceotto_AspuruGuzik_Curseofdimensionality_2011}
The present ``\emph{divide et impera}'' idea starts from the consideration
that a full-dimensional classical trajectory, once projected onto
a sub-dimensional space, is more likely to provide a useful spectroscopic
signal and for such a reduced dimensionality trajectory, a clear spectroscopic
signal can be obtained in a much shorter amount of time with respect
to the full-dimensional case, as we have recently shown.\citep{ceotto_conte_DCSCIVR_2017}
Thus, according to this divide-and-conquer strategy, after dividing
the full-dimensional space into mutual disjoint subspaces, a semiclassical
spectroscopic calculation is performed separately for each subspace.
While the classical trajectories are full-dimensional, the semiclassical
calculations employ subspace information for calculating each partial
spectrum. Composition of the projected spectra provides the full-dimensional
one. Considering that nuclear spectra of high dimensional systems
are often too crowded for an unambiguous interpretation, this ``divide-and-conquer''
strategy will also allow to better read and understand the physics
behind the spectra and help the interpretation of experimental results. 

In this paper we introduce some new features that significantly enhance
the accuracy of our divide-and-conquer semiclassical initial value
representation (DC SCIVR) method. Accuracy of results is estimated
by comparison to exact values for systems up to 30 degrees of freedom
(DOFs). In Section (\ref{sec:Theory}) we first recall the basics
of time averaged semiclassical spectral density calculations,\citep{Kaledin_Miller_Timeaveraging_2003,Kaledin_Miller_TAmolecules_2003}
and then we describe in details the DC-SCIVR approach and two new
subspace-separation criteria. In Section (\ref{sec:Results}) we test
the performance of DC SCIVR on strongly coupled Morse oscillators,
real molecular systems like H\textsubscript{2}O, CH\textsubscript{2}O,
CH\textsubscript{4}, CH\textsubscript{2}D\textsubscript{2}, the
very challenging Zundel cation (H\textsubscript{5}O\textsubscript{2}\textsuperscript{+})
and, finally, the benzene molecule, which is, at the best of our knowledge,
the highest dimensional molecular system for which exact quantum vibrational
calculations have been performed.\citep{Halverson_Poirier_Benzene_2015}
A summary and some conclusions end the paper.

\section{A Divide and Conquer Strategy for Semiclassical Dynamics\label{sec:Theory}}

This Section recalls the derivation of the DC-SCIVR expression for
spectroscopic calculations. We start from the SC-IVR power spectrum
formulation and its multiple coherent state time averaging implementation
(MC-SCIVR), and then move to the ``divide and conquer'' working
formula. Finally, we present three different techniques for partitioning
the full-dimensional vibrational space into suitable lower-dimensional
subspaces.\citep{ceotto_conte_DCSCIVR_2017}

\subsection{The SC-IVR time averaged spectral density}

We start by writing the power spectrum $\text{I}\left(E\right)$ of
a molecular system, characterized by the Hamiltonian $\hat{H}$, as
the Fourier transform of the survival amplitude\citep{Heller_SCspectroscopy_1981}
of a given and arbitrary reference state $\left|\chi\right\rangle $
\begin{equation}
\text{I}\left(E\right)\equiv\frac{1}{2\pi\hbar}\int_{-\infty}^{+\infty}\left\langle \chi\left|e^{-i\hat{H}t/\hbar}\right|\chi\right\rangle e^{iEt/\hbar}dt.\label{eq:power_spectrum}
\end{equation}
In semiclassical (SC) molecular dynamics, the quantum time-evolution
operator $e^{-i\hat{H}t/\hbar}$ of Eq. (\ref{eq:power_spectrum})
is substituted by the stationary phase approximation to its Feynman
Path Integral representation.\citep{feynman_pathintegral_1965} In
the position representation, the semiclassical propagator is a matrix
whose elements are obtained as products of a complex action exponential
and a stationary-phase pre-exponential factor, summed over all classical
trajectories that connect the two endpoints.\citep{Berry_Mount_Semiclassical_1972,Liu_Miller_linearizedSCIVR_2007,Takahashi_Takatsuka_PhaseQuantization_2007,Tao_Miller_Tdepsampling_2011,Garashchuk_Prezhdo_Bohmian_2011,VanVleck_SCpropagator_1928,Gutzwiller_SCpropagator_1967,Baranger_Schellhaass_2001,Buchholz_Ceotto_MixedSC_2016,Yamamoto_Miller_Fluxcorrelation_2002}
The search for these trajectories is hampered by the rigid double-boundary
condition. In the SC-IVR dynamics, introduced by Miller and later
also developed by Heller, Herman, Kluk, and Kay,\citep{Miller_S-Matrix_1970,Miller_Atom-Diatom_1970,Heller_FrozenGaussian_1981,Heller_SCspectroscopy_1981,Kay_Multidim_1994,Kay_Numerical_1994,Miller_Addingquantumtoclassical_2001,Miller_PNAScomplexsystems_2005}
the propagator is instead formulated in terms of classical trajectories
determined by initial conditions $\left(\mathbf{p}\left(0\right),\mathbf{q}\left(0\right)\right)$
so that Eq. (\ref{eq:power_spectrum}) becomes
\begin{equation}
\left\langle \chi\left|e^{-i\hat{H}t/\hbar}\right|\chi\right\rangle \approx\left(\frac{1}{2\pi\hbar}\right)^{F}\iintop d\mathbf{p}\left(0\right)d\mathbf{q}\left(0\right)C_{t}\left(\mathbf{p}\left(0\right),\mathbf{q}\left(0\right)\right)e^{\frac{i}{\hbar}S_{t}\left(\mathbf{p}\left(0\right),\mathbf{q}\left(0\right)\right)}\left\langle \chi\right.\left|\mathbf{p}\left(t\right)\mathbf{q}\left(t\right)\left\rangle \right\langle \mathbf{p}\left(0\right)\mathbf{q}\left(0\right)\right|\left.\chi\right\rangle ,\label{eq:HHKK_prop}
\end{equation}
where $F$ is the number of degrees of freedom, $S_{t}\left(\mathbf{p}\left(0\right),\mathbf{q}\left(0\right)\right)$
is the classical action, and $C_{t}\left(\mathbf{p}\left(0\right),\mathbf{q}\left(0\right)\right)$
indicates the pre-exponential stationary-phase factor. If $\left|\mathbf{p}\left(t\right),\mathbf{q}\left(t\right)\right\rangle $
is represented as a coherent state\citep{Heller_FrozenGaussian_1981,Heller_SCspectroscopy_1981,Heller_Cellulardynamics_1991,Shalashilin_Child_Coherentstates_2001}
of the type
\begin{equation}
\left\langle \mathbf{x}|\mathbf{p}\left(t\right),\mathbf{q}\left(t\right)\right\rangle =\left(\frac{\text{det}\left(\Gamma\right)}{\pi^{F}}\right)^{1/4}e^{-\left(\mathbf{x}-\mathbf{q}\left(t\right)\right)^{T}\Gamma\left(\mathbf{x}-\mathbf{q}\left(t\right)\right)/2+i\mathbf{p}^{T}\left(t\right)\left(\mathbf{x}-\mathbf{q}\left(t\right)\right)/\hbar},\label{eq:coherent_state}
\end{equation}
where $\Gamma$ is a diagonal width matrix with coefficients usually
equal to the square root of the vibrational frequencies for bound
states calculations, then the pre-exponential factor becomes

\begin{equation}
C_{t}\left(\mathbf{p}\left(0\right),\mathbf{q}\left(0\right)\right)=\sqrt{\text{det}\left|\frac{1}{2}\left(\frac{\partial\mathbf{q}\left(t\right)}{\partial\mathbf{q}\left(0\right)}+\frac{\partial\mathbf{p}\left(t\right)}{\partial\mathbf{p}\left(0\right)}-i\hbar\Gamma\frac{\partial\mathbf{q}\left(t\right)}{\partial\mathbf{p}\left(0\right)}+\frac{i}{\Gamma\hbar}\frac{\partial\mathbf{p}\left(t\right)}{\partial\mathbf{q}\left(0\right)}\right)\right|},\label{eq:prefactor}
\end{equation}
and Eq. (\ref{eq:HHKK_prop}) is commonly known as the Herman-Kluk
survival amplitude of the Hamiltonian $\hat{H}$. However, for complex
systems, the phase space integration of Eq. (\ref{eq:HHKK_prop})
requires too many trajectories to be feasible. To overcome this limitation,
Miller and Kaledin introduced a time averaged version of the semiclassical
propagator (TA-SCIVR),\citep{Kaledin_Miller_Timeaveraging_2003,Kaledin_Miller_TAmolecules_2003}
which significantly reduces the computational overhead
\begin{align}
\text{I}\left(E\right) & =\frac{1}{\left(2\pi\hbar\right)^{F}}\int\int d\mathbf{q}\left(0\right)d\mathbf{p}\left(0\right)\frac{\text{Re}}{\pi\hbar T}\int_{0}^{T}dt_{1}\int_{t_{1}}^{+\infty}dt_{2}e^{i\left(S_{t_{2}}\left(\mathbf{p}\left(0\right),\mathbf{q}\left(0\right)\right)+Et_{2}\right)/\hbar}\nonumber \\
\times & \langle\chi|\mathbf{p}(t_{2}),\mathbf{q}(t_{2})\rangle e^{-i\left(S_{t_{1}}\left(\mathbf{p}\left(0\right),\mathbf{q}\left(0\right)\right)+Et_{1}\right)/\hbar}\langle\mathbf{p}(t_{1}),\mathbf{q}(t_{1})|\chi\rangle C_{t_{2}}(\mathbf{p}(t_{1}),\mathbf{q}(t_{1})),\label{eq:two_times}
\end{align}
where $t_{1}$ is the additional time averaging variable and $t_{2}$
is the original Fourier transform variable. In Eq. (\ref{eq:two_times})
the integrand is time averaged by taking into account different portions
of time length $t_{2}-t_{1}$ of the same trajectory started in $\left(\mathbf{p}\left(0\right),\mathbf{q}\left(0\right)\right)$.
Considering that the pre-exponential factor is of the type $e^{i\omega t}$
for a harmonic $\omega$-frequency system, Eq. (\ref{eq:prefactor})
can be reasonably approximated as $C_{t}\left(\mathbf{p}\left(0\right),\mathbf{q}\left(0\right)\right)=e^{i\phi_{t}}$,
where $\phi_{t}=\mbox{phase}\left[C_{t}\left(\mathbf{p}\left(0\right),\mathbf{q}\left(0\right)\right)\right]$,
leading to the computationally more convenient separable approximation
version of TA-SCIVR \citep{Kaledin_Miller_Timeaveraging_2003,Kaledin_Miller_TAmolecules_2003}
\begin{equation}
\text{I}\left(E\right)=\left(\frac{1}{2\pi\hbar}\right)^{F}\iintop d\mathbf{p}\left(0\right)d\mathbf{q}\left(0\right)\frac{1}{2\pi\hbar T}\left|\intop_{0}^{T}e^{\frac{i}{\hbar}\left[S_{t}\left(\mathbf{p}\left(0\right),\mathbf{q}\left(0\right)\right)+Et+\phi_{t}\right]}\langle\chi|\mathbf{p}\left(t\right),\mathbf{q}\left(t\right)\rangle dt\right|^{2}.\label{eq:separable}
\end{equation}
Eq. (\ref{eq:separable}) is more amenable than (\ref{eq:two_times})
to phase space Monte Carlo integration, given the positive-definite
integrand, and it has been tested with excellent results on several
molecular systems. However, TA-SCIVR still requires thousands of trajectories
per degree of freedom to reach convergence.\citep{Kaledin_Miller_Timeaveraging_2003,Kaledin_Miller_TAmolecules_2003,Tamascelli_Ceotto_GPU_2014,DiLiberto_Ceotto_Prefactors_2016}
To further reduce the computational effort, the multiple coherent
time averaged SCIVR (MC SCIVR) has been introduced.\citep{Ceotto_AspuruGuzik_Curseofdimensionality_2011,Ceotto_AspuruGuzik_Firstprinciples_2011,Ceotto_AspuruGuzik_Multiplecoherent_2009,Ceotto_AspuruGuzik_PCCPFirstprinciples_2009,Ceotto_Tantardini_Copper100_2010,Conte_Ceotto_NH3_2013}
In the MC-SCIVR\textcolor{black}{{} formulation, the reference state
$\left|\chi\right\rangle $} is written as a combination of coherent
states placed at the classical phase space points $\left(\mathbf{p}_{\text{eq}}^{i},\mathbf{q}_{\text{eq}}^{i}\right)$,\textit{
}i.e. $\left|\chi\right\rangle =\sum_{i=1}^{N_{\text{states}}}\left|\mathbf{p}_{\text{eq}}^{i},\mathbf{q}_{\text{eq}}^{i}\right\rangle $.
$\mathbf{q}_{\text{eq}}^{i}$ is an equilibrium position, while $\mathbf{p}_{\text{eq}}^{i}$
is obtained in a harmonic fashion as $\left(p_{j,\text{eq}}^{i}\right)^{2}/2m=\hbar\omega_{j}\left(n+1/2\right)$,
where \textit{j} is a generic normal mode, $\omega_{j}$ is the associated
frequency, and $m$ is unitary in mass-scaled normal mode coordinates.
Eq.(\ref{eq:separable}) has been shown to be quite accurate with
respect to exact quantum mechanical simulations for the several molecules
tested, even when applied to systems as complex as glycine.\citep{DiLiberto_Ceotto_Prefactors_2016,Conte_Ceotto_NH3_2013,Tamascelli_Ceotto_GPU_2014,Gabas_Ceotto_Glycine_2017}

\subsection{The ``Divide-and-Conquer'' strategy applied to semiclassical dynamics}

In this Section we provide a more detailed explanation of the ``divide-and-conquer''
strategy previously introduced elsewhere.\citep{ceotto_conte_DCSCIVR_2017}
The idea is to calculate the power spectrum $\text{I}\left(E\right)$
of Eq. (\ref{eq:power_spectrum}) as composition of partial spectra
$\widetilde{\text{I}}\left(E\right)$ each one calculated in a reduced
$\text{M}$-dimensional phase space $\left(\tilde{\mathbf{p}},\tilde{\mathbf{q}}\right)$
of the full $\text{N}_{vib}$-dimensional space $\left(\mathbf{p},\mathbf{q}\right)\equiv\left(p_{1},q_{1,}...,\tilde{p}_{i+1},\tilde{q}_{i+1},...,\tilde{p}_{i+M},\tilde{q}_{i+M},...,p_{N_{vib}},q_{N_{vib}}\right)$.
In quantum mechanics, where operators can be represented by matrices,
the projection of an operator onto a sub-space is obtained by a preliminary
suitable singular value decomposition (SVD),\citep{Hinsen_Kneller_SingValueDecomp_2000}
followed by a subsequent matrix multiplication between the full-dimensional
operator and the projector. Semiclassically operators are represented
in phase space coordinates and a suitable SVD is the one involving
the displacement matrix $\mathbf{D}$ for the $\text{M}$-dimensional
subspace.\citep{Hinsen_Kneller_SingValueDecomp_2000,Harland_Roy_SCIVRconstrained_2003}
In our case, $\mathbf{D}$ is a $\mbox{N}_{\mbox{vib}}\times\mbox{M}$
dimensional matrix and a singular-value decomposition is obtained
when $\mathbf{D}=\mathbf{U}\mathbf{\Sigma}\mathbf{V}$, where $\mathbf{U}$
is a $\mbox{N}_{\text{vib}}\times\mbox{M}$ matrix, $\Sigma$ is a
$\mbox{M}\times\mbox{M}$ one, and $V$ a $\mbox{M}\times\mbox{M}$
one. The matrix $\Delta=\mathbf{U}\mathbf{U}^{T}$ is the projector
onto the M-dimensional subspace. Eventually, any matrix $\mathbf{A}$
is projected onto the reduced M-dimensional subspace by taking $\tilde{\mathbf{A}}=\mathbf{\Delta}\mathbf{A}\mathbf{\Delta}$
and retaining the $\mbox{M}\times\mbox{M}$ sub-block of non-zero
elements. Similarly, any vector $\mathbf{q}$ is projected by taking
$\tilde{\mathbf{q}}=\mathbf{\Delta}\mathbf{q}$. Given these considerations,
the projected power spectrum can be written as
\begin{equation}
\widetilde{I}\left(E\right)=\left(\frac{1}{2\pi\hbar}\right)^{M}\iintop d\tilde{\mathbf{p}}\left(0\right)d\tilde{\mathbf{q}}\left(0\right)\frac{1}{2\pi\hbar T}\left|\int_{0}^{T}e^{\frac{i}{\hbar}\left[\tilde{S}_{t}\left(\tilde{\mathbf{p}}\left(0\right),\tilde{\mathbf{q}}\left(0\right)\right)+Et+\tilde{\phi}_{t}\right]}\langle\tilde{\boldsymbol{\chi}}|\tilde{\mathbf{p}}\left(t\right),\tilde{\mathbf{q}}\left(t\right)\rangle dt\right|^{2},\label{eq:separable_projected}
\end{equation}
where the M-dimensional coherent state in the M-dimensional sub-space
is 
\begin{equation}
\langle\mathbf{\tilde{x}|\tilde{\mathbf{p}}\left(t\right),\tilde{\mathbf{q}}\left(t\right)\rangle}=\left(\frac{\mbox{det}(\tilde{\Gamma})}{\pi^{M}}\right)^{\frac{1}{4}}e^{-\left(\mathbf{\tilde{x}}-\mathbf{\tilde{q}}\left(t\right)\right)^{T}\mathbf{\tilde{\Gamma}}\left(\mathbf{\tilde{x}}-\mathbf{\tilde{q}}\left(t\right)\right)/2+i\mathbf{\tilde{p}}^{T}\left(t\right)\left(\mathbf{\tilde{x}}-\mathbf{\tilde{q}}\left(t\right)\right)/\hbar}\label{eq:projected_coherent}
\end{equation}
where the matrix $\tilde{\Gamma}=\mathbf{U}\mathbf{U}^{T}\Gamma\mathbf{U}^{T}\mathbf{U}$
is the projected Gaussian width matrix. $\langle\mathbf{\tilde{x}}|\tilde{\chi}\rangle$
is obtained in a similar way. The phase space integration is now limited
to $\int\int d\mathbf{\tilde{p}}\left(0\right)d\tilde{\mathbf{q}}\left(0\right)$,
\textit{i.e.} a 2M dimensional space. This greatly reduces the computational
cost and the number of trajectories necessary to converge the Monte
Carlo integration. Furthermore, the sampling of the initial conditions
of the full-dimensional trajectories can be done according to a Husimi
distribution in the subspace with the external degrees of freedom
at equilibrium. The representation of the phase $\tilde{\phi}_{t}$
in reduced dimensionality is approximated as $\tilde{\phi}_{t}=\mbox{phase}\left[\tilde{C}_{t}\left(\tilde{\mathbf{p}}\left(0\right),\tilde{\mathbf{q}}\left(0\right)\right)\right]$,
where the pre-exponential factor is calculated according to Eq.(\ref{eq:prefactor})
and each matrix block is of the type $\partial\tilde{\boldsymbol{q}}\left(t\right)/\partial\tilde{\boldsymbol{q}}\left(0\right)$,
and so on. The only component of Eq.(\ref{eq:separable_projected})
that cannot be projected onto the sub-space using the SVD is the classical
action 
\begin{equation}
\tilde{S}_{t}\left(\tilde{\mathbf{p}}\left(0\right),\tilde{\mathbf{q}}\left(0\right)\right)=\int_{0}^{T}\left[\frac{1}{2}m\dot{\tilde{{\bf q}}}^{2}\left(t\right)+V_{S}\left(\tilde{\boldsymbol{q}}\left(t\right)\right)\right]dt\label{eq:projected_action}
\end{equation}
since the expression of the ``projected'' potential $V_{S}\left(\tilde{\boldsymbol{q}}\left(t\right)\right)$
cannot be directly obtained. More specifically, the projected potential
$V_{S}\left(\tilde{\boldsymbol{q}}\left(t\right)\right)$ should be
the potential such that an M-dimensional trajectory starting with
initial conditions $\left(\tilde{\mathbf{p}}\left(0\right),\tilde{\mathbf{q}}\left(0\right)\right)$
visits at all times $t$ the same phase-space points $\left(\tilde{\mathbf{p}}\left(t\right),\tilde{\mathbf{q}}\left(t\right)\right)$
obtained upon projection of the full-dimensional trajectory. However,
the potential $V_{S}\left(\tilde{\boldsymbol{q}}\left(t\right)\right)$
is known only for systems characterized by a separable potential.
In an effort to find a general and suitable expression for $V_{S}\left(\tilde{\boldsymbol{q}}\left(t\right)\right)$,
we notice that the full-dimensional trajectory is continuous with
continuous first derivatives for the full-dimensional molecular potential
$V\left(\boldsymbol{q}\left(t\right)\right)$, and we deduce that
the M-dimensional trajectory and $V_{S}\left(\tilde{\mathbf{q}}\left(t\right)\right)$
have the same features. In a straightforward way, we initially define
the sub-dimensional potential as 
\begin{equation}
V_{S}\left(\tilde{\mathbf{q}}\left(t\right)\right)\equiv V\left(\tilde{\mathbf{q}}\left(t\right);\mathbf{q}_{N_{vib}-M}\left(t\right)\right)\label{eq:parameters}
\end{equation}
where the positions $\mathbf{q}_{N_{vib}-M}\left(t\right)$ belonging
to the other subspaces have been downgraded to parameters. Then, we
introduce a time-dependent field such that
\begin{equation}
V_{S}\left(\tilde{\mathbf{q}}\left(t\right)\right)=V\left(\tilde{\mathbf{q}}\left(t\right);\mathbf{q}_{N_{vib}-M}^{eq}\right)+\lambda\left(t\right),\label{eq:external_field}
\end{equation}
since it is more intuitive and convenient to represent the reduced
dimensionality potential in terms of the conditioned full-dimensional
one (with the parametric coordinates in their equilibrium positions)
plus an external time-dependent field. In agreement with our previous
work,\citep{ceotto_conte_DCSCIVR_2017} we take the following expression
for $\lambda\left(t\right)$
\begin{equation}
\lambda\left(t\right)=V\left(\tilde{\mathbf{q}}\left(t\right);\mathbf{q}_{N_{vib}-M}\left(t\right)\right)-\left[V\left(\tilde{\mathbf{q}}\left(t\right);\mathbf{q}_{N_{vib}-M}^{eq}\right)+V\left(\mathbf{q}_{M}^{eq};\mathbf{q}_{N_{vib}-M}\left(t\right)\right)\right]\label{eq:lambda}
\end{equation}
which is exact in the separable limit. To verify this, we consider
for simplicity a two dimensional separable potential of the type $V\left(q_{1}\left(t\right),q_{2}\left(t\right)\right)=V_{1}\left(q_{1}\left(t\right)\right)+V_{2}\left(q_{2}\left(t\right)\right)$
but the procedure is readily generalizable to separable potentials
of any dimensionality. In the 2D case, using Eqs (\ref{eq:external_field})
and (\ref{eq:lambda}), we obtain $V_{S}\left(q_{1}\left(t\right)\right)=V_{1}\left(q_{1}\left(t\right)\right)-V_{1}\left(q_{1}^{eq}\right)$
which is exact. We also notice that in Eq. (\ref{eq:lambda}) an additional
last term (the value of the instantaneous full-dimensional potential
with the subspace coordinates at equilibrium) has been introduced
with respect to Eq. (\ref{eq:parameters}). It provides a linear term
in the action and consequently shifts the spectrum by a constant,
allowing to match on the same scale each partial spectrum $\tilde{\text{I}}\left(E\right)$
and to obtain the full-dimensional spectrum $\text{I}\left(E\right)$
as a composition of the several $\tilde{\text{I}}\left(E\right)$.
In this last aspect the DC-SCIVR procedure is somewhat similar to
the one employed by Wehrle, Sul\v{c} and Vaní\v{c}ek in their reduced-dimensionality
emission spectra simulations.\citep{Wehrle_Vanicek_Oligothiophenes_2014}
There, they exploited conservation of energy to derive a projected
Lagrangian whose potential energy was made only of a constant term
that had the effect to shift the total spectra. Finally, the full-dimensional
DC-SCIVR zero point energy (ZPE) can simply be regained by summing
up the partial ZPE contributions of each subspace. 

To test the effectiveness of our scheme for $V_{S}\left(\tilde{\mathbf{q}}\left(t\right)\right)$,
we consider two strongly coupled monodimensional Morse oscillators,
whose analytical potential will be explicitly reported in Section
\ref{subsec:Coupled-Morse-Oscillators}. 
\begin{figure}
\begin{centering}
\includegraphics[scale=0.4]{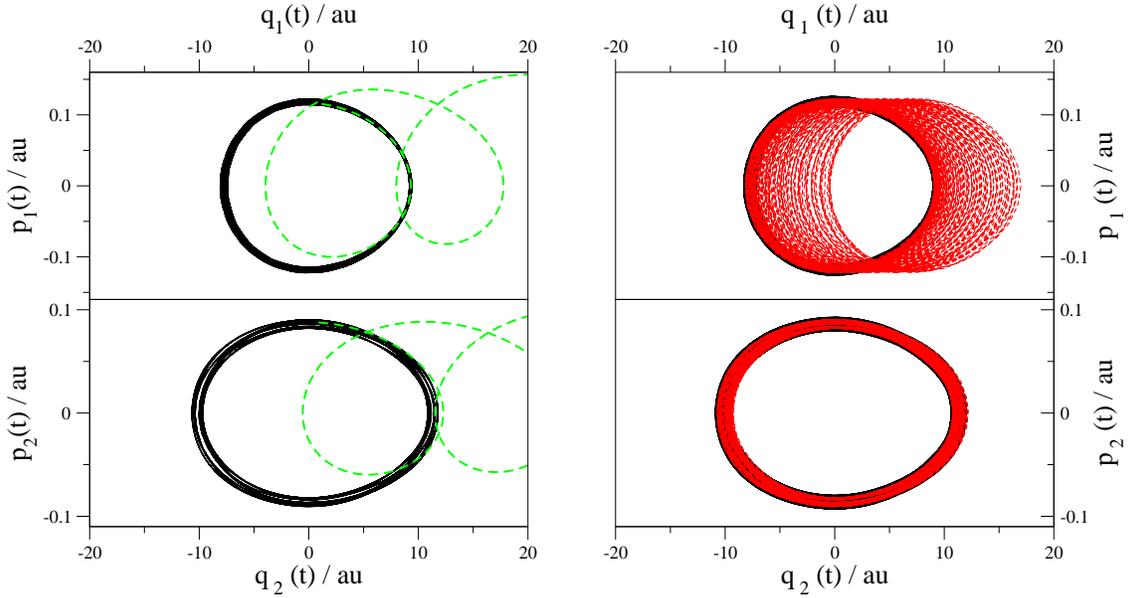}
\par\end{centering}
\caption{\label{fig:phase_space}Mass-scaled phase space plot for the two strongly
coupled Morse oscillators of Eq.(\ref{eq:CoupledMorse_strong}). Left
panel: Black continuous line for the exact, green dashed line for
the potential $V_{S}\left(\tilde{\mathbf{q}}\left(t\right)\right)=V\left(\tilde{\mathbf{q}}\left(t\right);\mathbf{q}_{N_{vib}-M}^{eq}\right)$.
Right panel: Black continuous line for the exact, red dashed line
for the potential of Eq. (\ref{eq:external_field}).}
\end{figure}
Fig.(\ref{fig:phase_space}) reports in the left panel the phase space
plots for a classical trajectory with energy equal to that of the
ground state. The black continuous line is for the $\left(\tilde{\mathbf{p}}\left(t\right),\tilde{\mathbf{q}}\left(t\right)\right)$
values obtained from the full-dimensional vector, which becomes \small$\left(\mathbf{p}\left(t\right),\mathbf{q}\left(t\right)\right)\equiv\left(p_{1}\left(t\right),q_{1}\left(t\right),p_{2}\left(t\right),q_{2}\left(t\right)\right)$
\normalsize in this specific case, evolved according to the full-dimensional
potential $V\left(\boldsymbol{q}\left(t\right)\right)$. The dashed
green line is for the classical trajectory starting at $\left(\tilde{\mathbf{p}}\left(0\right),\tilde{\mathbf{q}}\left(0\right)\right)$
and evolved according to the approximate potential $V_{S}\left(\tilde{\mathbf{q}}\left(t\right)\right)=V\left(\tilde{\mathbf{q}}\left(t\right);\mathbf{q}_{N_{vib}-M}^{eq}\right)$,
\textit{i.e} without any $\lambda\left(t\right)$ correction. Such
a potential is really unfit to describe the projected trajectory motion,
since the green curve diverges after a few time-steps, as it were
describing an unbound system. Two different phase-space plots for
the same Morse oscillators appear on the right panel of Fig.(\ref{fig:phase_space}).
Again, the black continuous line is for the exact projected trajectory,
while the dashed red line is for the classical trajectory starting
at $\left(\tilde{\mathbf{p}}\left(0\right),\tilde{\mathbf{q}}\left(0\right)\right)$
and evolved according to the approximate potential $V_{S}\left(\tilde{\mathbf{q}}\left(t\right)\right)$
of Eq.(\ref{eq:external_field}). In this case, the trajectory phase-space
plot is typical of a bound system. For one of the two dimensions,
the phase space exact and approximate trajectories can be hardly distinguished.
For the other degree of freedom, despite a phase accumulation, the
frequency of the approximate trajectory motion is very similar to
the exact one. 

\subsection{Vibrational space decomposition into mutually disjoint subspaces}

It is now important to define an appropriate strategy for the decomposition
of the full-dimensional space into mutually disjoint and convenient
subspaces. The identification of relevant DOFs for spectroscopic calculations
is a long-standing issue in spectroscopy, and several techniques to
determine the ``effective modes'' have been proposed.\citep{picconi_santoro_quantumclass_2013,cederbaum_Burghardt_shorttimeconinters_2005}
We present here three possible strategies: one is based on the time
evolution of the Hessian matrix, and the other two on the evolution
of the monodromy matrix. In all cases, a preliminary test trajectory
is classically evolved starting from the atomic equilibrium positions
and with initial kinetic energy equal to the harmonic zero point energy
(ZPE) and distributed among the vibrational modes proportionally to
their harmonic frequencies.

\subsubsection{The Hessian space-decomposition method}

\begin{figure}
\begin{centering}
\includegraphics[scale=0.7]{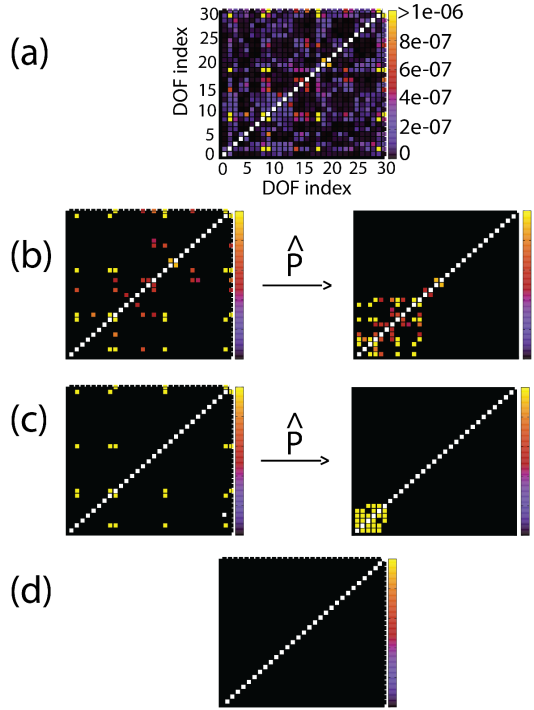}
\par\end{centering}
\caption{\label{fig:threshold_effect}Hessian matrix elements for a system
of 30 degrees of freedom (benzene) greater than a given threshold
value $\epsilon$. The greater the value of $\epsilon$, the less
dense is the matrix. Diagonal elements are out of scale and reported
as white pixels. Panel (a) shows as pixels only the coupling elements
that are greater than $\epsilon=0$ a.u. Panels (b), (c) and (d) are
similar respectively for $\epsilon=4.5\cdot10^{-7}$a.u., $\epsilon=9\cdot10^{-7}$a.u.
and $\epsilon=6\cdot10^{-6}$a.u. In (b) and (c), the matrix elements
have been conveniently arranged after permutations ($\hat{\mathbf{P}}$)
into sub-blocks. Each sub-block determines a subspace.}
\end{figure}

We recall a decomposition strategy that has been recently presented\citep{ceotto_conte_DCSCIVR_2017}
for the computation of molecular vibrational spectra. The full mass-scaled
Hessian matrix is calculated at each time-step and the time averaged
value of each Hessian matrix element is obtained, i.e. $\bar{H}_{ij}=\sum_{k=1}^{N}H_{ij}\left(t_{k}\right)/N$,
with $N$ the number of time steps. If $\bar{|H}_{ij}|\geq\epsilon$,
where $\epsilon$ is an arbitrarily fixed threshold parameter, then
the degrees of freedom $i$ and $j$ are considered as belonging to
the same subspace. If $|\bar{H}_{ij}|<\epsilon$, then $i$ and $j$
can still belong to the same subspace if there exists a third degree
of freedom $k$ such that $\bar{H}_{ik}$ and $\bar{H}_{jk}$ are
bigger than $\epsilon$. In that case, $i$ and $j$ (and also $k$)
are collected into the same subspace. In Fig.(\ref{fig:threshold_effect})
we report how the division into subspaces is affected by the chosen
value of $\epsilon$. Clearly for $\epsilon=0$, all degrees of freedom
are on the same full-dimensional space as shown in Fig.(\ref{fig:threshold_effect})(a).
 By gradually increasing the value of $\epsilon$, the subspaces become
more and more fragmented as illustrated in Fig.(\ref{fig:threshold_effect})(b)
and (c). Finally, for $\epsilon$ bigger than a certain value, the
full-dimensional space is broken down into a direct sum of mono-dimensional
subspaces, as in Fig.(\ref{fig:threshold_effect})(d). In our simulations
we usually choose a value of $\epsilon$ such that it maximizes the
dimensionality of the biggest subspace provided that a spectroscopic
signal can be collected and the curse of dimensionality does not kick
in. This strategy is very advantageous in terms of computational effort,
since the partition of the degrees of freedom into subspaces is instantaneous
after the classical trajectory is run and the Hessian matrices calculated.
However, there is no evidence that this strategy makes the partial
spectra $\widetilde{\text{I}}\left(E\right)$ of Eq.(\ref{eq:separable_projected})
the most accurate with respect to the full-dimensional spectrum $\text{I}\left(E\right)$
of Eq.(\ref{eq:separable}).

\subsubsection{Wehrle-Sul\v{c}-Vaní\v{c}ek (WSV) space-decomposition method}

An alternative decomposition approach (still based on dynamically
averaged quantities and an arbitrary threshold) has been recently
introduced by Vaní\v{c}ek and co-workers.\citep{Wehrle_Vanicek_Oligothiophenes_2014}
In fact, to quantify the coupling between various DOFs still in a
dynamical way one can utilize the stability matrix. This is a $2N_{vib}$
dimensional matrix also called monodromy matrix and defined as
\begin{equation}
\boldsymbol{M}\left(t\right)\equiv\left(\begin{array}{cc}
\partial\boldsymbol{p}\left(t\right)/\partial\boldsymbol{p}\left(0\right) & \partial\boldsymbol{p}\left(t\right)/\partial\boldsymbol{q}\left(0\right)\\
\partial\boldsymbol{q}\left(t\right)/\partial\boldsymbol{p}\left(0\right) & \partial\boldsymbol{q}\left(t\right)/\partial\boldsymbol{q}\left(0\right)
\end{array}\right)=\left(\begin{array}{cc}
\boldsymbol{M}_{\boldsymbol{pp}} & \boldsymbol{M}_{\boldsymbol{pq}}\\
\boldsymbol{M}_{\boldsymbol{qp}} & \boldsymbol{M}_{\boldsymbol{qq}}
\end{array}\right)\label{eq:Monodromy}
\end{equation}
It may be employed to measure how the classical energy is exchanged
in time between the DOFs and, by virtue of Liouville's theorem, its
determinant is always equal to 1. 

In their paper, Vaní\v{c}ek and co-workers define the following quantity
$\mathbf{B}$ to measure the amount of coupling between the vibrational
degrees of freedom in a dynamical fashion

\begin{equation}
B_{ij}=\left|\frac{\beta_{ij}}{\beta_{ii}}\right|,\;\text{with\:}\beta_{ij}=\frac{1}{T}\int_{0}^{T}dt\,(|M_{q_{i}q_{j}}(t)|+|M_{q_{i}p_{j}}(t)|+|M_{p_{i}q_{j}}(t)|+|M_{p_{i}p_{j}}(t)|),\label{eq:B_vanicek}
\end{equation}
where $\left|M_{ij}\left(t\right)\right|$ are the absolute values
of the monodromy matrix elements of Eq. (\ref{eq:Monodromy}). After
an arbitrary parameter $\epsilon_{B}$ is chosen, if the test $max\left\{ B_{ij},B_{ji}\right\} \geq\epsilon_{B}$
is passed, then modes i and j go into the same subspace, following
a procedure very similar to the one employed for our Hessian criterion
but with the difference that more than a single threshold is used.
In our calculations with the WSV method, given an $N_{vib}$ vibrational
space, the bigger $M-$dimensional subspace is determined through
a fixed value of $\epsilon_{B}$. For the remaining $N_{vib}-M$ DOFs,
a different value of $\epsilon_{B}$ is chosen to obtain the biggest
subspace between the remaining DOFs, and so on and so forth until
all DOFs are grouped. 

One might wonder if other dynamical quantities fit in the same general
scheme made of a trajectory average followed by a comparison versus
a threshold value. In this regard, the interested reader may find
tests and a thorough discussion of several ways to define $\mathbf{B}$
on the basis of alternative averaged quantities (like, for instance,
the correlation matrix of the wavepacket) in Wehrle's doctoral thesis.\citep{Wehrle_PhDThesis_2015} 

\subsubsection{Jacobi space-decomposition method}

We here introduce a new approach to determine a subspace partition
which leads to a more accurate calculation of $\widetilde{\text{I}}\left(E\right)$.
Since in DC SCIVR the coherent state overlap $\langle\chi|\mathbf{p}\left(t\right),\mathbf{q}\left(t\right)\rangle$
is already written in terms of direct mono-dimensional overlaps and
the action $\tilde{S}_{t}\left(\tilde{\mathbf{p}}\left(0\right),\tilde{\mathbf{q}}\left(0\right)\right)$
is approximated according to Eqs. (\ref{eq:projected_action}), (\ref{eq:external_field})
and (\ref{eq:lambda}), the best strategy is one that minimizes the
error in decomposing the full-dimensional pre-exponential factor into
a direct product of lower-dimensional ones so that $C_{t}\left(\mathbf{p}\left(0\right),\mathbf{q}\left(0\right)\right)\approx\prod_{i}^{N_{sub}}$
$\tilde{C}_{t,i}\left(\tilde{\mathbf{p}}\left(0\right),\tilde{\mathbf{q}}\left(0\right)\right)$,
where $N_{sub}$ is the number of subspaces. To understand how to
better proceed, we take a two-dimensional separable system. The pre-exponential
factor (\ref{eq:prefactor}), using Eq. (\ref{eq:Monodromy}), can
be written as 
\begin{equation}
C_{t}\left(\mathbf{p}\left(0\right),\mathbf{q}\left(0\right)\right)=\sqrt{\text{det}\left|\frac{1}{2}\left(\boldsymbol{M}_{\boldsymbol{qq}}+\boldsymbol{M_{pp}}-i\hbar\boldsymbol{\Gamma}\boldsymbol{M_{qp}}+\frac{i}{\boldsymbol{\Gamma}\hbar}\boldsymbol{M}_{\boldsymbol{pq}}\right)\right|}\label{eq:prefactor2}
\end{equation}
In the case of a two-dimensional separable system, the matrix components
of Eq.(\ref{eq:Monodromy}) are diagonal matrices
\begin{equation}
\boldsymbol{M}_{\boldsymbol{pp}}=\left(\begin{array}{cc}
M_{p_{1}p_{1}} & 0\\
0 & M_{p_{2}p_{2}}
\end{array}\right);\;\;\boldsymbol{M}_{\boldsymbol{pq}}=\left(\begin{array}{cc}
M_{p_{1}q_{1}} & 0\\
0 & M_{p_{2}q_{2}}
\end{array}\right)\:...\text{ etc}.\label{eq:Monodromy_2d}
\end{equation}
Since the determinant of a block diagonal matrix is equal to the product
of the block determinants, in the case of a separable system the pre-exponential
factor of Eq.(\ref{eq:prefactor2}) is given by the product of the
pre-exponential factors of each dimension. This consideration suggests
that the best sub-space division is the one that minimizes the off-diagonal
terms of the monodromy components in Eq. (\ref{eq:Monodromy_2d}).
The elements of the monodromy matrix can be rearranged into the Jacobian
matrix
\begin{equation}
\mathbf{J}\left(t\right)=\left(\begin{array}{cc}
\partial\mathbf{q}_{t}/\partial\mathbf{q}_{0} & \partial\mathbf{q}_{t}/\partial\mathbf{p}_{0}\\
\partial\mathbf{p}_{t}/\partial\mathbf{q}_{0} & \partial\mathbf{p}_{t}/\partial\mathbf{p}_{0}
\end{array}\right)\label{eq:Jacobian}
\end{equation}
and, in the case of a separable system, the determinant of the full-dimensional
Jacobian, $\mathbf{J}\left(t\right)$, is given by the product of
the determinants of each sub-space Jacobian $\tilde{{\bf J}_{i}}\left(t\right)$,
i.e. $\text{det}\left(\mathbf{J}\left(t\right)\right)=\prod_{i}^{N_{sub}}$$\text{det}\left(\mathbf{\widetilde{J}}_{i}\left(t\right)\right)$.
By virtue of Liouville's theorem $\text{det}\left(\mathbf{J}\left(t\right)\right)=1$
at anytime, i.e. $d\mathbf{p}\left(t\right)d\mathbf{q}\left(t\right)=d\mathbf{p}\left(0\right)d\mathbf{q}\left(0\right)$,
and, for a separable system, $\text{det}\left(\mathbf{\mathbf{\widetilde{J}}}_{i}\left(t\right)\right)=1$
for the generic $i\text{-th}$ subspace, so that $d\mathbf{\widetilde{p}}_{t}^{i}d\mathbf{\widetilde{q}}_{t}^{i}=d\mathbf{\widetilde{p}}_{0}^{i}d\mathbf{\widetilde{q}}_{0}^{i}$.
However, in general, $d\mathbf{\widetilde{p}}_{t}d\mathbf{\widetilde{q}}_{t}\neq d\mathbf{\widetilde{p}}_{0}d\mathbf{\widetilde{q}}_{0}$
and we need to look for the subspace partition which provides subspace
Jacobians $\mathbf{\widetilde{J}}_{i}\left(t\right)$ with the closest
determinants to one. Since the Jacobian is time dependent, the search
for the more suitable subspace division and for the best grouping
of the vibrational modes within the different subspaces also depends
on time. The chosen set of \textit{M} vibrational modes for a \textit{M}-dimensional
subspace is the one that makes the $\mathbf{\widetilde{J}}_{M}\left(t\right)$
determinant the closest to unity more often during the time evolution
of the test trajectory, and we will refer to this procedure as the
``Jacobi criterion'' from now on. The selection of the best subspace
dimensionality is instead performed in a hierarchical way starting
from the full-dimensional space and then proceeding through the remaining
degrees of freedom. More specifically, once the best \textit{M}-dimensional
grouping has been determined for each subspace of dimensionality $M\leq N_{vib}$,
we choose the one for which the determinant of $\tilde{{\bf J}}_{M}\left(t\right)$
(averaged over the trajectory) is the closest to unity. Clearly, $M$
is acceptable if it permits to achieve Monte Carlo convergence in
TA-SCIVR calculations in the subspace, so it cannot be too big, otherwise
the curse of dimensionality still kicks in. The same procedure is
then iteratively applied for the remaining degrees of freedom until
all of them have been grouped in various subspaces. The final result
is a separation of the full-dimensional space into subspaces, where
each subspace preserves Liouville's theorem with the best possible
accuracy. The main drawback of the method is that it comes at a higher
computational cost than the two previously described.

In the next Section, we will apply Eqs (\ref{eq:external_field})
and (\ref{eq:separable_projected}) to several systems and compare
our results with available quantum mechanical vibrational eigenvalues.

\section{Results and Discussion\label{sec:Results}}

\subsection{A model system: two strongly coupled Morse oscillators\label{subsec:Coupled-Morse-Oscillators}}

To test the accuracy of Eq.(\ref{eq:separable_projected}), we consider
a coupled system of the type
\begin{equation}
V\left(q_{1},q_{2}\right)=D\sum_{i=1}^{2}\left[1-e^{-\alpha_{i}\left(q_{i}-q_{i}^{eq}\right)}\right]^{2}+c\left(q_{1}-q_{1}^{eq}\right)^{2}\left(q_{2}-q_{2}^{eq}\right)^{2}\label{eq:CoupledMorse_strong}
\end{equation}
where the coupling is biquadratic, the dissociation energy $D=0.2\:a.u.$
is the same for each oscillator, $\alpha_{i}=\omega_{i}\sqrt{\mu/2D}$,
$c=10^{-7}\mu^{2}$, and $q_{1}^{eq}=q_{2}^{eq}=0$. The reduced mass
$\mu$ is that of the $\text{H}_{2}$ molecule, \textit{i.e}. $\mu=918.975\:\text{a.u.}$,
and the harmonic frequencies are 3,000 and 1,700 wavenumbers. The
oscillators are strongly coupled as shown by the deviation of the
vibrational eigenvalues from the uncoupled ones. In this case there
are two monodimensional subspaces and, as anticipated, we sample the
initial phase space conditions for the $\left(\tilde{\mathbf{p}}\left(t\right),\tilde{\mathbf{q}}\left(t\right)\right)$
trajectories according to a Husimi distribution for the internal degree
of freedom using a Box-Muller sampling centered at ($\tilde{p}_{1}^{eq}=\sqrt{\omega_{1}}$,$\tilde{q}_{1}^{eq}$)
or ($\tilde{p}_{2}^{eq}=\sqrt{\omega_{2}}$, $\tilde{q}_{2}^{eq}$),
with the other (external) degree of freedom initially set at equilibrium.
The projection of the reference state on the subspaces is $\left|\tilde{\chi}\right\rangle =\left|\sqrt{\omega_{i}},q_{i}^{eq}\right\rangle ,\,\,\,i=\left\{ 1,2\right\} $.
The potential of Eq. (\ref{eq:CoupledMorse_strong}) provides quite
a stringent test for the DC-SCIVR approach because of the artificial
strong coupling. We simulate the full-dimensional and the partial-dimensional
spectra both with single trajectories using the MC-SCIVR approach
and with many trajectories by means of Husimi-sampled TA-SCIVR calculations.
In this latter instance, we perform 10,000 trajectories 50,000 a.u.
long per subspace.

\begin{figure}
\begin{centering}
\includegraphics[scale=0.7]{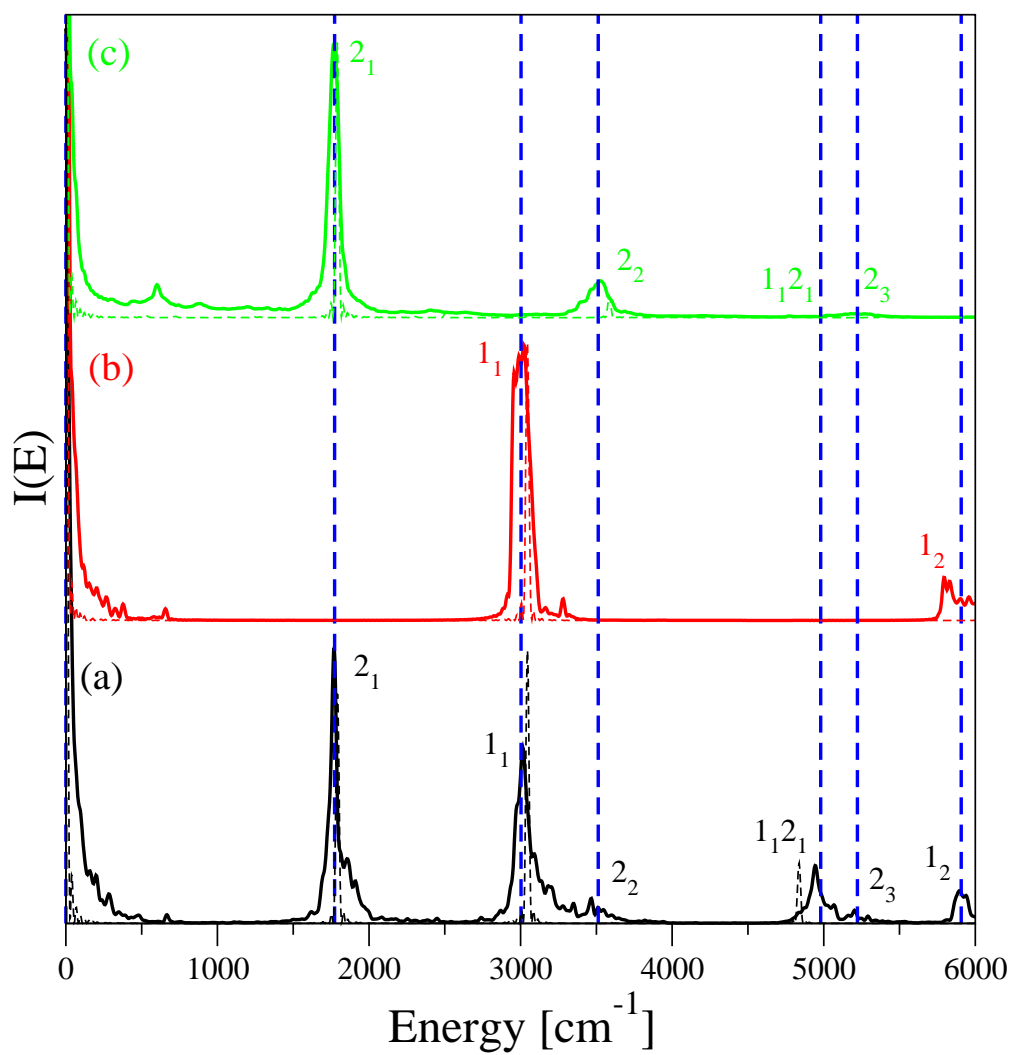}
\par\end{centering}
\caption{\label{fig:copledMorsesx2y2}DC-SC-IVR spectra for the Morse oscillators
of Eq. (\ref{eq:CoupledMorse_strong}). Dashed lines are for the MC-SCIVR
simulations and continuous ones for 10000-trajectory simulations.
(a) black line for the full-dimensional TA-SCIVR spectrum; (b) red
line for the DC-SCIVR spectrum of mode 1; (c) green line, the same
of (b) for mode 2. Vertical dashed blue lines indicate the exact values
calculated by a Discrete Variable Representation (DVR) approach.\citep{colbert_miller_dvr_1992}}
\end{figure}
 The MC-SCIVR spectrum is losing accuracy only at high energies, since
such energy range is not well sampled by MC SCIVR. In the partial
spectra $\tilde{\text{I}}\left(E\right)$ in Fig. (\ref{fig:copledMorsesx2y2})
the overtones generated by the quantum contribution from the other
subspace are much less intense and barely detectable. Nevertheless,
the main spectroscopic features, \textit{i.e}. fundamentals and most
of the overtones, are faithfully reproduced.

\subsection{Small molecules: $\text{\textbf{H}}_{2}\text{\textbf{O}}$, $\text{\textbf{CH}}_{2}\text{\textbf{O}}$,
$\text{\textbf{CH}}\text{\ensuremath{_{4}}}$, and $\text{\textbf{CH}}_{2}\text{\textbf{D}}_{2}$}

We choose H\textsubscript{2}O, CH\textsubscript{2}O, CH\textsubscript{4},
and CH\textsubscript{2}D\textsubscript{2} as test cases for DC SCIVR,
since these are molecular systems accessible to full-dimensional SCIVR
calculations, as it has been shown in the past.\citep{Kaledin_Miller_Timeaveraging_2003,Kaledin_Miller_TAmolecules_2003,Tamascelli_Ceotto_GPU_2014,DiLiberto_Ceotto_Prefactors_2016}
We perform full-dimensional SCIVR and DC-SCIVR calculations using
30,000 a.u. long classical trajectories, which is a typical dynamics
length for semiclassical calculations on molecules.\citep{Tamascelli_Ceotto_GPU_2014,Ceotto_AspuruGuzik_PCCPFirstprinciples_2009,Gabas_Ceotto_Glycine_2017} 

\begin{figure}
\begin{centering}
\includegraphics[scale=0.6]{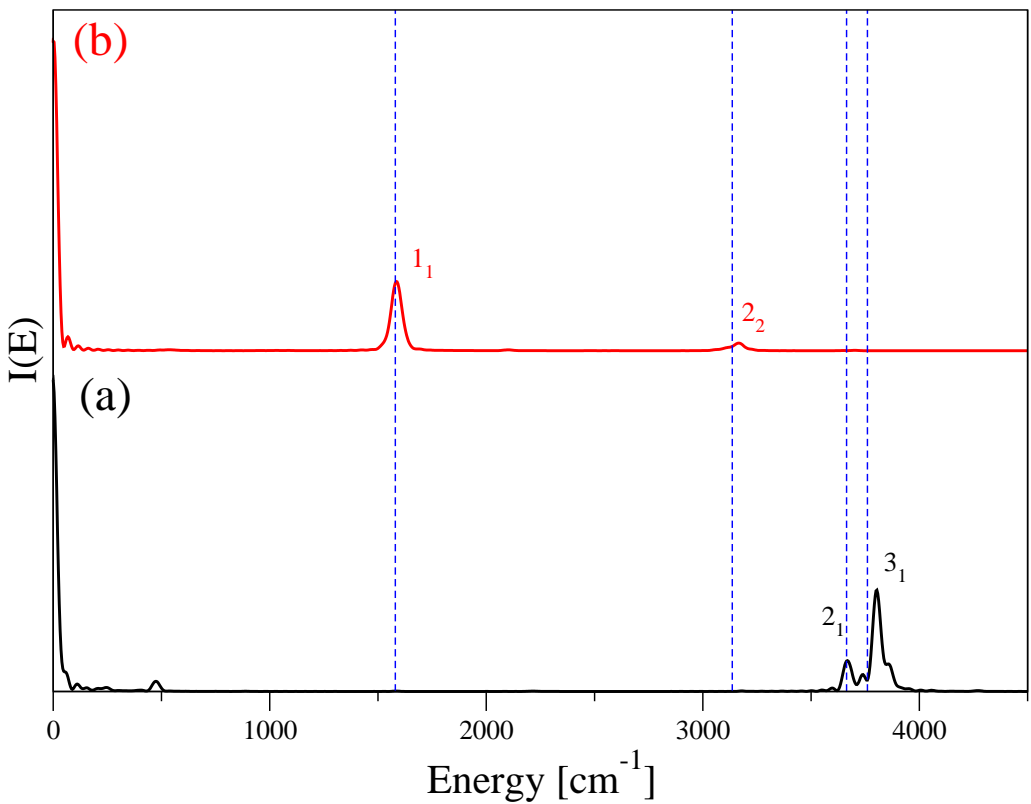}
\par\end{centering}
\caption{\label{fig:H2O_spectra}DC-SCIVR vibrational spectra of H\protect\textsubscript{2}O.
The black line in panel (a) reports the two-dimensional subspace spectrum
and the red line in panel (b) the monodimensional one. Vertical blue
dashed lines are the full-dimensional TA-SCIVR values. }

\end{figure}

Starting from H\textsubscript{2}O, which is the smallest of these
systems, we generate 12,000 classical trajectories on the potential
energy surface of Partridge and Schwenke\citep{partridge_Schwenke_PESH2Omonomer_1997}
for the full dimensional TA-SCIVR calculations, while 4,000 trajectories
per degree of freedom are sufficient in the case of DC-SCIVR spectra.
As in the case of the Morse oscillators, the reference state of each
M-dimensional subspace is $\left|\chi\right\rangle =\prod_{i}^{M}\left|\sqrt{\omega_{i}},q_{i}^{eq}\right\rangle $
, where $\omega_{i}$ is the harmonic frequency of the i-th normal
mode of vibration included in the subspace. Harmonic frequencies are
listed in the ``HO'' column of Table (\ref{tab:freq_H2O}). By employing
the three different subspace partition criteria previously illustrated,
we find that the three vibrational degrees of freedom of water should
always be grouped into two different subspaces. However, in the case
of the Hessian approach modes 1 and 2 (the bending and symmetric stretch
respectively) are separated from mode 3 (the asymmetric stretch),
while the Jacobi and WSV methods suggest to collect together modes
2 and 3, leaving mode 1 alone. In Figure (\ref{fig:H2O_spectra})
the DC-SCIVR spectra of water obtained with the Jacobi criterion are
presented, while Table (\ref{tab:freq_H2O}) reports the detailed
computed energy levels and compares them with full-dimensional SCIVR
estimates and exact values. 

\begin{table}
\centering{}\caption{\label{tab:freq_H2O}Vibrational energy levels of water. The first
and second columns show the vibrational state label and the exact
results respectively; the third column reports the full-dimensional
TA-SCIVR eigenvalues. Column four shows the DC-SCIVR results with
the Jacobi subspace criterion (DC SCIVR$_{\text{Jacobian}}$); column
five refers to frequencies based on the WSV method (DC SCIVR$_{\text{WSV}}$);
in column six results obtained by employing the Hessian matrix criterion
(DC SCIVR$_{\text{Hess}}$) are listed. The last column reports the
harmonic estimates. All values are in $\text{cm}{}^{-1}$. MAE stands
for Mean Absolute Error and it is calculated with respect to the exact
values,\citep{partridge_Schwenke_PESH2Omonomer_1997} and for DC-SCIVR
simulations also with respect to the full-dimensional TA-SCIVR values.
Values for DC SCIVR$_{\text{Jacobi}}$ and DC SCIVR$_{\text{WSV}}$
are exactly the same because they are based on exactly the same partition
of the vibrational modes into the two work subspaces.}
\begin{tabular}{ccccccc}
Mode & Exact\citep{partridge_Schwenke_PESH2Omonomer_1997} & TA SCIVR & DC SCIVR$_{\text{Jacobi}}$ & DC SCIVR$_{\text{WSV}}$ & DC SCIVR$_{\text{Hess}}$ & HO\tabularnewline
\hline 
$1_{1}$ & 1595 & 1580 & 1584 & 1584 & 1581 & 1649\tabularnewline
\hline 
$1_{2}$ & 3152 & 3136 & 3164 & 3164 & 3154 & 3298\tabularnewline
\hline 
$2_{1}$ & 3657 & 3664 & 3668 & 3668 & 3656 & 3833\tabularnewline
\hline 
$3_{1}$ & 3756 & 3760 & 3802 & 3802 & 3824 & 3944\tabularnewline
\hline 
MAE Exact &  & 11 & 20 & 20 & 21 & 141\tabularnewline
\hline 
MAE SCIVR &  &  & 20 & 20 & 23 & \tabularnewline
\hline 
\end{tabular}
\end{table}
 First of all we observe that DC-SCIVR estimates generally account
pretty well for the anharmonicity of water. This can be appreciated
by comparing the mean absolute deviations from quantum exact values
of the DC-SCIVR estimates (\textasciitilde{} 20 cm\textsuperscript{-1})
to the mean deviation of the harmonic frequencies (\textasciitilde{}
140 cm\textsuperscript{-1}). In spite of the anharmonicity and intermode
coupling of water, all separation criteria offer rather accurate estimates.
Only in the case of the asymmetric stretch fundamental frequency the
partition procedure overestimates the quantum value, which is anyway
very accurately regained by the full-dimensional semiclassical approximation.

Moving to CH\textsubscript{2}O, we sample 24,000 classical trajectories
to have the full-dimensional SCIVR calculation converged on the potential
energy surface constructed by Martin et al.,\citep{martin_taylo_PESch2o_1993}.
To keep the same overall computational cost, we take 4,000 trajectories
per degree of freedom when calculating the partial spectra. The dimensionality
of each subspace for the DC-SCIVR calculations is chosen by employing
the three criteria introduced in Section (\ref{sec:Theory}). In the
case of the Hessian matrix criterion, we find that for a value of
$\epsilon=3.0\cdot10^{-7}$ the full six-dimensional vibrational space
is partitioned into a three-dimensional, a bi-dimensional and a mono-dimensional
subspace. When using the WSV approach, the biggest subspace dimensionality
is\textbf{ }four for a threshold value of $\epsilon_{B}=120$. When
employing the Jacobi criterion, the division turns out to be different.
\begin{figure}
\centering{}\includegraphics[scale=0.5]{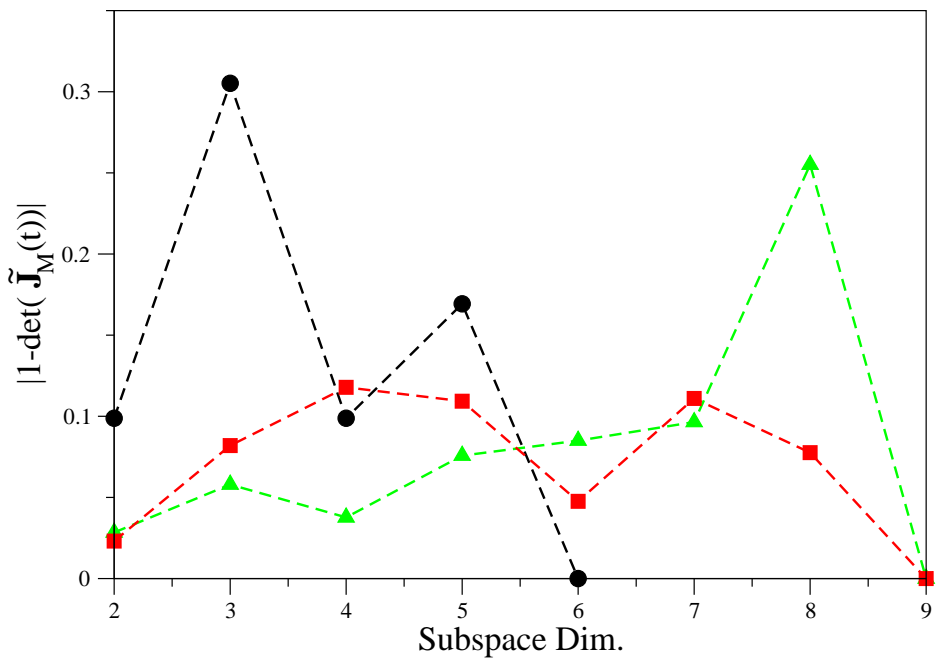}\caption{\label{fig:max_subspace_trend}Average values of $\left|1-\text{det}\left(\tilde{\mathbf{J}}_{M}\left(t\right)\right)\right|$
for the best grouping for different subspace dimensionalities M. Black
filled circles for CH\protect\textsubscript{2}O , red filled squares
for CH\protect\textsubscript{4}, and green filled triangles for CH\protect\textsubscript{2}D\protect\textsubscript{2}.}
\end{figure}
 Figure (\ref{fig:max_subspace_trend}) shows the displacement of
the determinant of the reduced-dimensional Jacobian matrix, i.e. $\text{det}\left(\tilde{\mathbf{J}}_{i}\left(t\right)\right)$
calculated on the basis of the projected trajectories $\tilde{\mathbf{p}}\left(t\right),\tilde{\mathbf{q}}\left(t\right)$,
from unity for different choices of the subspace dimensionality M
in the case of CH\textsubscript{2}O, CH\textsubscript{4}, and CH\textsubscript{2}D\textsubscript{2}.
Clearly, there is no approximation for the full-dimensional analyses.
For the CH\textsubscript{2}O molecule, the smaller deviation is obtained
for a maximum subspace dimensionality equal to 4, which is slightly
better than a bi-dimensional choice. After fixing these four normal
modes into the same subspace, the other two left modes are taken in
the same subspace. Eventually the initial full-dimensional space is
divided into 4- and 2- dimensional ones. The corresponding spectra
are reported in Fig.(\ref{fig:H2CO_spectra}).
\begin{figure}
\centering{}\includegraphics[scale=0.5]{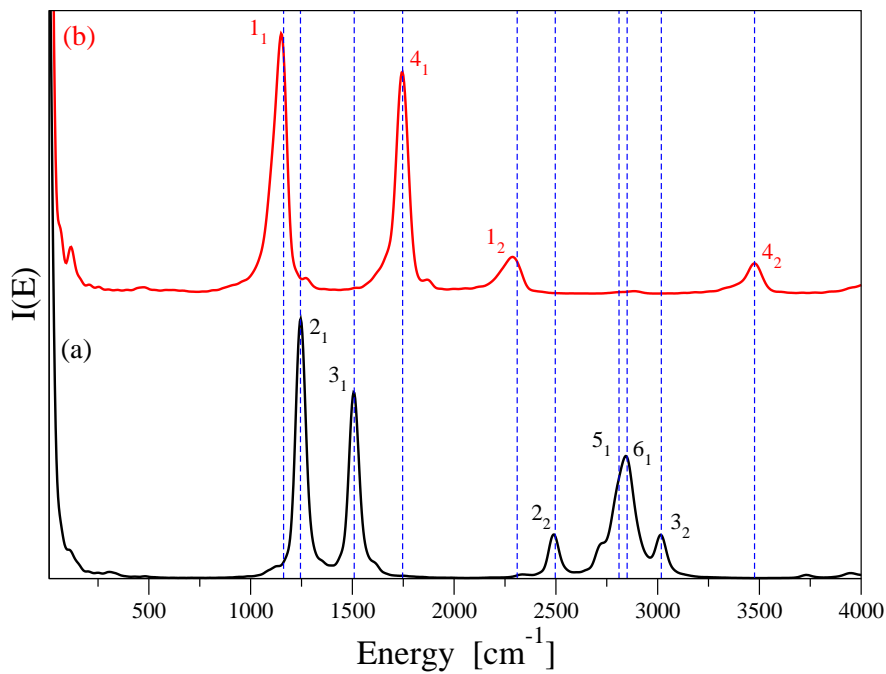}\caption{\label{fig:H2CO_spectra}DC-SCIVR vibrational spectra of CH\protect\textsubscript{2}O.
The black line in panel (a) reports the four-dimensional subspace
spectrum and the red line in panel (b) the bi-dimensional one. Vertical
blue dashed lines are the full-dimensional TA-SCIVR values. }
\end{figure}
 As a comparison, the full-dimensional TA-SCIVR values are reported
as vertical blue dashed lines. All vibrational features are faithfully
reproduced, including overtones. It may be noticed that the signals
of the fifth and sixth fundamentals sum up to a broader peak in the
4-dimensional spectrum. They can be separated by inserting the parity
symmetry into the reference state when performing the 4-dimensional
simulation. This common practice in semiclassical calculations permits
to enhance the signal of one vibration at a time.\citep{Kaledin_Miller_TAmolecules_2003,Ceotto_AspuruGuzik_Curseofdimensionality_2011}
To have a more detailed comparison Table (\ref{tab:freq_CH2O_new})
shows DC-SCIVR results, the exact ones,\citep{carter_handy_exactCH2O_1995}
and the full-dimensional SCIVR frequencies.

\begin{table}
\centering{}\caption{\label{tab:freq_CH2O_new}The same as in Table(\ref{tab:freq_H2O})
this time for the vibrational energy levels of CH\protect\textsubscript{2}O.}
\begin{tabular}{ccccccc}
Mode & Exact\citep{carter_handy_exactCH2O_1995} & TA SCIVR & DC SCIVR$_{\text{Jacobi}}$ & DC SCIVR$_{\text{WSV}}$ & DC SCIVR$_{\text{Hess}}$ & HO\tabularnewline
\hline 
$1_{1}$ & 1171 & 1162 & 1154 & 1154 & 1192 & 1192\tabularnewline
\hline 
$2_{1}$ & 1253 & 1245 & 1246 & 1246 & 1244 & 1275\tabularnewline
\hline 
$3_{1}$ & 1509 & 1509 & 1508 & 1508 & 1508 & 1544\tabularnewline
\hline 
$4_{1}$ & 1750 & 1747 & 1746 & 1746 & 1755 & 1780\tabularnewline
\hline 
$1_{2}$ & 2333 & 2310 & 2288 & 2288 & 2286 & 2384\tabularnewline
\hline 
$2_{2}$ & 2502 & 2497 & 2490 & 2490 & 2423 & 2550\tabularnewline
\hline 
$5_{1}$ & 2783 & 2810 & 2816 & 2816 & 2836 & 2930\tabularnewline
\hline 
$6_{1}$ & 2842 & 2850 & 2845 & 2845 & 2864 & 2996\tabularnewline
\hline 
$3_{2}$ & 3016 & 3018 & 3016 & 3016 & 3024 & 3088\tabularnewline
\hline 
$4_{2}$ & 3480 & 3476 & 3478 & 3478 & 3486 & 3560\tabularnewline
\hline 
MAE Exact &  & 9 & 12 & 12 & 25 & 66\tabularnewline
\hline 
MAE SCIVR &  &  & 6 & 6 & 19 & \tabularnewline
\hline 
\end{tabular}
\end{table}
 To help the reader to better appreciate the level of accuracy for
each semiclassical approximation, we report in the last lines the
Mean Absolute Error (MAE). The DC-SCIVR deviation with respect to
the exact value is $12\text{cm}^{-1}$ for the Jacobi and WSV approaches,
and $25\text{cm}^{-1}$ for the Hessian one. These values are comparable
with the full-dimensional TA-SCIVR one of $9\text{cm}^{-1}$. Conversely,
a harmonic estimate is almost three times less accurate than the DC-SCIVR
ones. When comparing the approximate DC-SCIVR results with the TA-SCIVR
ones, the deviation is on average really small, respectively $6\text{cm}^{-1}$,
$6\text{cm}^{-1}$ and $19\text{cm}^{-1}$ for the Jacobi, WSV, and
Hessian criteria.

In the case of the CH\textsubscript{4} molecule, we employ the potential
energy surface (PES) by Lee \textit{et al}.\citep{lee_taylor_PESch4_1995}
Given the highly chaotic regime for the classical trajectories of
this PES, about 95\% of the trajectories have been rejected due to
the deviation of the full-dimensional monodromy matrix determinant
from unity. By employing an amount of 180,000 trajectories, we still
have enough trajectories left for TA-SCIVR Monte Carlo convergence.
When dividing the space into subspaces, we keep the number of trajectories
per degree of freedom equal to 20,000, in order to have for the overall
DC-SCIVR calculation the same total amount of trajectories. We have
recently shown \citep{ceotto_conte_DCSCIVR_2017} that when a value
of $\epsilon=4.8\cdot10^{-7}$ is employed for the Hessian criterion,
the nine-dimensional vibrational space of methane is decomposed into
six-dimensional and three-dimensional ones. When applying the WSV
criterion with $\epsilon_{B}=85$, we also obtain a six-dimensional
and a three dimensional subspace. Finally, even on the basis of the
Jacobi criterion the better choice for the maximum dimensional subspace
is six, as shown in Fig.(\ref{fig:max_subspace_trend}). We then hierarchically
apply the same criterion for the remaining vibrational modes and find
out that a division into a bi-dimensional plus a mono-dimensional
subspace is preferred with respect to a single three-dimensional one.
Eventually, the nine-dimensional vibrational space is partitioned
into six-, two- and mono-dimensional ones.
\begin{figure}
\centering{}\includegraphics[scale=0.5]{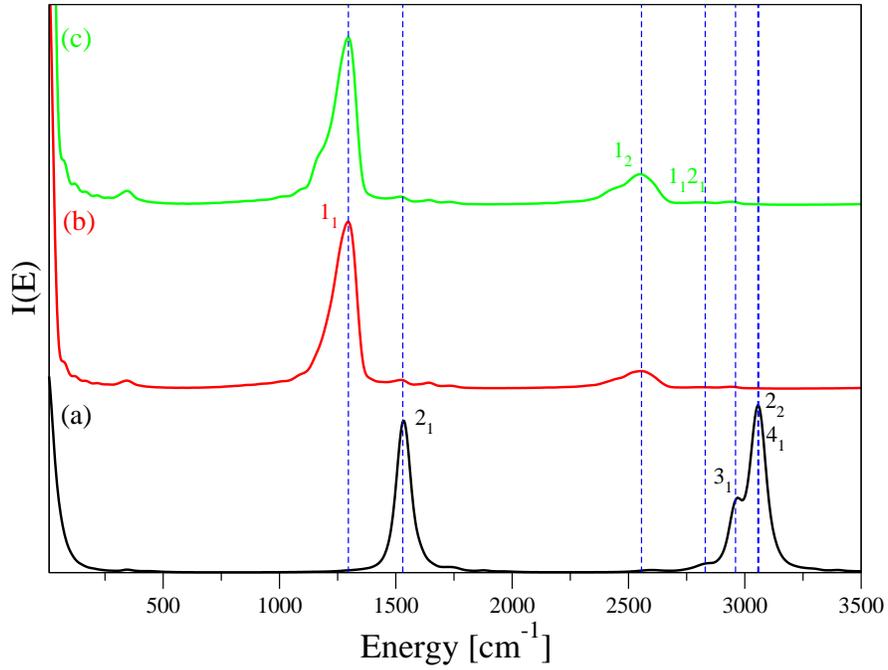}\caption{\label{fig:CH4_spectra}DC-SCIVR vibrational spectra of methane. Black
line in panel (a) reports the six-dimensional subspace partial spectrum,
the red line in panel (b) the bi-dimensional one, and the green line
in panel (c) the mono-dimensional one. Vertical blue dashed lines
indicate the full-dimensional TA-SCIVR values.}
\end{figure}
 Fig.(\ref{fig:CH4_spectra}) reports the partial spectra of the three
subspaces. Given the degeneracy of some of methane vibrations, the
nine vibrational modes are labeled in four groups. Since degenerate
modes can be projected onto different subspaces, spectral contributions
to the same peak may be observed in Fig.(\ref{fig:CH4_spectra}) from
different spectra. The full-dimensional TA-SCIVR peaks are once again
well reproduced, including overtones and combination of overtones.
Vibrations 4\textsubscript{1} and 2\textsubscript{2} have been separated
by including the parity symmetry into the reference state. For a detailed
comparison, we report in Table (\ref{tab:freq_CH4}) our vibrational
eigenvalues and compare them with the exact ones.
\begin{table}
\centering{}\caption{\label{tab:freq_CH4}The same as in Table(\ref{tab:freq_H2O}) but
for the vibrational energy levels of CH\protect\textsubscript{4}. }
\begin{tabular}{ccccccc}
Mode & Exact\citep{Carter_Bowman_Methane_1999} & TA SCIVR & DC SCIVR$_{\text{Jacobi}}$ & DC SCIVR$_{\text{WSV}}$ & DC SCIVR$_{\text{Hess}}$\citep{ceotto_conte_DCSCIVR_2017} & HO\tabularnewline
\hline 
$1_{1}$ & 1313 & 1300 & 1296 & 1308 & 1300 & 1345\tabularnewline
\hline 
$2_{1}$ & 1535 & 1529 & 1530 & 1530 & 1532 & 1570\tabularnewline
\hline 
$1_{2}$ & 2624 & 2594 & 2556 & 2588 & 2606 & 2690\tabularnewline
\hline 
$1_{1}2_{1}$ & 2836 & 2825 & 2830 & 2832 & 2834 & 2915\tabularnewline
\hline 
$3_{1}$ & 2949 & 2948 & 2960 & 2933 & 2964 & 3036\tabularnewline
\hline 
$2_{2}$ & 3067 & 3048 & 3060 & 3044 & 3050 & 3140\tabularnewline
\hline 
$4_{1}$ & 3053 & 3048 & 3056 & 3038 & 3044 & 3157\tabularnewline
\hline 
MAE Exact &  & 12 & 17 & 15 & 11 & 68\tabularnewline
\hline 
MAE SCIVR &  &  & 11 & 7 & 7 & \tabularnewline
\hline 
\end{tabular}
\end{table}
 On average, the full-dimensional TA SCIVR is quite accurate, i.e.
there is only a $12\text{cm}^{-1}$ difference from the exact frequency.
The DC-SCIVR accuracy using the Jacobi criterion is slightly worse
(MAE = $17\text{cm}^{-1}$), and it is comparable when using either
the WSV or the Hessian criterion. These deviations are about six times
more accurate than a crude harmonic approximation. Finally, a comparison
among the different semiclassical approaches shows that in this case
the Hessian criterion provides slightly more accurate results than
the Jacobi ones. However, it is the overtone excitation $1_{2}$ which
is responsible for the slightly worse accuracy of the Jacobi criterion
with respect to the Hessian one. If one did not consider this term
on the MAE calculation, the Jacobi DC SCIVR estimate would be on average
within $9\text{cm}^{-1}$ of the exact one and only $6\text{cm}^{-1}$
away from the TA-SCIVR value.

Finally, we look at the lower symmetry molecule CH\textsubscript{2}D\textsubscript{2},
where some of the typical degenerations of methane have been removed.
We employ the same PES as in the case of CH\textsubscript{4} and
experience a comparable percentage of trajectory rejection for the
monodromy matrix evolution in a chaotic potential. As above, we choose
to employ 180,000 trajectories. Using the Hessian matrix criterion
at a value $\epsilon=2\cdot10^{-7}$ we obtain a decomposition of
the full nine-dimensional space into a six-dimensional and a three-dimensional
one. According to the WSV criterion, at a value $\epsilon_{B}=180$,
we obtain a decomposition of the full nine-dimensional space into
a four-dimensional, a three-dimensional and a bi-dimensional one.
In the Jacobi approach reported in Fig.(\ref{fig:max_subspace_trend}),
we look at the green triangle profile and conclude that a four dimensional
subspace is the first step in the hierarchical determination of the
subspaces. Then, among the remaining five dimensional modes, the Jacobi
analysis leads to a partition into a three- and a two-dimensional
subspace. Eventually, the nine-dimensional space is divided into four,
three and two dimensional subspaces.

\begin{figure}
\centering{}\includegraphics[scale=0.5]{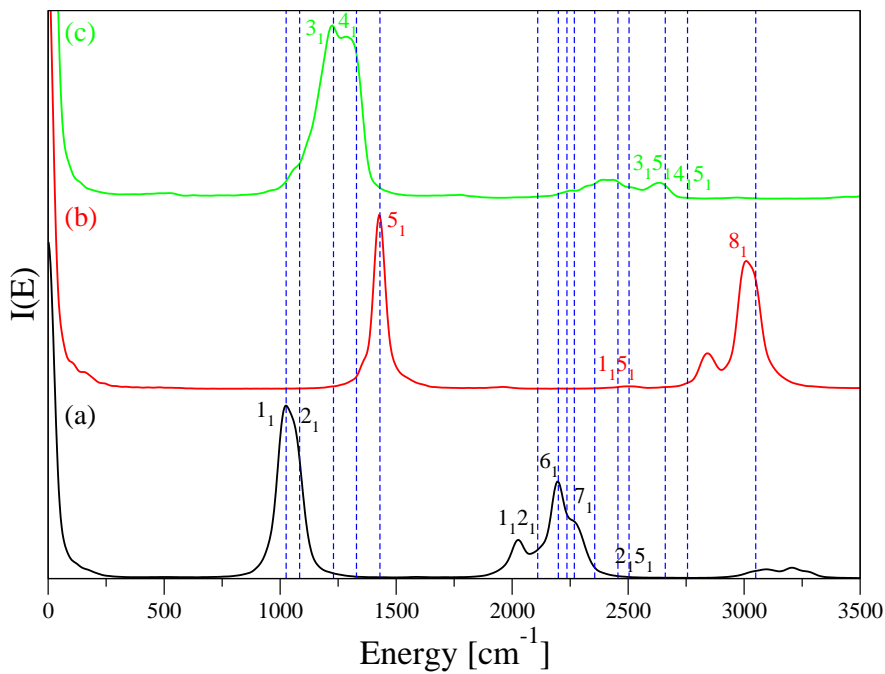}\caption{\label{fig:CH2D2_spectra}DC-SCIVR vibrational spectra of the CH\protect\textsubscript{2}D\protect\textsubscript{2}
molecule. The black line in panel (a) reports the partial spectrum
for the 4-dimensional subspace, the red line in panel (b) the three-dimensional
one and the green line in panel (c) the bi-dimensional one. Vertical
blue dashed lines are the full-dimensional TA-SCIVR values.}
\end{figure}
 Fig.(\ref{fig:CH2D2_spectra}) reports the partial spectra for the
four-dimensional (a), the three-dimensional (b), and the two dimensional
(c) subspaces. By comparison with the dashed vertical lines representing
the full-dimensional semiclassical results we can observe that some
accuracy is lost for the combined overtones (see the $1_{1}2_{1}$
peak) with respect to the typical accuracy of the fundamental peaks,
as it was noticed for the strongly coupled Morse oscillators.

\begin{table}
\centering{}\caption{\label{tab:freq_CH2D2_new}The same as in Table(\ref{tab:freq_H2O})
but for the vibrational energy levels of CH\protect\textsubscript{2}D\protect\textsubscript{2}.}
\begin{tabular}{ccccccc}
Mode & Exact\citep{Carter_Bowman_Methane_1999} & TA SCIVR & DC SCIVR$_{\text{Jacobi}}$ & DC SCIVR$_{\text{WSV}}$ & DC SCIVR$_{\text{Hess}}$ & HO\tabularnewline
\hline 
$1_{1}$ & 1034 & 1026 & 1028 & 1020 & 1038 & 1053\tabularnewline
\hline 
$2_{1}$ & 1093 & 1084 & 1072 & 1098 & 1086 & 1116\tabularnewline
\hline 
$3_{1}$ & 1238 & 1230 & 1234 & 1212 & 1230 & 1266\tabularnewline
\hline 
$4_{1}$ & 1332 & 1329 & 1320 & 1326 & 1316 & 1360\tabularnewline
\hline 
$5_{1}$ & 1436 & 1430 & 1430 & 1420 & 1434 & 1471\tabularnewline
\hline 
$1_{1}2_{1}$ & 2128 & 2110 & 2089 & 2080 & 2114 & 2169\tabularnewline
\hline 
$6_{1}$ & 2211 & 2199 & 2195 & 2192 & 2137 & 2236\tabularnewline
\hline 
$1_{1}3_{1}$ & 2242 & 2236 & 2250 & 2231 & 2210 & 2319\tabularnewline
\hline 
$7_{1}$ & 2294 & 2268 & 2274 & 2250 & 2274 & 2336\tabularnewline
\hline 
$1_{1}4_{1}$ & 2368 & 2356 & / & / & 2400 & 2413\tabularnewline
\hline 
$1_{1}5_{1}$ & 2474 & 2456 & 2485 & 2436 & 2484 & 2524\tabularnewline
\hline 
$2_{1}5_{1}$ & 2519 & 2504 & 2516 & 2494 & 2510 & 2587\tabularnewline
\hline 
$3_{1}5_{1}$ & 2674 & 2660 & 2661 & 2672 & 2627 & 2737\tabularnewline
\hline 
$4_{1}5_{1}$ & 2769 & 2756 & 2754 & 2734 & / & 2831\tabularnewline
\hline 
$8_{1}$ & 3008 & 3050 & 3000 & 3012 & 3026 & 3103\tabularnewline
\hline 
MAE Exact &  & 14 & 13 & 21 & 21 & 47\tabularnewline
\hline 
MAE SCIVR &  &  & 12 & 15 & 19 & \tabularnewline
\hline 
\end{tabular}
\end{table}
 Table (\ref{tab:freq_CH2D2_new}) shows the computed DC-SCIVR energy
levels which are compared with both the exact values \citep{Carter_Bowman_Methane_1999}
and the full-dimensional TA-SCIVR ones. For this system, the MAEs
relative to the exact values are more accurate for the TA-SCIVR and
the Jacobian DC SCIVR than for the standard Hessian criterion. When
comparing the different semiclassical approaches the expected order
is found, i.e. from the more accurate TA SCIVR to the less accurate
DC SCIVR.

\subsection{A complex and strongly anharmonic molecular system: $\text{\textbf{H}}_{5}\text{\textbf{O\ensuremath{_{2}^{+}}}}$}

We keep proceeding in the application of DC-SCIVR to larger and larger
molecules and face the challenge represented by the Zundel cation.
H\textsubscript{5}O\textsubscript{2}\textsuperscript{+} with its
15 vibrational degrees of freedom has attracted the interest of many,
mainly due to the vibrational features related to the motion of the
shared proton. Specifically, a doublet is found in the vibrational
pre-dissociation spectra of Zundel ions in the region of the O-H-O
stretch associated to the proton transfer (\textasciitilde{}1000 cm\textsuperscript{-1}).
Furthermore, two neatly separated bending signals are present owing
to the water bending - proton transfer interaction.\citep{vendrell_Meyer_zundelspectra_2007,rossi_manolopoulos_TRPDM_2014}
Consequently in our investigation we focus our attention on the proton
transfer doublet, the water bendings, and, in addition, the four high-frequency
free OH stretchings which are well detected by experimental spectra.\citep{hammer_carter_expzundelspectrum_2005}
We benchmark our DC-SCIVR simulations against the MCTDH calculations
of Meyer et al. \citep{vendrell_Meyer_isotopeeffects_2009,vendrell_Meyer_isotopeffects2_2009,vendrell_Meyer_Jacobianparametriz_2009,vendrell_Meyer_Zundeldynamics_2007,vendrell_Meyer_ZundelHamiltonian_2007,vendrell_meyer_zundelquantumdynamics_2008,vendrell_Meyer_zundelspectra_2007}
and also compare them with the VCI estimates of Bowman and collaborators.\citep{mccoy_bowman_VCIzundel_2005} 

We propagate the test classical trajectory on an accurate H\textsubscript{5}O\textsubscript{2}\textsuperscript{+}
PES.\citep{huang_Bowman_ZundelPES_2005} The trajectory is characterized
by a strongly roto-vibrationally coupled motion leading to monodromy
matrix instability and to a couple of hindrances to the application
of our semiclassical techniques. For this reason, a Jacobi-based subspace
partition is not feasible and we have to rely on the Hessian method
to determine our work subspaces. Also, the coupling is responsible
for an exaggerated broadening of the spectral features. This latter
drawback can be overcome by removing the Cartesian angular momentum
every few steps along the dynamics of the trajectories employed in
our calculations. The associated loss in energy may partially affect
the frequency accuracy (an artificial shift towards their harmonic
counterparts is anticipated for the high frequencies) but it is compensated
by the Husimi distribution of energies around the harmonic zero-point
one employed for the initial conditions. Finally, due to the monodromy
matrix instability, the original Herman-Kluk prefactor cannot be employed,
so we approximate it by means of a reliable second order iterative
approximation that depends only on the Hessian matrix.\citep{DiLiberto_Ceotto_Prefactors_2016}
As expected, not only peaks in the spectra still have good accuracy
but they are also much narrower thus decreasing the uncertainty of
our results. 

The Hessian criterion suggests us to enroll the normal modes associated
to the free OH stretchings of the two water molecules into a four
dimensional subspace, while all the other degrees of freedom are grouped
into mono-dimensional subspaces. For this reason, we assign the two
water bendings to two separate mono-dimensional subspaces, and the
same fate applies to the mode associated to the shared proton motion.
The only exception concerns the O-O stretching mode which is collected
with a wagging state into a bi-dimensional subspace. This choice is
driven by previous studies that have provided evidence of the occurrence
of a combined state interacting with the shared proton motion.\citep{vendrell_Meyer_zundelspectra_2007,vendrell_Meyer_isotopeffects2_2009}
We run 2,000 full-dimensional classical trajectories per degree of
freedom, i.e. 2,000 for the mono-dimensional subspaces, 4,000 for
the bi-dimensional one, and 8,000 for the four-dimensional subspace.
For each subspace, the initial kinetic energy is given in the usual
harmonic fashion to the four OH stretches and to the modes enrolled
in the subspace under investigation. No energy is instead given to
the other modes.

\begin{figure}
\centering{}\includegraphics[scale=0.5]{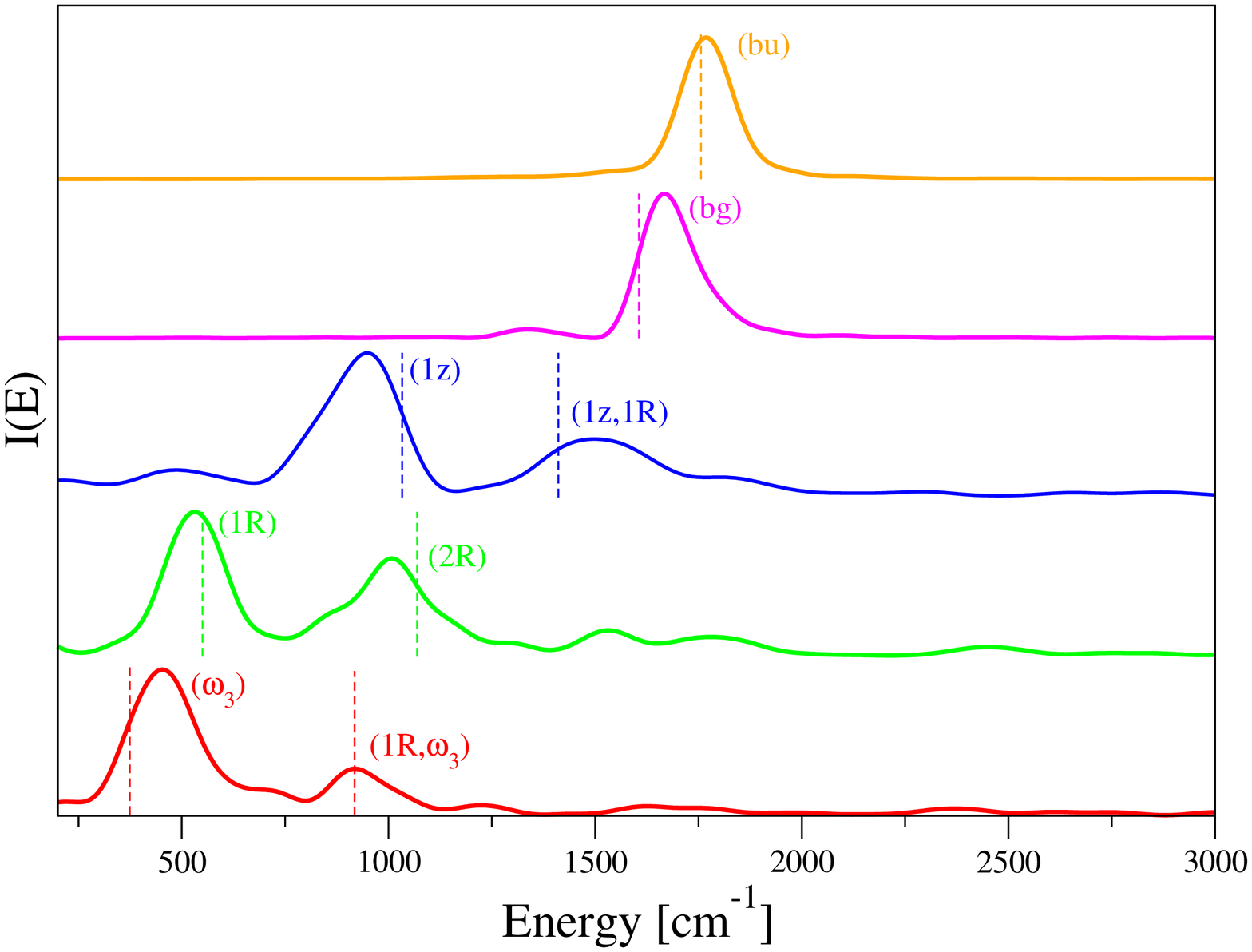}\caption{\label{fig:spectra_Zundel_1-1}Vibrational spectra of the Zundel cation.
Starting from the top, orange, magenta, and blue lines report the
spectra of the mono-dimensional subspaces associate to the $\text{\ensuremath{\left(bu\right)}}$,
$\text{\ensuremath{\left(bg\right)}}$, and $\text{\ensuremath{\left(1z\right)}}$
excitations; the green and red lines build up together the bi-dimensional
subspace. The zero point energy value has been shifted to the origin
in each subspace to help the reader in comparing the different frequencies.
The vertical lines indicate the MCTDH reference.\citep{vendrell_Meyer_zundelspectra_2007}}
\end{figure}

\begin{figure}
\centering{}\includegraphics[scale=0.5]{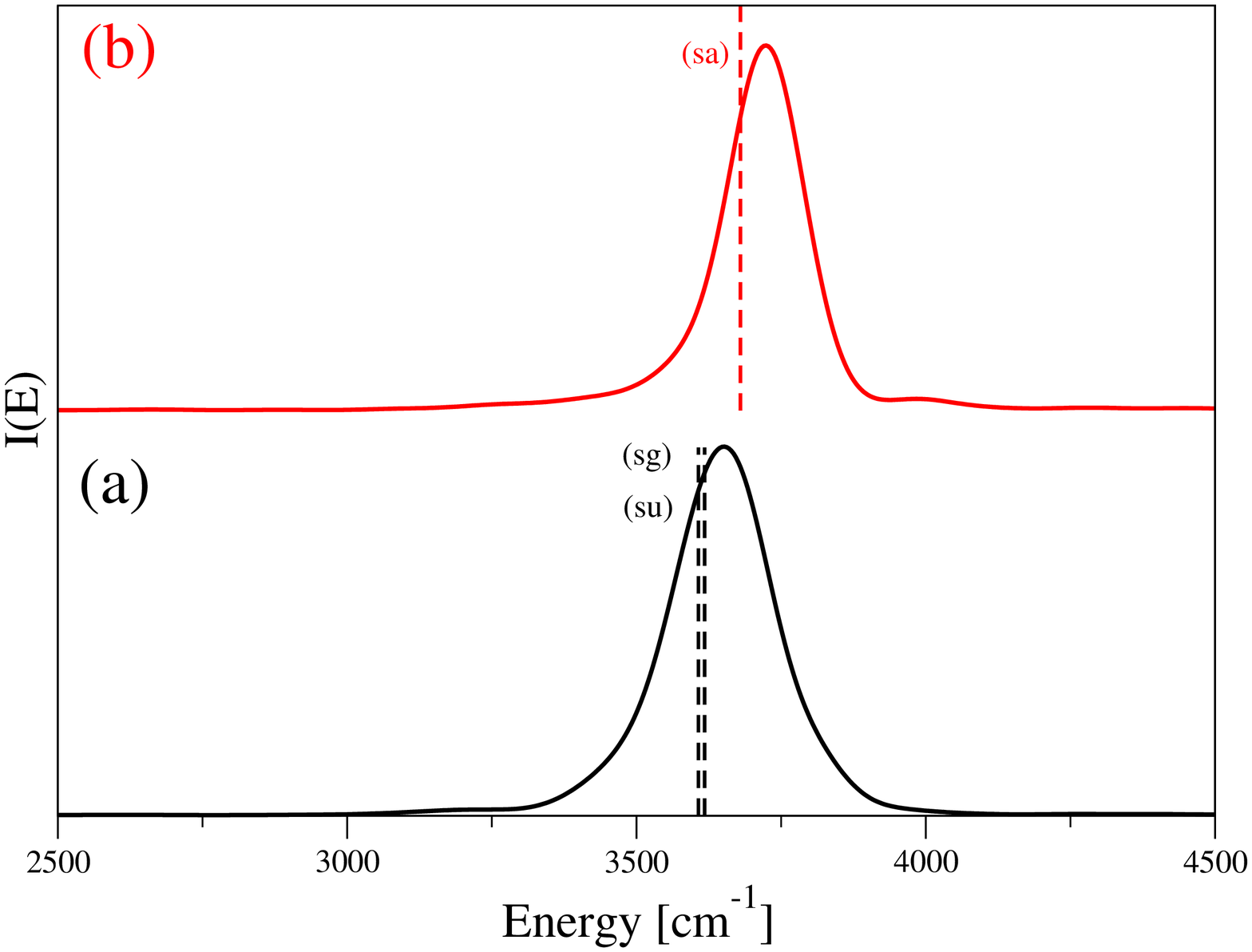}\caption{\label{fig:spectra_Zundel_2-1}Vibrational spectra of the Zundel cation
in the free OH stretching region. Starting from the bottom, panel
(a) reports the spectrum of $\text{\ensuremath{\left(sg\right)}}$
and $\text{\ensuremath{\left(su\right)}}$, panel (b) refers to $\text{\ensuremath{\left(sa\right)}}$
excitations. The zero point energy value has been shifted to the origin
to help the reader in evaluating the frequencies of the peaks. The
vertical lines indicate the MCTDH estimates.\citep{vendrell_Meyer_zundelspectra_2007}}
\end{figure}

Figure (\ref{fig:spectra_Zundel_1-1}) reports the main excitations
below 2,000 wavenumbers. To remove any spurious noise effect, we add
a Gaussian filter of type $e^{-\alpha t^{2}}$ in the Fourier transform,
with $\alpha=3\cdot10^{-8}\:\text{a.u.}$ The orange and magenta lines
refer to the two water bendings $\text{\ensuremath{\left(bu\right)}}$
and $\text{\ensuremath{\left(bg\right)}}$; the blue line shows the
signal of the shared proton motion $\text{\ensuremath{\left(1z\right)}}$
and a mixed excitation$\text{\ensuremath{\left(1z,1R\right)}}$. Finally,
on the bottom of the Figure are the spectra associated to the bi-dimensional
subspace. The usual procedure based on selecting the parity of the
semiclassical reference state permits to separate the overlapping
features of this bidimensional subspace. Specifically, in green the
fundamental for the O-O stretch $\text{\ensuremath{\left(1R\right)}}$
and its overtone $\text{\ensuremath{\left(2R\right)}}$ are detected,
while in red the excitation $\omega_{3}$ of the wagging state (assigned
on the basis of the MCTDH benchmark) and the combined excitation $\text{\ensuremath{\left(1R,\omega_{3}\right)}}$
stand out. In Figure (\ref{fig:spectra_Zundel_2-1}) are instead illustrated
the DC-SCIVR spectra of the free OH stretchings. In panel (a) the
spectra of the $\text{\ensuremath{\left(sg\right)}}$ and $\text{\ensuremath{\left(su\right)}}$
excitations are reported, while panel (b) shows the signal of the
two remaining OH stretchings labeled as $\text{\ensuremath{\left(sa\right)}}$.


Table (\ref{tab:Zundel_freq}) shows our computed energy levels, labeled
with the usual nomenclature for the Zundel cation reported in the
literature.\citep{vendrell_Meyer_zundelspectra_2007,vendrell_Meyer_isotopeffects2_2009}
Our DC-SCIVR estimates are pretty accurate with the exception of the
combined excitation $\text{\ensuremath{\left(1z,1R\right)}}$ which
is rather off-the-mark, but anyway better than the VCI value. A certain
degree of inaccuracy arises also for the (1z) signal. As anticipated,
the high frequency estimates are blue shifted with respect to the
benchmark values, an effect the instantaneous removal of the Cartesian
angular momentum may have largely contributed to. Overall, the average
deviation from MCTDH results is 46 wavenumbers that decreases to 38
if $\text{\ensuremath{\left(1z,1R\right)}}$ is not considered. These
values are not far from those found for smaller molecules and are
satisfactory given the high complexity of the Zundel cation.

\begin{table}
\begin{centering}
\caption{\label{tab:Zundel_freq}Vibrational energy levels of the Zundel cation
reported in cm$\text{\ensuremath{^{-1}}}$. The first column presents
the label of the excitation according to Ref.\citep{vendrell_Meyer_zundelspectra_2007}
The second column contains the experimental values, the third and
fourth ones show the MCTDH results from two different works,\citep{vendrell_Meyer_zundelspectra_2007,vendrell_Meyer_isotopeffects2_2009}
while in the fifth column our DC-SCIVR estimates are reported. Column
six contains the VCI energy levels\citep{mccoy_bowman_VCIzundel_2005}
and, finally, in the last column are the harmonic estimates of the
fundamental excitations. The last row reports the mean absolute error
of the DC-SCIVR estimates with respect to the benchmark MCTDH values
of Ref. \onlinecite{vendrell_Meyer_zundelspectra_2007}. }
\begin{tabular}{ccccccc}
Label & Exp\citep{hammer_carter_expzundelspectrum_2005} & MCTDH\citep{vendrell_Meyer_zundelspectra_2007} & MCTDH\citep{vendrell_Meyer_isotopeffects2_2009} & DC SCIVR & VCI\citep{mccoy_bowman_VCIzundel_2005} & HO\tabularnewline
\hline 
$\text{\ensuremath{\left(\omega_{3}\right)}}$$^{\text{a}}$ &  & 374 & 386 & 452 &  & \tabularnewline
\hline 
(1R) &  & 550 &  & 532 &  & 630\tabularnewline
\hline 
$\text{\ensuremath{\left(1R,\omega_{3}\right)}}$ & 928 & 918 & 913 & 920 &  & \tabularnewline
\hline 
(1z) & 1047 & 1033 & 1050 & 952 & 1070 & 861\tabularnewline
\hline 
(2R) &  & 1069 &  & 1008 &  & \tabularnewline
\hline 
$\text{\ensuremath{\left(1z,1R\right)}}$ & 1470 & 1411 & 1392 & 1520 & 1600 & \tabularnewline
\hline 
bg &  & 1606 &  & 1668 & 1604 & 1720\tabularnewline
\hline 
bu & 1763 & 1756 & 1756 & 1768 & 1781 & 1770\tabularnewline
\hline 
sg &  & 3607 &  & 3650 & 3610 & 3744\tabularnewline
\hline 
su & 3603 & 3614 & 3618 & 3650 & 3625 & 3750\tabularnewline
\hline 
sa & 3683 & 3689 & 3680 & 3720 & 3698 & 3832\tabularnewline
\hline 
MAE &  &  &  & 46 &  & \tabularnewline
\hline 
\end{tabular}
\par\end{centering}
$^{\text{a}}$This assignment of the $\omega_{3}$ wagging excitation
is done upon comparison to the benchmark MCTDH values.
\end{table}

\subsection{``Divide-and-Conquer'' semiclassical dynamics for a high dimensional
molecule: vibrational power spectrum of benzene}

Halverson and Poirier have recently calculated the vibrational frequencies
of benzene using a DVR approach. They pushed the limits of ``exact''
vibrational state calculations up to thirty dimensions.\citep{Halverson_Poirier_Benzene_2015}
In their method, the DVR basis set and grid has been conveniently
selected using phase-space localized basis sets (PSLBs) and truncated
Harmonic functions (HOB).\citep{avila_carrington_punedbases_2012,avila_carrington_exact12D_2011,carter_sharma_multimode_2012}
They were able to obtain all the relevant (about a million) vibrational
energy levels of benzene within a given energy threshold. They employed
a quartic force field modeling for the PES.\citep{handy_laming_benzenepes_1992}

We employ the same surface for a direct comparison between the present
DC-SCIVR method and the exact DVR one. First we study how to best
partition the 30-dimensional space. Using the Hessian-based approach
and $\epsilon=9\cdot10^{-7}$, the full-dimensional space is separated
into one eight-dimensional, eight bi-dimensional and six mono-dimensional
subspaces. When employing the WSV criterion and $\epsilon_{B}=5.6\cdot10^{3}$,
the full-dimensional space is partitioned into one ten-dimensional,
two seven-dimensional and one six-dimensional subspace. When using
the Jacobian-based criterion, the computational search for space decomposition
is much more computationally expensive since all possible combinations
of the 30 vibrational modes into groups of $M$ should be tested.
We restrict instead our search to $6\leq M\leq10$, since the Hessian
criterion shows that when the biggest subspace is eight-dimensional
then the results are quite accurate. We cannot rule out that there
may be a better choice for $M>10$. However, the potential little
improvement in the accuracy of the results does not justify the additional
huge computational overhead.

\begin{figure}
\centering{}\includegraphics[scale=0.5]{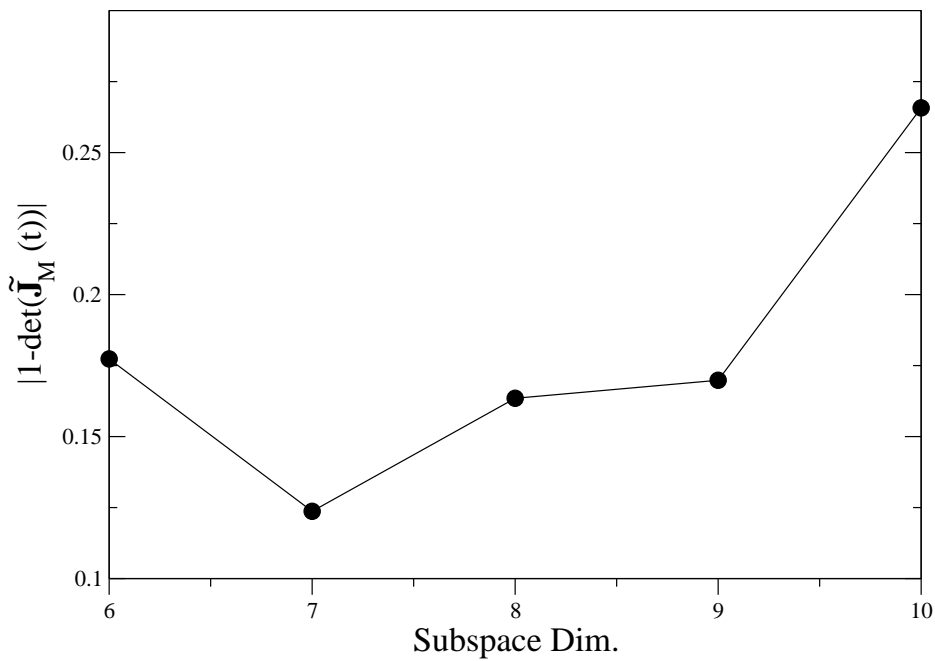}\caption{\label{fig:C6H6_subspaces}Values of $\left|1-\text{det}\left(\tilde{\mathbf{J}}_{M}\left(t\right)\right)\right|$
for different choices of the subspace dimensionality M for the $\text{C}_{6}\text{H}_{6}$
molecule.}
\end{figure}
 Fig.(\ref{fig:C6H6_subspaces}) shows the result of this search and
points to a seven-dimensional subspace for the first partition. The
same procedure is repeated and involves the remaining 23 modes. The
second subspace found is a six-dimensional one. The third search (among
the remaining 17 modes) leads to a ten-dimensional subspace. The remaining
seven modes are collected together within the same subspace. Eventually,
the full thirty dimensional vibrational space has been partitioned
into a ten-dimensional, two seven-dimensional, and one six-dimensional
subspace. Whatever the method employed for partitioning the space,
we run 1000 trajectories per degree of freedom to calculate the frequencies.
Each trajectory is 30,000 a.u. long. To remove any spurious noise
effect, in the Fourier transform we add the same Gaussian filter used
for the Zundel cation. As usual, the reference state of each M-dimensional
subspace is written as $\left|\chi\right\rangle =\prod_{i}^{M}\left|\sqrt{\omega_{i}},q_{i}^{eq}\right\rangle $
, where $\omega_{i}$ are the harmonic frequencies that we report
under the columns ``HO'' in Table (\ref{tab:Benzene_freq}). For
the evolution of the pre-exponential factor (\ref{eq:prefactor})
and its phase calculation we use a recently introduced iterative second-order
approximation.\citep{DiLiberto_Ceotto_Prefactors_2016} This approximation
allows for the calculation of the pre-exponential factor without explicitly
calculate the monodromy matrix elements, and it can be safely employed
for strongly chaotic and high dimensional systems, as in the case
of the benzene molecule.
\begin{figure}
\centering{}\includegraphics[scale=0.6]{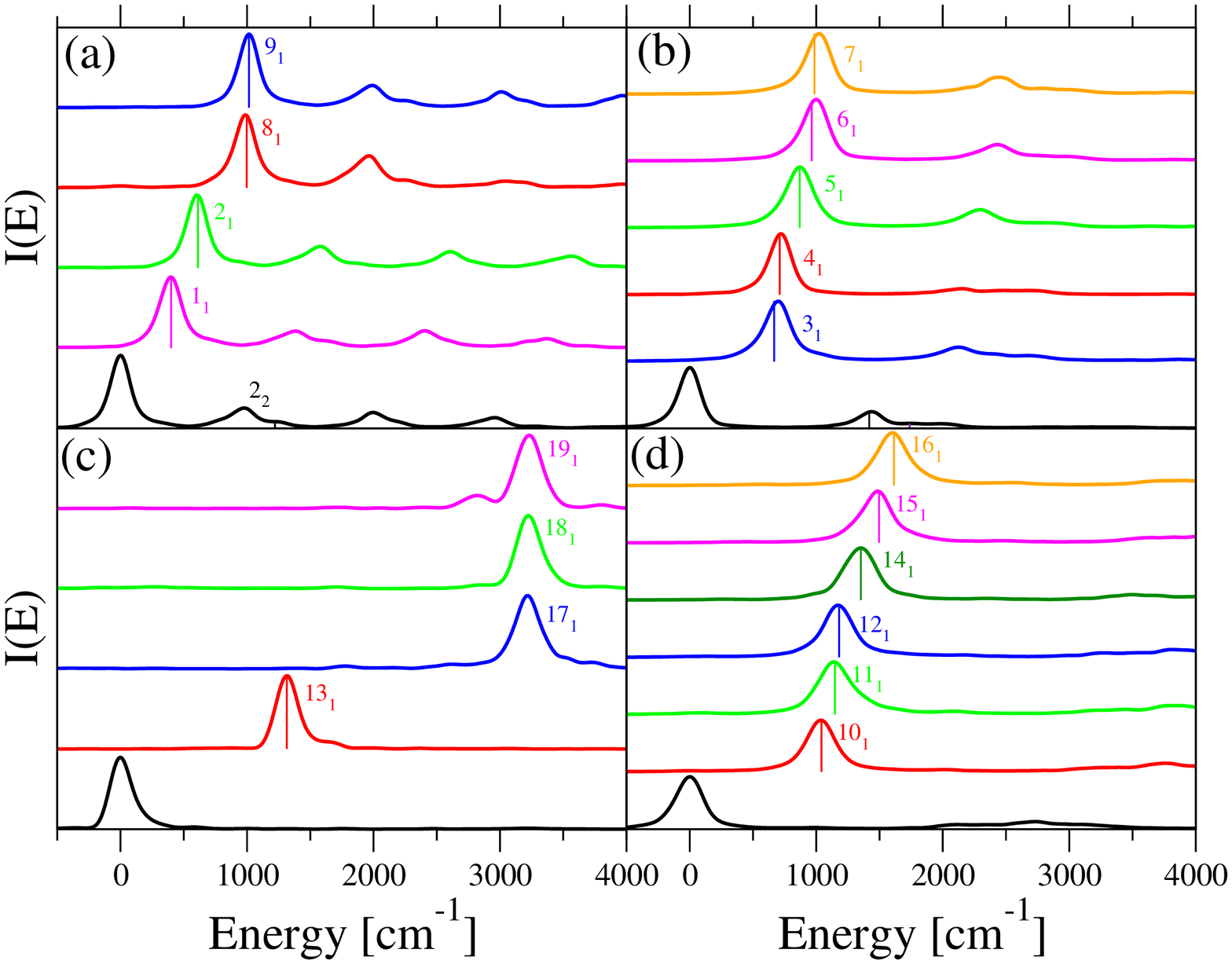}\caption{\label{fig:spectra_benzene}Vibrational spectra of C\protect\textsubscript{6}H\protect\textsubscript{6}
as obtained upon partition of the full-dimensional space according
to the Jacobian criterion. Panel (a) reports the features of the six-dimensional
subspace. Panels (b) and (c) contain the spectra of the two seven-dimensional
subspaces, while panel (d) refers to the 10-dimensional subspace.
The zero point energy value has been shifted to the origin to help
the reader in evaluating the frequencies of the other peaks. The vertical
lines indicate the exact levels from Poirier's EQD calculations.\citep{Halverson_Poirier_Benzene_2015}}
\end{figure}
 Fig. (\ref{fig:spectra_benzene}) shows our computed spectra. Panel
(a) reports the six-dimensional subspace, panels (b) and (c) the seven-dimensional
ones, and panel (d) the 10-dimensional subspace.

\begin{table}
\caption{\label{tab:Benzene_freq}Benzene DC-SCIVR vibrational frequencies
compared with available quantum results (EQD). Degenerate frequencies
are not replicated. Values are given in $\text{cm}^{-1}.$}

\centering{}%
\begin{tabular}{ccccccccccc}
State & HO & DC SCIVR$_{\text{WSV}}$ & DC SCIVR$_{\text{Jacobi}}$ & EQD &  & State & HO & DC SCIVR$_{\text{WSV}}$ & DC SCIVR$_{\text{Jacobi}}$ & EQD\tabularnewline
\hline 
1\textsubscript{1} & 407 & 432 & 399 & 399.4554 &  & 11\textsubscript{1} & 1167 & 1150 & 1144 & 1147.751\tabularnewline
\hline 
2\textsubscript{1} & 613 & 610 & 606 & 611.4227 &  & 12\textsubscript{1} & 1192 & 1189 & 1175 & 1180.374\tabularnewline
\hline 
3\textsubscript{1} & 686 & 610 & 696 & 666.9294 &  & 2\textsubscript{2} & 1226 & 1223 & 1228 & 1221.27\tabularnewline
\hline 
4\textsubscript{1} & 718 & 742 & 719 & 710.7318 &  & 13\textsubscript{1} & 1295 & 1330 & 1314 & 1315.612\tabularnewline
\hline 
5\textsubscript{1} & 866 & 865 & 869 & 868.9106 &  & 14\textsubscript{1} & 1390 & 1375 & 1352 & 1352.563\tabularnewline
\hline 
6\textsubscript{1} & 989 & 990 & 997 & 964.0127 &  & 4\textsubscript{2} & 1436 & 1410 & 1437 & 1418.58\tabularnewline
\hline 
7\textsubscript{1} & 1011 & 1038 & 1020 & 985.8294 &  & 15\textsubscript{1} & 1512 & 1464 & 1492 & 1496.231\tabularnewline
\hline 
8\textsubscript{1} & 1008 & 1002 & 990 & 997.6235 &  & 16\textsubscript{1} & 1639 & 1614 & 1602 & 1614.455\tabularnewline
\hline 
9\textsubscript{1} & 1024 & 1014 & 1014 & 1015.64 &  & 5\textsubscript{2} & 1732 & / & 1752 & 1737.51\tabularnewline
\hline 
10\textsubscript{1} & 1058 & 1042 & 1042 & 1040.98 &  & MAE &  & 15 & 9 & \tabularnewline
\hline 
\end{tabular}
\end{table}
We follow Halverson and Poirier in their labeling of vibrational states.
Table (\ref{tab:Benzene_freq}) reports our computed energy levels
compared with the available exact ones. We find an excellent agreement
with a MAE of only 9 wavenumbers when adopting Jacobi's criterion.
With the WSV approach, the MAE increases to $15\text{cm}^{-1}$. As
we have recently reported,\citep{ceotto_conte_DCSCIVR_2017} the Hessian
criterion leads to still acceptable but less accurate results, with
a MAE of 19 wavenumbers. 

Despite the increase in dimensionality, we conclude that moving from
the three smaller molecular systems of the previous section to benzene,
the MAE referred to the exact results is anyway limited to 10-20 $\text{cm}{}^{-1}$,
a proof of the reliability of DC SCIVR and of the accuracy of the
new Jacobian criterion.

\section{Summary and Conclusions\label{sec:Conclusions}}

All quantum mechanical methods suffer from the curse of dimensionality.
In this paper we have illustrated a method to deal with it and to
obtain vibrational frequencies almost as accurate as in standard SCIVR
simulations, \emph{i.e.} just a few wavenumbers away from the exact
quantum values. More specifically, a ``\emph{divide et impera}''
strategy has been adopted, in which spectra are calculated in partial
dimensionality even if they are still based on full-dimensional classical
trajectories. The method does not take advantage in any way of molecular
symmetry. 

\begin{figure}
\centering{}\includegraphics[scale=0.5]{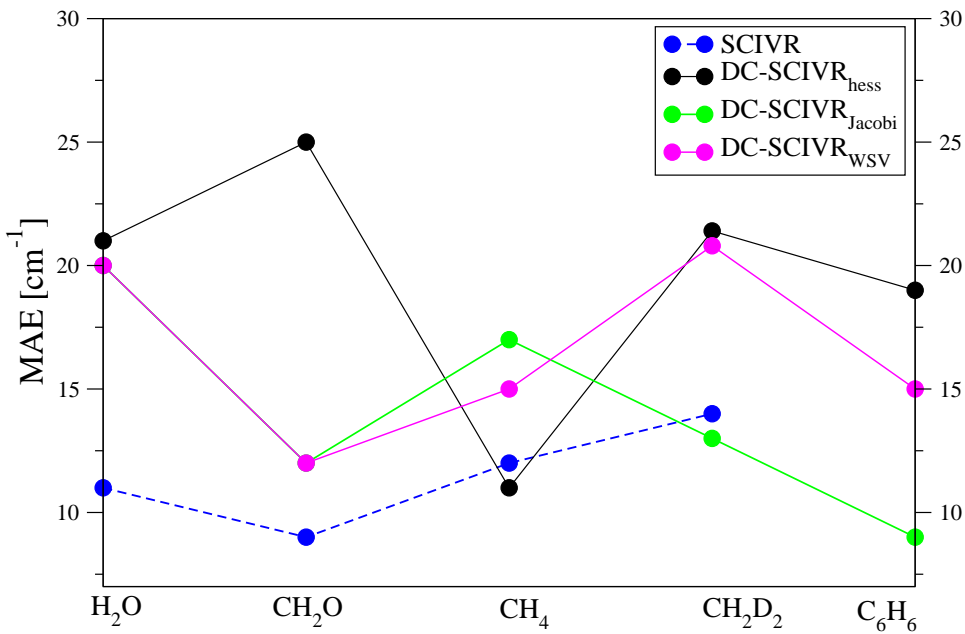}\caption{Trend of the mean absolute error (MAE) with respect to exact results
for the different molecules investigated. Results refer to full-dimensional
SC-IVR calculations (blue), DC SCIVR with Hessian matrix criterion
(black), DC-SCIVR with Jacobian matrix criterion (green), and DC-SCIVR
with WSV subspace partition (magenta).\label{fig:Trend_MAE_molecules} }
\end{figure}

We have shown how crucial the choice of the criterium for the decomposition
of the full-dimensional space into mutually disjoint subspaces can
be. In particular, the partition procedure based on the Jacobian matrix
is the one that usually minimizes the error in approximating the full-dimensional
pre-exponential factor as the direct product of several reduced dimensionality
ones. This is evident from Fig. (\ref{fig:Trend_MAE_molecules}) where
DC-SCIVR\textsubscript{Jacobi} is clearly the overall more accurate
way to decompose the vibrational space. The exception of CH\textsubscript{4}
is due to a not very accurate estimate of a single overtone which
we have anyway included in the MAE calculation, while the Jacobian-based
partition strategy remains the most accurate even for CH\textsubscript{4}
as far as fundamental frequencies are concerned. The apparent better
accuracy of DC-SCIVR\textsubscript{Jacobi} with respect to the full-dimensional
calculation for CH\textsubscript{2}D\textsubscript{2} is to be ascribed
instead to an accidental compensation of errors between the semiclassical
and subspace-partition approximations. Another key advantage of the
Jacobian-based approach lies on its less noisy spectra with better
resolved peaks, which is going to be more and more evident and helpful
as the dimensionality of the system increases. Remarkably, the Jacobi
criterion provides an internally consistent method to check the reliability
of the subspace partition. In fact, not always an increase in the
subspace dimensionality leads to more accurate vibrational frequencies.
On the contrary, spectra can be noisier or it could be even impossibile
to collect a sensible spectral signal. The partitioning schemes here
developed can be also adopted for on-the-fly DC-SCIVR calculations.
In fact, upon calculation of the test trajectory and of the associated
Hessians and monodromy matrix elements by means of ab initio molecular
dynamics, it is possible to determine the best subspace partition
by following exactly the same procedures and at no additional cost
with respect to DC-SCIVR simulations based on analytical potential
energy surfaces.

DC-SCIVR, like other semiclassical and classical methods, is based
on the Fourier transform of a survival amplitude. According to Nyquist's
theorem, for a total evolution time T a peak width equal to $ 2\pi /T $
should be expected. In our simulations, though, other factors contribute
to increase the width of the spectral features. The ro-vibrational
coupling generate a vibrational angular momentum which perturbs the
pure vibrational motion. Furthermore, when a Gaussian filter is employed,
peaks may be substantially enlarged (as in the case of H\textsubscript{5}O\textsubscript{2}\textsuperscript{+}
and benzene). The full width at half maximum (FWHM) of the peaks provides
a measure of the uncertainty of our results and benchmark values are
always within this uncertainty bar. A potential drawback related to
the width of the peaks is that it may hinder the resolution of spectral
features very close to each other. A practical way to overcome this
issue, that is largely adopted in semiclassical dynamics, consists
in employing a proper combination of coherent states able to introduce
a parity symmetry\citep{Kaledin_Miller_TAmolecules_2003,Tamascelli_Ceotto_GPU_2014}
which permits to distinguish among spectral features belonging to
different vibrational modes.

A known issue of semiclassical spectra is represented by the so-called
\textquotedblleft ghost\textquotedblright{} peaks. These are unphysical
features that can be generally distinguished from the true fundamental
transitions because of their much lower intensity. As shown in Figure
3, this is not a specific drawback of DC-SCIVR simulations since full-dimensional
calculations present the same issue. The adoption of a combination
of coherent states able to account for the parity symmetry further
enhances this discrepancy in the intensities making the identification
of the true vibrational features even more favored.

DC SCIVR can be employed to simulate all kinds of spectroscopies relative
to the nuclear motion, such as IR, Raman, absorption/emission dipole,
vibro-electronic, and photodetachment spectra. It will allow to read
each part of the spectra in a wider molecular context up to the nanoscale,
with inclusion of non-trivial long-range quantum interactions. The
calculation of partial spectra representations has not only the advantage
to accelerate the Monte Carlo integration by virtue of the reduced
dimensionality of each subspace and to get better resolved spectra,
but also simplifies the identification of each peak. Another potential
application of DC SCIVR is in the field of mixed (hybrid) semiclassical
methods\citep{Grossmann_SChybrid_2006,Buchholz_Ceotto_MixedSC_2016,Ceotto_Buchholz_MixedSC_2017}
due to the possibility to assign different degrees of freedom to the
different semiclassical techniques employed.

In conclusion, we think that semiclassical molecular dynamics is a
very convenient approach for quantum mechanical simulations of nuclear
vibrational spectroscopy. Future challenges, concerning the study
of vibrational features of large molecules involved in biological
mechanisms and technological processes, will be tackled in a novel
quantum-mechanical fashion thanks to DC SCIVR and the implementation
of the newly proposed subspace-separation criterion. 

\section*{Acknowledgments}

Professor Bill Poirier is gratefully acknowledged for providing the
potential energy surface of benzene and the results of his quantum
simulations. Profs. Jiri Vaní\v{c}ek and Frank Grossmann, and Dr.
Max Buchholz are warmly thanked for their comments on a preliminary
draft of the paper, and an anonymous referee is thanked for suggesting
the water and Zundel cation applications. We acknowledge financial
support from the European Research Council (ERC) under the European
Union\textquoteright s Horizon 2020 research and innovation programme
(grant agreement No {[}647107{]} \textendash{} SEMICOMPLEX \textendash{}
ERC-2014-CoG). We thank Università degli Studi di Milano for further
computational time at CINECA (Italian Supercomputing Center) and the
Regione Lombardia award under the LISA initiative (grant GREENTI)
for the availability of high performance computing resources.

\newpage

\bibliographystyle{aipnum4-1}
\bibliography{SEMICOMPLEX}

\begin{thebibliography}{121}%
\makeatletter
\providecommand \@ifxundefined [1]{%
 \@ifx{#1\undefined}
}%
\providecommand \@ifnum [1]{%
 \ifnum #1\expandafter \@firstoftwo
 \else \expandafter \@secondoftwo
 \fi
}%
\providecommand \@ifx [1]{%
 \ifx #1\expandafter \@firstoftwo
 \else \expandafter \@secondoftwo
 \fi
}%
\providecommand \natexlab [1]{#1}%
\providecommand \enquote  [1]{``#1''}%
\providecommand \bibnamefont  [1]{#1}%
\providecommand \bibfnamefont [1]{#1}%
\providecommand \citenamefont [1]{#1}%
\providecommand \href@noop [0]{\@secondoftwo}%
\providecommand \href [0]{\begingroup \@sanitize@url \@href}%
\providecommand \@href[1]{\@@startlink{#1}\@@href}%
\providecommand \@@href[1]{\endgroup#1\@@endlink}%
\providecommand \@sanitize@url [0]{\catcode `\\12\catcode `\$12\catcode
  `\&12\catcode `\#12\catcode `\^12\catcode `\_12\catcode `\%12\relax}%
\providecommand \@@startlink[1]{}%
\providecommand \@@endlink[0]{}%
\providecommand \url  [0]{\begingroup\@sanitize@url \@url }%
\providecommand \@url [1]{\endgroup\@href {#1}{\urlprefix }}%
\providecommand \urlprefix  [0]{URL }%
\providecommand \Eprint [0]{\href }%
\providecommand \doibase [0]{http://dx.doi.org/}%
\providecommand \selectlanguage [0]{\@gobble}%
\providecommand \bibinfo  [0]{\@secondoftwo}%
\providecommand \bibfield  [0]{\@secondoftwo}%
\providecommand \translation [1]{[#1]}%
\providecommand \BibitemOpen [0]{}%
\providecommand \bibitemStop [0]{}%
\providecommand \bibitemNoStop [0]{.\EOS\space}%
\providecommand \EOS [0]{\spacefactor3000\relax}%
\providecommand \BibitemShut  [1]{\csname bibitem#1\endcsname}%
\let\auto@bib@innerbib\@empty
\bibitem [{\citenamefont {Bowman}, \citenamefont {Carrington},\ and\
  \citenamefont {Meyer}(2008)}]{Bowman_Meyer_Polyatomic_2008}%
  \BibitemOpen
  \bibfield  {author} {\bibinfo {author} {\bibfnamefont {J.~M.}\ \bibnamefont
  {Bowman}}, \bibinfo {author} {\bibfnamefont {T.}~\bibnamefont {Carrington}},
  \ and\ \bibinfo {author} {\bibfnamefont {H.-D.}\ \bibnamefont {Meyer}},\
  }\href@noop {} {\bibfield  {journal} {\bibinfo  {journal} {Molecular
  Physics}\ }\textbf {\bibinfo {volume} {106}},\ \bibinfo {pages} {2145}
  (\bibinfo {year} {2008})}\BibitemShut {NoStop}%
\bibitem [{\citenamefont {Avila}\ and\ \citenamefont
  {Carrington~Jr}(2011{\natexlab{a}})}]{Avila_Carrington_C2H4_2011}%
  \BibitemOpen
  \bibfield  {author} {\bibinfo {author} {\bibfnamefont {G.}~\bibnamefont
  {Avila}}\ and\ \bibinfo {author} {\bibfnamefont {T.}~\bibnamefont
  {Carrington~Jr}},\ }\href@noop {} {\bibfield  {journal} {\bibinfo  {journal}
  {J. Chem. Phys.}\ }\textbf {\bibinfo {volume} {135}},\ \bibinfo {pages}
  {064101} (\bibinfo {year} {2011}{\natexlab{a}})}\BibitemShut {NoStop}%
\bibitem [{\citenamefont {Avila}\ and\ \citenamefont
  {Carrington~Jr}(2011{\natexlab{b}})}]{avila_carrington_exact12D_2011}%
  \BibitemOpen
  \bibfield  {author} {\bibinfo {author} {\bibfnamefont {G.}~\bibnamefont
  {Avila}}\ and\ \bibinfo {author} {\bibfnamefont {T.}~\bibnamefont
  {Carrington~Jr}},\ }\href@noop {} {\bibfield  {journal} {\bibinfo  {journal}
  {J. Chem. Phys.}\ }\textbf {\bibinfo {volume} {134}},\ \bibinfo {pages}
  {054126} (\bibinfo {year} {2011}{\natexlab{b}})}\BibitemShut {NoStop}%
\bibitem [{\citenamefont {Avila}\ and\ \citenamefont
  {Carrington~Jr}(2012)}]{avila_carrington_punedbases_2012}%
  \BibitemOpen
  \bibfield  {author} {\bibinfo {author} {\bibfnamefont {G.}~\bibnamefont
  {Avila}}\ and\ \bibinfo {author} {\bibfnamefont {T.}~\bibnamefont
  {Carrington~Jr}},\ }\href@noop {} {\bibfield  {journal} {\bibinfo  {journal}
  {J. Chem. Phys.}\ }\textbf {\bibinfo {volume} {137}},\ \bibinfo {pages}
  {174108} (\bibinfo {year} {2012})}\BibitemShut {NoStop}%
\bibitem [{\citenamefont {Thomas}\ and\ \citenamefont
  {Carrington~Jr}(2015)}]{Thomas_Carrington_SevenAtoms_2015}%
  \BibitemOpen
  \bibfield  {author} {\bibinfo {author} {\bibfnamefont {P.~S.}\ \bibnamefont
  {Thomas}}\ and\ \bibinfo {author} {\bibfnamefont {T.}~\bibnamefont
  {Carrington~Jr}},\ }\href@noop {} {\bibfield  {journal} {\bibinfo  {journal}
  {The Journal of Physical Chemistry A}\ }\textbf {\bibinfo {volume} {119}},\
  \bibinfo {pages} {13074} (\bibinfo {year} {2015})}\BibitemShut {NoStop}%
\bibitem [{\citenamefont {Barone}(2005)}]{Barone_AutomatedVPT2_2005}%
  \BibitemOpen
  \bibfield  {author} {\bibinfo {author} {\bibfnamefont {V.}~\bibnamefont
  {Barone}},\ }\href {\doibase http://dx.doi.org/10.1063/1.1824881} {\bibfield
  {journal} {\bibinfo  {journal} {J. Chem. Phys.}\ }\textbf {\bibinfo {volume}
  {122}},\ \bibinfo {pages} {014108} (\bibinfo {year} {2005})}\BibitemShut
  {NoStop}%
\bibitem [{\citenamefont {Puzzarini}, \citenamefont {Biczysko},\ and\
  \citenamefont {Barone}(2010)}]{Puzzarini_Barone_Open-shell_2010}%
  \BibitemOpen
  \bibfield  {author} {\bibinfo {author} {\bibfnamefont {C.}~\bibnamefont
  {Puzzarini}}, \bibinfo {author} {\bibfnamefont {M.}~\bibnamefont {Biczysko}},
  \ and\ \bibinfo {author} {\bibfnamefont {V.}~\bibnamefont {Barone}},\ }\href
  {\doibase 10.1021/ct900594h} {\bibfield  {journal} {\bibinfo  {journal} {J.
  Chem. Theory Comput.}\ }\textbf {\bibinfo {volume} {6}},\ \bibinfo {pages}
  {828} (\bibinfo {year} {2010})}\BibitemShut {NoStop}%
\bibitem [{\citenamefont {Puzzarini}, \citenamefont {Biczysko},\ and\
  \citenamefont {Barone}(2011)}]{Puzzarini_Barone_Uracil_2011}%
  \BibitemOpen
  \bibfield  {author} {\bibinfo {author} {\bibfnamefont {C.}~\bibnamefont
  {Puzzarini}}, \bibinfo {author} {\bibfnamefont {M.}~\bibnamefont {Biczysko}},
  \ and\ \bibinfo {author} {\bibfnamefont {V.}~\bibnamefont {Barone}},\ }\href
  {\doibase 10.1021/ct200552m} {\bibfield  {journal} {\bibinfo  {journal} {J.
  Chem. Theory Comput.}\ }\textbf {\bibinfo {volume} {7}},\ \bibinfo {pages}
  {3702} (\bibinfo {year} {2011})}\BibitemShut {NoStop}%
\bibitem [{\citenamefont {Biczysko}\ \emph {et~al.}(2012)\citenamefont
  {Biczysko}, \citenamefont {Bloino}, \citenamefont {Carnimeo}, \citenamefont
  {Panek},\ and\ \citenamefont {Barone}}]{Biczysko_Barone_abinitioIRgly_2012}%
  \BibitemOpen
  \bibfield  {author} {\bibinfo {author} {\bibfnamefont {M.}~\bibnamefont
  {Biczysko}}, \bibinfo {author} {\bibfnamefont {J.}~\bibnamefont {Bloino}},
  \bibinfo {author} {\bibfnamefont {I.}~\bibnamefont {Carnimeo}}, \bibinfo
  {author} {\bibfnamefont {P.}~\bibnamefont {Panek}}, \ and\ \bibinfo {author}
  {\bibfnamefont {V.}~\bibnamefont {Barone}},\ }\href {\doibase
  http://dx.doi.org/10.1016/j.molstruc.2011.10.012} {\bibfield  {journal}
  {\bibinfo  {journal} {J. Mol. Struct.}\ }\textbf {\bibinfo {volume} {1009}},\
  \bibinfo {pages} {74 } (\bibinfo {year} {2012})}\BibitemShut {NoStop}%
\bibitem [{\citenamefont {Bludsky}\ \emph {et~al.}(2000)\citenamefont
  {Bludsky}, \citenamefont {Chocholousova}, \citenamefont {Vacek},
  \citenamefont {Huisken},\ and\ \citenamefont
  {Hobza}}]{Bludsky_Hobza_Anharmonicgly_2000}%
  \BibitemOpen
  \bibfield  {author} {\bibinfo {author} {\bibfnamefont {O.}~\bibnamefont
  {Bludsky}}, \bibinfo {author} {\bibfnamefont {J.}~\bibnamefont
  {Chocholousova}}, \bibinfo {author} {\bibfnamefont {J.}~\bibnamefont
  {Vacek}}, \bibinfo {author} {\bibfnamefont {F.}~\bibnamefont {Huisken}}, \
  and\ \bibinfo {author} {\bibfnamefont {P.}~\bibnamefont {Hobza}},\ }\href
  {\doibase http://dx.doi.org/10.1063/1.1288914} {\bibfield  {journal}
  {\bibinfo  {journal} {J. Chem. Phys.}\ }\textbf {\bibinfo {volume} {113}},\
  \bibinfo {pages} {4629} (\bibinfo {year} {2000})}\BibitemShut {NoStop}%
\bibitem [{\citenamefont {Bloino}, \citenamefont {Baiardi},\ and\ \citenamefont
  {Biczysko}(2016)}]{bloino_Biczysko_vibronic_2016}%
  \BibitemOpen
  \bibfield  {author} {\bibinfo {author} {\bibfnamefont {J.}~\bibnamefont
  {Bloino}}, \bibinfo {author} {\bibfnamefont {A.}~\bibnamefont {Baiardi}}, \
  and\ \bibinfo {author} {\bibfnamefont {M.}~\bibnamefont {Biczysko}},\
  }\href@noop {} {\bibfield  {journal} {\bibinfo  {journal} {Int. J. Quantum
  Chem.}\ }\textbf {\bibinfo {volume} {116}},\ \bibinfo {pages} {1543}
  (\bibinfo {year} {2016})}\BibitemShut {NoStop}%
\bibitem [{\citenamefont {Petersen}\ \emph {et~al.}(2005)\citenamefont
  {Petersen}, \citenamefont {Wang}, \citenamefont {Blake}, \citenamefont
  {Metiu},\ and\ \citenamefont {Voth}}]{petersen_Voth_pocket_2005}%
  \BibitemOpen
  \bibfield  {author} {\bibinfo {author} {\bibfnamefont {M.~K.}\ \bibnamefont
  {Petersen}}, \bibinfo {author} {\bibfnamefont {F.}~\bibnamefont {Wang}},
  \bibinfo {author} {\bibfnamefont {N.~P.}\ \bibnamefont {Blake}}, \bibinfo
  {author} {\bibfnamefont {H.}~\bibnamefont {Metiu}}, \ and\ \bibinfo {author}
  {\bibfnamefont {G.~A.}\ \bibnamefont {Voth}},\ }\href@noop {} {\bibfield
  {journal} {\bibinfo  {journal} {J. Phys. Chem. B}\ }\textbf {\bibinfo
  {volume} {109}},\ \bibinfo {pages} {3727} (\bibinfo {year}
  {2005})}\BibitemShut {NoStop}%
\bibitem [{\citenamefont {Vanommeslaeghe}\ \emph {et~al.}(2010)\citenamefont
  {Vanommeslaeghe}, \citenamefont {Hatcher}, \citenamefont {Acharya},
  \citenamefont {Kundu}, \citenamefont {Zhong}, \citenamefont {Shim},
  \citenamefont {Darian}, \citenamefont {Guvench}, \citenamefont {Lopes},
  \citenamefont {Vorobyov} \emph
  {et~al.}}]{vanommeslaeghe_Mackerell_charmm_2010}%
  \BibitemOpen
  \bibfield  {author} {\bibinfo {author} {\bibfnamefont {K.}~\bibnamefont
  {Vanommeslaeghe}}, \bibinfo {author} {\bibfnamefont {E.}~\bibnamefont
  {Hatcher}}, \bibinfo {author} {\bibfnamefont {C.}~\bibnamefont {Acharya}},
  \bibinfo {author} {\bibfnamefont {S.}~\bibnamefont {Kundu}}, \bibinfo
  {author} {\bibfnamefont {S.}~\bibnamefont {Zhong}}, \bibinfo {author}
  {\bibfnamefont {J.}~\bibnamefont {Shim}}, \bibinfo {author} {\bibfnamefont
  {E.}~\bibnamefont {Darian}}, \bibinfo {author} {\bibfnamefont
  {O.}~\bibnamefont {Guvench}}, \bibinfo {author} {\bibfnamefont
  {P.}~\bibnamefont {Lopes}}, \bibinfo {author} {\bibfnamefont
  {I.}~\bibnamefont {Vorobyov}},  \emph {et~al.},\ }\href@noop {} {\bibfield
  {journal} {\bibinfo  {journal} {J. Comput. Chem.}\ }\textbf {\bibinfo
  {volume} {31}},\ \bibinfo {pages} {671} (\bibinfo {year} {2010})}\BibitemShut
  {NoStop}%
\bibitem [{\citenamefont {Wang}\ \emph {et~al.}(2004)\citenamefont {Wang},
  \citenamefont {Wolf}, \citenamefont {Caldwell}, \citenamefont {Kollman},\
  and\ \citenamefont {Case}}]{wang_Case_gaffFF_2004}%
  \BibitemOpen
  \bibfield  {author} {\bibinfo {author} {\bibfnamefont {J.}~\bibnamefont
  {Wang}}, \bibinfo {author} {\bibfnamefont {R.~M.}\ \bibnamefont {Wolf}},
  \bibinfo {author} {\bibfnamefont {J.~W.}\ \bibnamefont {Caldwell}}, \bibinfo
  {author} {\bibfnamefont {P.~A.}\ \bibnamefont {Kollman}}, \ and\ \bibinfo
  {author} {\bibfnamefont {D.~A.}\ \bibnamefont {Case}},\ }\href@noop {}
  {\bibfield  {journal} {\bibinfo  {journal} {J. Comput. Chem.}\ }\textbf
  {\bibinfo {volume} {25}},\ \bibinfo {pages} {1157} (\bibinfo {year}
  {2004})}\BibitemShut {NoStop}%
\bibitem [{\citenamefont {Mathias}\ \emph {et~al.}(2011)\citenamefont
  {Mathias}, \citenamefont {Ivanov}, \citenamefont {Witt}, \citenamefont
  {Baer},\ and\ \citenamefont {Marx}}]{marx_mathias_IRfluxional_2011}%
  \BibitemOpen
  \bibfield  {author} {\bibinfo {author} {\bibfnamefont {G.}~\bibnamefont
  {Mathias}}, \bibinfo {author} {\bibfnamefont {S.~D.}\ \bibnamefont {Ivanov}},
  \bibinfo {author} {\bibfnamefont {A.}~\bibnamefont {Witt}}, \bibinfo {author}
  {\bibfnamefont {M.~D.}\ \bibnamefont {Baer}}, \ and\ \bibinfo {author}
  {\bibfnamefont {D.}~\bibnamefont {Marx}},\ }\href@noop {} {\bibfield
  {journal} {\bibinfo  {journal} {J. Chem. Theory. Comput.}\ }\textbf {\bibinfo
  {volume} {8}},\ \bibinfo {pages} {224} (\bibinfo {year} {2011})}\BibitemShut
  {NoStop}%
\bibitem [{\citenamefont
  {Gaigeot}(2010)}]{gaigeot_gaigeot_floppypeptides_2010}%
  \BibitemOpen
  \bibfield  {author} {\bibinfo {author} {\bibfnamefont {M.-P.}\ \bibnamefont
  {Gaigeot}},\ }\href@noop {} {\bibfield  {journal} {\bibinfo  {journal} {Phys.
  Chem. Chem. Phys.}\ }\textbf {\bibinfo {volume} {12}},\ \bibinfo {pages}
  {3336} (\bibinfo {year} {2010})}\BibitemShut {NoStop}%
\bibitem [{\citenamefont {Thomas}\ \emph {et~al.}(2013)\citenamefont {Thomas},
  \citenamefont {Brehm}, \citenamefont {Fligg}, \citenamefont {V{\"o}hringer},\
  and\ \citenamefont {Kirchner}}]{thomas_Kirchner_vibroAIMD_2013}%
  \BibitemOpen
  \bibfield  {author} {\bibinfo {author} {\bibfnamefont {M.}~\bibnamefont
  {Thomas}}, \bibinfo {author} {\bibfnamefont {M.}~\bibnamefont {Brehm}},
  \bibinfo {author} {\bibfnamefont {R.}~\bibnamefont {Fligg}}, \bibinfo
  {author} {\bibfnamefont {P.}~\bibnamefont {V{\"o}hringer}}, \ and\ \bibinfo
  {author} {\bibfnamefont {B.}~\bibnamefont {Kirchner}},\ }\href@noop {}
  {\bibfield  {journal} {\bibinfo  {journal} {Phys. Chem. Chem. Phys.}\
  }\textbf {\bibinfo {volume} {15}},\ \bibinfo {pages} {6608} (\bibinfo {year}
  {2013})}\BibitemShut {NoStop}%
\bibitem [{\citenamefont {Gomez~Llorente}\ and\ \citenamefont
  {Pollak}(1992)}]{Pollak_gomez_spectroscopy_1992}%
  \BibitemOpen
  \bibfield  {author} {\bibinfo {author} {\bibfnamefont {J.}~\bibnamefont
  {Gomez~Llorente}}\ and\ \bibinfo {author} {\bibfnamefont {E.}~\bibnamefont
  {Pollak}},\ }\href@noop {} {\bibfield  {journal} {\bibinfo  {journal} {Annu.
  Rev. Phys. Chem.}\ }\textbf {\bibinfo {volume} {43}},\ \bibinfo {pages} {91}
  (\bibinfo {year} {1992})}\BibitemShut {NoStop}%
\bibitem [{\citenamefont {Iftimie}, \citenamefont {Minary},\ and\ \citenamefont
  {Tuckerman}(2005)}]{Tuckerman_iftimie_AIMD_2005}%
  \BibitemOpen
  \bibfield  {author} {\bibinfo {author} {\bibfnamefont {R.}~\bibnamefont
  {Iftimie}}, \bibinfo {author} {\bibfnamefont {P.}~\bibnamefont {Minary}}, \
  and\ \bibinfo {author} {\bibfnamefont {M.~E.}\ \bibnamefont {Tuckerman}},\
  }\href@noop {} {\bibfield  {journal} {\bibinfo  {journal} {Proc. Natl. Acad.
  Sci.}\ }\textbf {\bibinfo {volume} {102}},\ \bibinfo {pages} {6654} (\bibinfo
  {year} {2005})}\BibitemShut {NoStop}%
\bibitem [{\citenamefont {Pratihar}\ \emph {et~al.}(2017)\citenamefont
  {Pratihar}, \citenamefont {Ma}, \citenamefont {Homayoon}, \citenamefont
  {Barnes},\ and\ \citenamefont {Hase}}]{hase_pratihar_directdynamics_2017}%
  \BibitemOpen
  \bibfield  {author} {\bibinfo {author} {\bibfnamefont {S.}~\bibnamefont
  {Pratihar}}, \bibinfo {author} {\bibfnamefont {X.}~\bibnamefont {Ma}},
  \bibinfo {author} {\bibfnamefont {Z.}~\bibnamefont {Homayoon}}, \bibinfo
  {author} {\bibfnamefont {G.~L.}\ \bibnamefont {Barnes}}, \ and\ \bibinfo
  {author} {\bibfnamefont {W.~L.}\ \bibnamefont {Hase}},\ }\href@noop {}
  {\bibfield  {journal} {\bibinfo  {journal} {J. Am. Chem. Soc.}\ }\textbf
  {\bibinfo {volume} {139}},\ \bibinfo {pages} {3570} (\bibinfo {year}
  {2017})}\BibitemShut {NoStop}%
\bibitem [{\citenamefont {Schlegel}\ \emph {et~al.}(2001)\citenamefont
  {Schlegel}, \citenamefont {Millam}, \citenamefont {Iyengar}, \citenamefont
  {Voth}, \citenamefont {Daniels}, \citenamefont {Scuseria},\ and\
  \citenamefont {Frisch}}]{schlegel_Frisch_AIMD_2001}%
  \BibitemOpen
  \bibfield  {author} {\bibinfo {author} {\bibfnamefont {H.~B.}\ \bibnamefont
  {Schlegel}}, \bibinfo {author} {\bibfnamefont {J.~M.}\ \bibnamefont
  {Millam}}, \bibinfo {author} {\bibfnamefont {S.~S.}\ \bibnamefont {Iyengar}},
  \bibinfo {author} {\bibfnamefont {G.~A.}\ \bibnamefont {Voth}}, \bibinfo
  {author} {\bibfnamefont {A.~D.}\ \bibnamefont {Daniels}}, \bibinfo {author}
  {\bibfnamefont {G.~E.}\ \bibnamefont {Scuseria}}, \ and\ \bibinfo {author}
  {\bibfnamefont {M.~J.}\ \bibnamefont {Frisch}},\ }\href@noop {} {\bibfield
  {journal} {\bibinfo  {journal} {J. Chem. Phys.}\ }\textbf {\bibinfo {volume}
  {114}},\ \bibinfo {pages} {9758} (\bibinfo {year} {2001})}\BibitemShut
  {NoStop}%
\bibitem [{\citenamefont
  {Heller}(1981{\natexlab{a}})}]{Heller_SCspectroscopy_1981}%
  \BibitemOpen
  \bibfield  {author} {\bibinfo {author} {\bibfnamefont {E.~J.}\ \bibnamefont
  {Heller}},\ }\href@noop {} {\bibfield  {journal} {\bibinfo  {journal} {Acc.
  Chem. Res.}\ }\textbf {\bibinfo {volume} {14}},\ \bibinfo {pages} {368}
  (\bibinfo {year} {1981}{\natexlab{a}})}\BibitemShut {NoStop}%
\bibitem [{\citenamefont {Herman}\ and\ \citenamefont
  {Kluk}(1984)}]{Herman_Kluk_SCnonspreading_1984}%
  \BibitemOpen
  \bibfield  {author} {\bibinfo {author} {\bibfnamefont {M.~F.}\ \bibnamefont
  {Herman}}\ and\ \bibinfo {author} {\bibfnamefont {E.}~\bibnamefont {Kluk}},\
  }\href {\doibase http://dx.doi.org/10.1016/0301-0104(84)80039-7} {\bibfield
  {journal} {\bibinfo  {journal} {Chem. Phys.}\ }\textbf {\bibinfo {volume}
  {91}},\ \bibinfo {pages} {27 } (\bibinfo {year} {1984})}\BibitemShut
  {NoStop}%
\bibitem [{\citenamefont {Antipov}, \citenamefont {Ye},\ and\ \citenamefont
  {Ananth}(2015)}]{Antipov_Nandini_Mixedqcl_2015}%
  \BibitemOpen
  \bibfield  {author} {\bibinfo {author} {\bibfnamefont {S.~V.}\ \bibnamefont
  {Antipov}}, \bibinfo {author} {\bibfnamefont {Z.}~\bibnamefont {Ye}}, \ and\
  \bibinfo {author} {\bibfnamefont {N.}~\bibnamefont {Ananth}},\ }\href@noop {}
  {\bibfield  {journal} {\bibinfo  {journal} {J. Chem. Phys.}\ }\textbf
  {\bibinfo {volume} {142}},\ \bibinfo {pages} {184102} (\bibinfo {year}
  {2015})}\BibitemShut {NoStop}%
\bibitem [{\citenamefont {Walton}\ and\ \citenamefont
  {Manolopoulos}(1995)}]{Walton_Manolopoulos_FrozenGaussianCO2_1995}%
  \BibitemOpen
  \bibfield  {author} {\bibinfo {author} {\bibfnamefont {A.~R.}\ \bibnamefont
  {Walton}}\ and\ \bibinfo {author} {\bibfnamefont {D.~E.}\ \bibnamefont
  {Manolopoulos}},\ }\href@noop {} {\bibfield  {journal} {\bibinfo  {journal}
  {Chem. Phys. Lett.}\ }\textbf {\bibinfo {volume} {244}},\ \bibinfo {pages}
  {448} (\bibinfo {year} {1995})}\BibitemShut {NoStop}%
\bibitem [{\citenamefont {Elran}\ and\ \citenamefont
  {Kay}(1999)}]{Elran_Kay_ImprovingHK_1999}%
  \BibitemOpen
  \bibfield  {author} {\bibinfo {author} {\bibfnamefont {Y.}~\bibnamefont
  {Elran}}\ and\ \bibinfo {author} {\bibfnamefont {K.}~\bibnamefont {Kay}},\
  }\href@noop {} {\bibfield  {journal} {\bibinfo  {journal} {J. Chem. Phys.}\
  }\textbf {\bibinfo {volume} {110}},\ \bibinfo {pages} {3653} (\bibinfo {year}
  {1999})}\BibitemShut {NoStop}%
\bibitem [{\citenamefont {Kay}(1994{\natexlab{a}})}]{Kay_Multidim_1994}%
  \BibitemOpen
  \bibfield  {author} {\bibinfo {author} {\bibfnamefont {K.~G.}\ \bibnamefont
  {Kay}},\ }\href@noop {} {\bibfield  {journal} {\bibinfo  {journal} {J. Chem.
  Phys.}\ }\textbf {\bibinfo {volume} {101}},\ \bibinfo {pages} {2250}
  (\bibinfo {year} {1994}{\natexlab{a}})}\BibitemShut {NoStop}%
\bibitem [{\citenamefont {Kay}(1994{\natexlab{b}})}]{Kay_Numerical_1994}%
  \BibitemOpen
  \bibfield  {author} {\bibinfo {author} {\bibfnamefont {K.~G.}\ \bibnamefont
  {Kay}},\ }\href@noop {} {\bibfield  {journal} {\bibinfo  {journal} {J. Chem.
  Phys.}\ }\textbf {\bibinfo {volume} {100}},\ \bibinfo {pages} {4432}
  (\bibinfo {year} {1994}{\natexlab{b}})}\BibitemShut {NoStop}%
\bibitem [{\citenamefont
  {Kay}(1994{\natexlab{c}})}]{Kay_Integralexpression_1994}%
  \BibitemOpen
  \bibfield  {author} {\bibinfo {author} {\bibfnamefont {K.~G.}\ \bibnamefont
  {Kay}},\ }\href@noop {} {\bibfield  {journal} {\bibinfo  {journal} {J. Chem.
  Phys.}\ }\textbf {\bibinfo {volume} {100}},\ \bibinfo {pages} {4377}
  (\bibinfo {year} {1994}{\natexlab{c}})}\BibitemShut {NoStop}%
\bibitem [{\citenamefont
  {Miller}(1970{\natexlab{a}})}]{Miller_Atom-Diatom_1970}%
  \BibitemOpen
  \bibfield  {author} {\bibinfo {author} {\bibfnamefont {W.~H.}\ \bibnamefont
  {Miller}},\ }\href {\doibase http://dx.doi.org/10.1063/1.1674275} {\bibfield
  {journal} {\bibinfo  {journal} {J. Chem. Phys.}\ }\textbf {\bibinfo {volume}
  {53}},\ \bibinfo {pages} {1949} (\bibinfo {year}
  {1970}{\natexlab{a}})}\BibitemShut {NoStop}%
\bibitem [{\citenamefont {Miller}(1970{\natexlab{b}})}]{Miller_S-Matrix_1970}%
  \BibitemOpen
  \bibfield  {author} {\bibinfo {author} {\bibfnamefont {W.~H.}\ \bibnamefont
  {Miller}},\ }\href {\doibase http://dx.doi.org/10.1063/1.1674535} {\bibfield
  {journal} {\bibinfo  {journal} {J. Chem. Phys.}\ }\textbf {\bibinfo {volume}
  {53}},\ \bibinfo {pages} {3578} (\bibinfo {year}
  {1970}{\natexlab{b}})}\BibitemShut {NoStop}%
\bibitem [{\citenamefont {Church}, \citenamefont {Antipov},\ and\ \citenamefont
  {Ananth}(2017)}]{Nandini_Church_Mixedqcl_2015}%
  \BibitemOpen
  \bibfield  {author} {\bibinfo {author} {\bibfnamefont {M.}~\bibnamefont
  {Church}}, \bibinfo {author} {\bibfnamefont {S.~V.}\ \bibnamefont {Antipov}},
  \ and\ \bibinfo {author} {\bibfnamefont {N.}~\bibnamefont {Ananth}},\
  }\href@noop {} {\bibfield  {journal} {\bibinfo  {journal} {J. Chem. Phys.}\
  }\textbf {\bibinfo {volume} {146}},\ \bibinfo {pages} {234104} (\bibinfo
  {year} {2017})}\BibitemShut {NoStop}%
\bibitem [{\citenamefont {Zhang}\ and\ \citenamefont
  {Pollak}(2004)}]{Zhang_Pollak_Deeptunneling_2004}%
  \BibitemOpen
  \bibfield  {author} {\bibinfo {author} {\bibfnamefont {D.~H.}\ \bibnamefont
  {Zhang}}\ and\ \bibinfo {author} {\bibfnamefont {E.}~\bibnamefont {Pollak}},\
  }\href@noop {} {\bibfield  {journal} {\bibinfo  {journal} {Phys. Rev. Lett.}\
  }\textbf {\bibinfo {volume} {93}},\ \bibinfo {pages} {140401} (\bibinfo
  {year} {2004})}\BibitemShut {NoStop}%
\bibitem [{\citenamefont
  {Miller}(2001)}]{Miller_Addingquantumtoclassical_2001}%
  \BibitemOpen
  \bibfield  {author} {\bibinfo {author} {\bibfnamefont {W.~H.}\ \bibnamefont
  {Miller}},\ }\href@noop {} {\bibfield  {journal} {\bibinfo  {journal} {J.
  Phys. Chem. A}\ }\textbf {\bibinfo {volume} {105}},\ \bibinfo {pages} {2942}
  (\bibinfo {year} {2001})}\BibitemShut {NoStop}%
\bibitem [{\citenamefont {Kay}(2005)}]{Kay_Atomsandmolecules_2005}%
  \BibitemOpen
  \bibfield  {author} {\bibinfo {author} {\bibfnamefont {K.~G.}\ \bibnamefont
  {Kay}},\ }\href@noop {} {\bibfield  {journal} {\bibinfo  {journal} {Annu.
  Rev. Phys. Chem.}\ }\textbf {\bibinfo {volume} {56}},\ \bibinfo {pages} {255}
  (\bibinfo {year} {2005})}\BibitemShut {NoStop}%
\bibitem [{\citenamefont {Shalashilin}\ and\ \citenamefont
  {Child}(2004)}]{Shalashilin_Child_CCS_2004}%
  \BibitemOpen
  \bibfield  {author} {\bibinfo {author} {\bibfnamefont {D.~V.}\ \bibnamefont
  {Shalashilin}}\ and\ \bibinfo {author} {\bibfnamefont {M.~S.}\ \bibnamefont
  {Child}},\ }\href@noop {} {\bibfield  {journal} {\bibinfo  {journal} {Chem.
  Phys.}\ }\textbf {\bibinfo {volume} {304}},\ \bibinfo {pages} {103} (\bibinfo
  {year} {2004})}\BibitemShut {NoStop}%
\bibitem [{\citenamefont {Bonnet}\ and\ \citenamefont
  {Rayez}(1997)}]{bonnet1_rayez_quasiclscattering_1997}%
  \BibitemOpen
  \bibfield  {author} {\bibinfo {author} {\bibfnamefont {L.}~\bibnamefont
  {Bonnet}}\ and\ \bibinfo {author} {\bibfnamefont {J.}~\bibnamefont {Rayez}},\
  }\href@noop {} {\bibfield  {journal} {\bibinfo  {journal} {Chem. Phys.
  Lett.}\ }\textbf {\bibinfo {volume} {277}},\ \bibinfo {pages} {183} (\bibinfo
  {year} {1997})}\BibitemShut {NoStop}%
\bibitem [{\citenamefont {Bonnet}\ and\ \citenamefont
  {Rayez}(2004)}]{bonnet_rayez_Gaussweighting_2004}%
  \BibitemOpen
  \bibfield  {author} {\bibinfo {author} {\bibfnamefont {L.}~\bibnamefont
  {Bonnet}}\ and\ \bibinfo {author} {\bibfnamefont {J.-C.}\ \bibnamefont
  {Rayez}},\ }\href@noop {} {\bibfield  {journal} {\bibinfo  {journal} {Chem.
  Phys. Lett.}\ }\textbf {\bibinfo {volume} {397}},\ \bibinfo {pages} {106}
  (\bibinfo {year} {2004})}\BibitemShut {NoStop}%
\bibitem [{\citenamefont {Crespos}\ \emph {et~al.}(2017)\citenamefont
  {Crespos}, \citenamefont {Decock}, \citenamefont {Larregaray},\ and\
  \citenamefont {Bonnet}}]{crespos_Bonnet_H2Pdscattering_2017}%
  \BibitemOpen
  \bibfield  {author} {\bibinfo {author} {\bibfnamefont {C.}~\bibnamefont
  {Crespos}}, \bibinfo {author} {\bibfnamefont {J.}~\bibnamefont {Decock}},
  \bibinfo {author} {\bibfnamefont {P.}~\bibnamefont {Larregaray}}, \ and\
  \bibinfo {author} {\bibfnamefont {L.}~\bibnamefont {Bonnet}},\ }\href@noop {}
  {\bibfield  {journal} {\bibinfo  {journal} {J. Phys. Chem. C}\ }\textbf
  {\bibinfo {volume} {121}},\ \bibinfo {pages} {16854} (\bibinfo {year}
  {2017})}\BibitemShut {NoStop}%
\bibitem [{\citenamefont {Gu}\ and\ \citenamefont
  {Garashchuk}(2016)}]{Garashchuk_gu_quantumdynamics_2016}%
  \BibitemOpen
  \bibfield  {author} {\bibinfo {author} {\bibfnamefont {B.}~\bibnamefont
  {Gu}}\ and\ \bibinfo {author} {\bibfnamefont {S.}~\bibnamefont
  {Garashchuk}},\ }\href@noop {} {\bibfield  {journal} {\bibinfo  {journal} {J.
  Phys. Chem. A}\ }\textbf {\bibinfo {volume} {120}},\ \bibinfo {pages} {3023}
  (\bibinfo {year} {2016})}\BibitemShut {NoStop}%
\bibitem [{\citenamefont {Garashchuk}, \citenamefont {Rassolov},\ and\
  \citenamefont {Prezhdo}(2011)}]{Garashchuk_Prezhdo_Bohmian_2011}%
  \BibitemOpen
  \bibfield  {author} {\bibinfo {author} {\bibfnamefont {S.}~\bibnamefont
  {Garashchuk}}, \bibinfo {author} {\bibfnamefont {V.}~\bibnamefont
  {Rassolov}}, \ and\ \bibinfo {author} {\bibfnamefont {O.}~\bibnamefont
  {Prezhdo}},\ }\href@noop {} {\bibfield  {journal} {\bibinfo  {journal} {Rev.
  Comput. Chem.}\ }\textbf {\bibinfo {volume} {27}},\ \bibinfo {pages} {287}
  (\bibinfo {year} {2011})}\BibitemShut {NoStop}%
\bibitem [{\citenamefont {Conte}\ and\ \citenamefont
  {Pollak}(2010)}]{Conte_Pollak_ThawedGaussian_2010}%
  \BibitemOpen
  \bibfield  {author} {\bibinfo {author} {\bibfnamefont {R.}~\bibnamefont
  {Conte}}\ and\ \bibinfo {author} {\bibfnamefont {E.}~\bibnamefont {Pollak}},\
  }\href@noop {} {\bibfield  {journal} {\bibinfo  {journal} {Phys. Rev. E}\
  }\textbf {\bibinfo {volume} {81}},\ \bibinfo {pages} {036704} (\bibinfo
  {year} {2010})}\BibitemShut {NoStop}%
\bibitem [{\citenamefont {Conte}\ and\ \citenamefont
  {Pollak}(2012)}]{Conte_Pollak_ContinuumLimit_2012}%
  \BibitemOpen
  \bibfield  {author} {\bibinfo {author} {\bibfnamefont {R.}~\bibnamefont
  {Conte}}\ and\ \bibinfo {author} {\bibfnamefont {E.}~\bibnamefont {Pollak}},\
  }\href@noop {} {\bibfield  {journal} {\bibinfo  {journal} {J. Chem. Phys.}\
  }\textbf {\bibinfo {volume} {136}},\ \bibinfo {pages} {094101} (\bibinfo
  {year} {2012})}\BibitemShut {NoStop}%
\bibitem [{\citenamefont {Kondorskiy}\ and\ \citenamefont
  {Nanbu}(2015)}]{Kondorskiy_Nanbu_Nonadiabatic_2015}%
  \BibitemOpen
  \bibfield  {author} {\bibinfo {author} {\bibfnamefont {A.~D.}\ \bibnamefont
  {Kondorskiy}}\ and\ \bibinfo {author} {\bibfnamefont {S.}~\bibnamefont
  {Nanbu}},\ }\href@noop {} {\bibfield  {journal} {\bibinfo  {journal} {J.
  Chem. Phys.}\ }\textbf {\bibinfo {volume} {143}},\ \bibinfo {pages} {114103}
  (\bibinfo {year} {2015})}\BibitemShut {NoStop}%
\bibitem [{\citenamefont {Nakamura}\ \emph {et~al.}(2016)\citenamefont
  {Nakamura}, \citenamefont {Nanbu}, \citenamefont {Teranishi},\ and\
  \citenamefont {Ohta}}]{Nakamura_Ohta_SCDevelopment_2016}%
  \BibitemOpen
  \bibfield  {author} {\bibinfo {author} {\bibfnamefont {H.}~\bibnamefont
  {Nakamura}}, \bibinfo {author} {\bibfnamefont {S.}~\bibnamefont {Nanbu}},
  \bibinfo {author} {\bibfnamefont {Y.}~\bibnamefont {Teranishi}}, \ and\
  \bibinfo {author} {\bibfnamefont {A.}~\bibnamefont {Ohta}},\ }\href@noop {}
  {\bibfield  {journal} {\bibinfo  {journal} {Phys. Chem. Chem. Phys.}\
  }\textbf {\bibinfo {volume} {18}},\ \bibinfo {pages} {11972} (\bibinfo {year}
  {2016})}\BibitemShut {NoStop}%
\bibitem [{\citenamefont {Koda}(2015)}]{Koda_SCIVRWigner_2015}%
  \BibitemOpen
  \bibfield  {author} {\bibinfo {author} {\bibfnamefont {S.-I.}\ \bibnamefont
  {Koda}},\ }\href@noop {} {\bibfield  {journal} {\bibinfo  {journal} {J. Chem.
  Phys.}\ }\textbf {\bibinfo {volume} {143}},\ \bibinfo {pages} {244110}
  (\bibinfo {year} {2015})}\BibitemShut {NoStop}%
\bibitem [{\citenamefont {Koda}(2016)}]{Koda_Mixedsemiclassical_2016}%
  \BibitemOpen
  \bibfield  {author} {\bibinfo {author} {\bibfnamefont {S.-I.}\ \bibnamefont
  {Koda}},\ }\href@noop {} {\bibfield  {journal} {\bibinfo  {journal} {J. Chem.
  Phys.}\ }\textbf {\bibinfo {volume} {144}},\ \bibinfo {pages} {154108}
  (\bibinfo {year} {2016})}\BibitemShut {NoStop}%
\bibitem [{\citenamefont {Chapman}\ and\ \citenamefont
  {Cina}(2007)}]{Cina_Chapman_2007_SCsmallmolecules}%
  \BibitemOpen
  \bibfield  {author} {\bibinfo {author} {\bibfnamefont {C.~T.}\ \bibnamefont
  {Chapman}}\ and\ \bibinfo {author} {\bibfnamefont {J.~A.}\ \bibnamefont
  {Cina}},\ }\href@noop {} {\bibfield  {journal} {\bibinfo  {journal} {J. Chem.
  Phys.}\ }\textbf {\bibinfo {volume} {127}},\ \bibinfo {pages} {114502}
  (\bibinfo {year} {2007})}\BibitemShut {NoStop}%
\bibitem [{\citenamefont {Chapman}, \citenamefont {Cheng},\ and\ \citenamefont
  {Cina}(2011)}]{Cina_Chapman_2011_smallmolecules}%
  \BibitemOpen
  \bibfield  {author} {\bibinfo {author} {\bibfnamefont {C.~T.}\ \bibnamefont
  {Chapman}}, \bibinfo {author} {\bibfnamefont {X.}~\bibnamefont {Cheng}}, \
  and\ \bibinfo {author} {\bibfnamefont {J.~A.}\ \bibnamefont {Cina}},\
  }\href@noop {} {\bibfield  {journal} {\bibinfo  {journal} {J. Phys. Chem. A}\
  }\textbf {\bibinfo {volume} {115}},\ \bibinfo {pages} {3980} (\bibinfo {year}
  {2011})}\BibitemShut {NoStop}%
\bibitem [{\citenamefont {Cheng}\ and\ \citenamefont
  {Cina}(2014)}]{Cina_cheng_2014_variationalquantumclass}%
  \BibitemOpen
  \bibfield  {author} {\bibinfo {author} {\bibfnamefont {X.}~\bibnamefont
  {Cheng}}\ and\ \bibinfo {author} {\bibfnamefont {J.~A.}\ \bibnamefont
  {Cina}},\ }\href@noop {} {\bibfield  {journal} {\bibinfo  {journal} {J. Chem.
  Phys.}\ }\textbf {\bibinfo {volume} {141}},\ \bibinfo {pages} {034113}
  (\bibinfo {year} {2014})}\BibitemShut {NoStop}%
\bibitem [{\citenamefont {Grossmann}\ and\ \citenamefont
  {Xavier}(1998)}]{Grossmann_Xavier_SCderivation_1998}%
  \BibitemOpen
  \bibfield  {author} {\bibinfo {author} {\bibfnamefont {F.}~\bibnamefont
  {Grossmann}}\ and\ \bibinfo {author} {\bibfnamefont {A.~L.}\ \bibnamefont
  {Xavier}},\ }\href@noop {} {\bibfield  {journal} {\bibinfo  {journal} {Phys.
  Lett. A}\ }\textbf {\bibinfo {volume} {243}},\ \bibinfo {pages} {243}
  (\bibinfo {year} {1998})}\BibitemShut {NoStop}%
\bibitem [{\citenamefont {Harabati}, \citenamefont {Rost},\ and\ \citenamefont
  {Grossmann}(2004)}]{Harabati_Grossmann_LongtimeSCIVR_2004}%
  \BibitemOpen
  \bibfield  {author} {\bibinfo {author} {\bibfnamefont {C.}~\bibnamefont
  {Harabati}}, \bibinfo {author} {\bibfnamefont {J.~M.}\ \bibnamefont {Rost}},
  \ and\ \bibinfo {author} {\bibfnamefont {F.}~\bibnamefont {Grossmann}},\
  }\href@noop {} {\bibfield  {journal} {\bibinfo  {journal} {J. Chem. Phys.}\
  }\textbf {\bibinfo {volume} {120}},\ \bibinfo {pages} {26} (\bibinfo {year}
  {2004})}\BibitemShut {NoStop}%
\bibitem [{\citenamefont {Grossmann}(1999)}]{Grossmann_SConPES_1999}%
  \BibitemOpen
  \bibfield  {author} {\bibinfo {author} {\bibfnamefont {F.}~\bibnamefont
  {Grossmann}},\ }\href@noop {} {\bibfield  {journal} {\bibinfo  {journal}
  {Phys. Rev. A}\ }\textbf {\bibinfo {volume} {60}},\ \bibinfo {pages} {1791}
  (\bibinfo {year} {1999})}\BibitemShut {NoStop}%
\bibitem [{\citenamefont {Bonella}, \citenamefont {Montemayor},\ and\
  \citenamefont {Coker}(2005)}]{Bonella_Coker_Linearizedpathintegral_2005}%
  \BibitemOpen
  \bibfield  {author} {\bibinfo {author} {\bibfnamefont {S.}~\bibnamefont
  {Bonella}}, \bibinfo {author} {\bibfnamefont {D.}~\bibnamefont {Montemayor}},
  \ and\ \bibinfo {author} {\bibfnamefont {D.~F.}\ \bibnamefont {Coker}},\
  }\href@noop {} {\bibfield  {journal} {\bibinfo  {journal} {Proc. Natl. Ac.
  Sci.}\ }\textbf {\bibinfo {volume} {102}},\ \bibinfo {pages} {6715} (\bibinfo
  {year} {2005})}\BibitemShut {NoStop}%
\bibitem [{\citenamefont {Bonella}, \citenamefont {Ciccotti},\ and\
  \citenamefont {Kapral}(2010)}]{Bonella_Kapral_quantum-classical_2010}%
  \BibitemOpen
  \bibfield  {author} {\bibinfo {author} {\bibfnamefont {S.}~\bibnamefont
  {Bonella}}, \bibinfo {author} {\bibfnamefont {G.}~\bibnamefont {Ciccotti}}, \
  and\ \bibinfo {author} {\bibfnamefont {R.}~\bibnamefont {Kapral}},\
  }\href@noop {} {\bibfield  {journal} {\bibinfo  {journal} {Chem. Phys.
  Lett.}\ }\textbf {\bibinfo {volume} {484}},\ \bibinfo {pages} {399} (\bibinfo
  {year} {2010})}\BibitemShut {NoStop}%
\bibitem [{\citenamefont {Gottwald}\ and\ \citenamefont
  {Ivanov}()}]{Gottwald_Ivanov_2017}%
  \BibitemOpen
  \bibfield  {author} {\bibinfo {author} {\bibfnamefont {F.}~\bibnamefont
  {Gottwald}}\ and\ \bibinfo {author} {\bibfnamefont {S.~D.}\ \bibnamefont
  {Ivanov}},\ }\href@noop {} {\bibinfo  {journal}
  {https://arxiv.org/abs/1704.00477}\ }\BibitemShut {NoStop}%
\bibitem [{\citenamefont {Miller}(2005)}]{Miller_PNAScomplexsystems_2005}%
  \BibitemOpen
\bibfield  {journal} {  }\bibfield  {author} {\bibinfo {author} {\bibfnamefont
  {W.~H.}\ \bibnamefont {Miller}},\ }\href {\doibase 10.1073/pnas.0408043102}
  {\bibfield  {journal} {\bibinfo  {journal} {Proc. Natl. Acad. Sci. USA}\
  }\textbf {\bibinfo {volume} {102}},\ \bibinfo {pages} {6660} (\bibinfo {year}
  {2005})}\BibitemShut {NoStop}%
\bibitem [{\citenamefont {Ceotto}, \citenamefont {Tantardini},\ and\
  \citenamefont
  {Aspuru-Guzik}(2011)}]{Ceotto_AspuruGuzik_Curseofdimensionality_2011}%
  \BibitemOpen
  \bibfield  {author} {\bibinfo {author} {\bibfnamefont {M.}~\bibnamefont
  {Ceotto}}, \bibinfo {author} {\bibfnamefont {G.~F.}\ \bibnamefont
  {Tantardini}}, \ and\ \bibinfo {author} {\bibfnamefont {A.}~\bibnamefont
  {Aspuru-Guzik}},\ }\href {\doibase http://dx.doi.org/10.1063/1.3664731}
  {\bibfield  {journal} {\bibinfo  {journal} {J. Chem. Phys.}\ }\textbf
  {\bibinfo {volume} {135}},\ \bibinfo {pages} {214108} (\bibinfo {year}
  {2011})}\BibitemShut {NoStop}%
\bibitem [{\citenamefont {Ceotto}\ \emph
  {et~al.}(2009{\natexlab{a}})\citenamefont {Ceotto}, \citenamefont {Atahan},
  \citenamefont {Tantardini},\ and\ \citenamefont
  {Aspuru-Guzik}}]{Ceotto_AspuruGuzik_Multiplecoherent_2009}%
  \BibitemOpen
  \bibfield  {author} {\bibinfo {author} {\bibfnamefont {M.}~\bibnamefont
  {Ceotto}}, \bibinfo {author} {\bibfnamefont {S.}~\bibnamefont {Atahan}},
  \bibinfo {author} {\bibfnamefont {G.~F.}\ \bibnamefont {Tantardini}}, \ and\
  \bibinfo {author} {\bibfnamefont {A.}~\bibnamefont {Aspuru-Guzik}},\ }\href
  {\doibase http://dx.doi.org/10.1063/1.3155062} {\bibfield  {journal}
  {\bibinfo  {journal} {J. Chem. Phys.}\ }\textbf {\bibinfo {volume} {130}},\
  \bibinfo {pages} {234113} (\bibinfo {year} {2009}{\natexlab{a}})}\BibitemShut
  {NoStop}%
\bibitem [{\citenamefont {Ceotto}\ \emph
  {et~al.}(2009{\natexlab{b}})\citenamefont {Ceotto}, \citenamefont {Atahan},
  \citenamefont {Shim}, \citenamefont {Tantardini},\ and\ \citenamefont
  {Aspuru-Guzik}}]{Ceotto_AspuruGuzik_PCCPFirstprinciples_2009}%
  \BibitemOpen
  \bibfield  {author} {\bibinfo {author} {\bibfnamefont {M.}~\bibnamefont
  {Ceotto}}, \bibinfo {author} {\bibfnamefont {S.}~\bibnamefont {Atahan}},
  \bibinfo {author} {\bibfnamefont {S.}~\bibnamefont {Shim}}, \bibinfo {author}
  {\bibfnamefont {G.~F.}\ \bibnamefont {Tantardini}}, \ and\ \bibinfo {author}
  {\bibfnamefont {A.}~\bibnamefont {Aspuru-Guzik}},\ }\href {\doibase
  10.1039/B820785B} {\bibfield  {journal} {\bibinfo  {journal} {Phys. Chem.
  Chem. Phys.}\ }\textbf {\bibinfo {volume} {11}},\ \bibinfo {pages} {3861}
  (\bibinfo {year} {2009}{\natexlab{b}})}\BibitemShut {NoStop}%
\bibitem [{\citenamefont {Ceotto}, \citenamefont {Dell`~Angelo},\ and\
  \citenamefont {Tantardini}(2010)}]{Ceotto_Tantardini_Copper100_2010}%
  \BibitemOpen
  \bibfield  {author} {\bibinfo {author} {\bibfnamefont {M.}~\bibnamefont
  {Ceotto}}, \bibinfo {author} {\bibfnamefont {D.}~\bibnamefont
  {Dell`~Angelo}}, \ and\ \bibinfo {author} {\bibfnamefont {G.~F.}\
  \bibnamefont {Tantardini}},\ }\href@noop {} {\bibfield  {journal} {\bibinfo
  {journal} {J. Chem. Phys.}\ }\textbf {\bibinfo {volume} {133}},\ \bibinfo
  {pages} {054701} (\bibinfo {year} {2010})}\BibitemShut {NoStop}%
\bibitem [{\citenamefont {Conte}, \citenamefont {Aspuru-Guzik},\ and\
  \citenamefont {Ceotto}(2013)}]{Conte_Ceotto_NH3_2013}%
  \BibitemOpen
  \bibfield  {author} {\bibinfo {author} {\bibfnamefont {R.}~\bibnamefont
  {Conte}}, \bibinfo {author} {\bibfnamefont {A.}~\bibnamefont {Aspuru-Guzik}},
  \ and\ \bibinfo {author} {\bibfnamefont {M.}~\bibnamefont {Ceotto}},\ }\href
  {\doibase 10.1021/jz401603f} {\bibfield  {journal} {\bibinfo  {journal} {J.
  Phys. Chem. Lett.}\ }\textbf {\bibinfo {volume} {4}},\ \bibinfo {pages}
  {3407} (\bibinfo {year} {2013})}\BibitemShut {NoStop}%
\bibitem [{\citenamefont {Gabas}, \citenamefont {Conte},\ and\ \citenamefont
  {Ceotto}(2017)}]{Gabas_Ceotto_Glycine_2017}%
  \BibitemOpen
  \bibfield  {author} {\bibinfo {author} {\bibfnamefont {F.}~\bibnamefont
  {Gabas}}, \bibinfo {author} {\bibfnamefont {R.}~\bibnamefont {Conte}}, \ and\
  \bibinfo {author} {\bibfnamefont {M.}~\bibnamefont {Ceotto}},\ }\href@noop {}
  {\bibfield  {journal} {\bibinfo  {journal} {J. Chem. Theory Comput.}\
  }\textbf {\bibinfo {volume} {13}},\ \bibinfo {pages} {2378} (\bibinfo {year}
  {2017})}\BibitemShut {NoStop}%
\bibitem [{\citenamefont {Ceotto}, \citenamefont {Zhuang},\ and\ \citenamefont
  {Hase}(2013)}]{Ceotto_Hase_AcceleratedSC_2013}%
  \BibitemOpen
  \bibfield  {author} {\bibinfo {author} {\bibfnamefont {M.}~\bibnamefont
  {Ceotto}}, \bibinfo {author} {\bibfnamefont {Y.}~\bibnamefont {Zhuang}}, \
  and\ \bibinfo {author} {\bibfnamefont {W.~L.}\ \bibnamefont {Hase}},\
  }\href@noop {} {\bibfield  {journal} {\bibinfo  {journal} {J. Chem. Phys.}\
  }\textbf {\bibinfo {volume} {138}},\ \bibinfo {pages} {054116} (\bibinfo
  {year} {2013})}\BibitemShut {NoStop}%
\bibitem [{\citenamefont {Zhuang}\ \emph {et~al.}(2012)\citenamefont {Zhuang},
  \citenamefont {Siebert}, \citenamefont {Hase}, \citenamefont {Kay},\ and\
  \citenamefont {Ceotto}}]{Zhuang_Ceotto_Hessianapprox_2012}%
  \BibitemOpen
  \bibfield  {author} {\bibinfo {author} {\bibfnamefont {Y.}~\bibnamefont
  {Zhuang}}, \bibinfo {author} {\bibfnamefont {M.~R.}\ \bibnamefont {Siebert}},
  \bibinfo {author} {\bibfnamefont {W.~L.}\ \bibnamefont {Hase}}, \bibinfo
  {author} {\bibfnamefont {K.~G.}\ \bibnamefont {Kay}}, \ and\ \bibinfo
  {author} {\bibfnamefont {M.}~\bibnamefont {Ceotto}},\ }\href@noop {}
  {\bibfield  {journal} {\bibinfo  {journal} {J. Chem. Theory Comput.}\
  }\textbf {\bibinfo {volume} {9}},\ \bibinfo {pages} {54} (\bibinfo {year}
  {2012})}\BibitemShut {NoStop}%
\bibitem [{\citenamefont {Braams}\ and\ \citenamefont
  {Bowman}(2009)}]{Braams_Bowman_PermutInvariant_2009}%
  \BibitemOpen
  \bibfield  {author} {\bibinfo {author} {\bibfnamefont {B.~J.}\ \bibnamefont
  {Braams}}\ and\ \bibinfo {author} {\bibfnamefont {J.~M.}\ \bibnamefont
  {Bowman}},\ }\href {\doibase 10.1080/01442350903234923} {\bibfield  {journal}
  {\bibinfo  {journal} {Int. Rev. Phys. Chem.}\ }\textbf {\bibinfo {volume}
  {28}},\ \bibinfo {pages} {577} (\bibinfo {year} {2009})}\BibitemShut
  {NoStop}%
\bibitem [{\citenamefont {Jiang}\ and\ \citenamefont
  {Guo}(2014)}]{Jiang_Guo_NeuralNetworks_2014}%
  \BibitemOpen
  \bibfield  {author} {\bibinfo {author} {\bibfnamefont {B.}~\bibnamefont
  {Jiang}}\ and\ \bibinfo {author} {\bibfnamefont {H.}~\bibnamefont {Guo}},\
  }\href {\doibase http://dx.doi.org/10.1063/1.4887363} {\bibfield  {journal}
  {\bibinfo  {journal} {J. Chem. Phys.}\ }\textbf {\bibinfo {volume} {141}},\
  \bibinfo {pages} {034109} (\bibinfo {year} {2014})}\BibitemShut {NoStop}%
\bibitem [{\citenamefont {Conte}, \citenamefont {Qu},\ and\ \citenamefont
  {Bowman}(2015)}]{Conte_Bowman_Manybody_2015}%
  \BibitemOpen
  \bibfield  {author} {\bibinfo {author} {\bibfnamefont {R.}~\bibnamefont
  {Conte}}, \bibinfo {author} {\bibfnamefont {C.}~\bibnamefont {Qu}}, \ and\
  \bibinfo {author} {\bibfnamefont {J.~M.}\ \bibnamefont {Bowman}},\
  }\href@noop {} {\bibfield  {journal} {\bibinfo  {journal} {J. Chem. Theory
  Comp.}\ }\textbf {\bibinfo {volume} {11}},\ \bibinfo {pages} {1631} (\bibinfo
  {year} {2015})}\BibitemShut {NoStop}%
\bibitem [{\citenamefont {Homayoon}\ \emph {et~al.}(2015)\citenamefont
  {Homayoon}, \citenamefont {Conte}, \citenamefont {Qu},\ and\ \citenamefont
  {Bowman}}]{Homayoon_Bowman_H2-H2O_2015}%
  \BibitemOpen
  \bibfield  {author} {\bibinfo {author} {\bibfnamefont {Z.}~\bibnamefont
  {Homayoon}}, \bibinfo {author} {\bibfnamefont {R.}~\bibnamefont {Conte}},
  \bibinfo {author} {\bibfnamefont {C.}~\bibnamefont {Qu}}, \ and\ \bibinfo
  {author} {\bibfnamefont {J.~M.}\ \bibnamefont {Bowman}},\ }\href@noop {}
  {\bibfield  {journal} {\bibinfo  {journal} {J. Chem. Phys.}\ }\textbf
  {\bibinfo {volume} {143}},\ \bibinfo {pages} {084302} (\bibinfo {year}
  {2015})}\BibitemShut {NoStop}%
\bibitem [{\citenamefont {Paukku}\ \emph {et~al.}(2013)\citenamefont {Paukku},
  \citenamefont {Yang}, \citenamefont {Varga},\ and\ \citenamefont
  {Truhlar}}]{Paukku_Truhlar_N4_2013}%
  \BibitemOpen
  \bibfield  {author} {\bibinfo {author} {\bibfnamefont {Y.}~\bibnamefont
  {Paukku}}, \bibinfo {author} {\bibfnamefont {K.~R.}\ \bibnamefont {Yang}},
  \bibinfo {author} {\bibfnamefont {Z.}~\bibnamefont {Varga}}, \ and\ \bibinfo
  {author} {\bibfnamefont {D.~G.}\ \bibnamefont {Truhlar}},\ }\href {\doibase
  10.1063/1.4811653} {\bibfield  {journal} {\bibinfo  {journal} {J. Chem.
  Phys.}\ }\textbf {\bibinfo {volume} {139}},\ \bibinfo {pages} {044309}
  (\bibinfo {year} {2013})}\BibitemShut {NoStop}%
\bibitem [{\citenamefont {Conte}, \citenamefont {Houston},\ and\ \citenamefont
  {Bowman}(2015)}]{Conte_Bowman_CollisionsCH4-H2O_2015}%
  \BibitemOpen
  \bibfield  {author} {\bibinfo {author} {\bibfnamefont {R.}~\bibnamefont
  {Conte}}, \bibinfo {author} {\bibfnamefont {P.~L.}\ \bibnamefont {Houston}},
  \ and\ \bibinfo {author} {\bibfnamefont {J.~M.}\ \bibnamefont {Bowman}},\
  }\href@noop {} {\bibfield  {journal} {\bibinfo  {journal} {J. Phys. Chem. A}\
  }\textbf {\bibinfo {volume} {119}},\ \bibinfo {pages} {12304} (\bibinfo
  {year} {2015})}\BibitemShut {NoStop}%
\bibitem [{\citenamefont {Varga}\ \emph {et~al.}(2016)\citenamefont {Varga},
  \citenamefont {Meana-Paneda}, \citenamefont {Song}, \citenamefont {Paukku},\
  and\ \citenamefont {Truhlar}}]{Varga_Truhlar_PESN2O2_2016}%
  \BibitemOpen
  \bibfield  {author} {\bibinfo {author} {\bibfnamefont {Z.}~\bibnamefont
  {Varga}}, \bibinfo {author} {\bibfnamefont {R.}~\bibnamefont {Meana-Paneda}},
  \bibinfo {author} {\bibfnamefont {G.}~\bibnamefont {Song}}, \bibinfo {author}
  {\bibfnamefont {Y.}~\bibnamefont {Paukku}}, \ and\ \bibinfo {author}
  {\bibfnamefont {D.~G.}\ \bibnamefont {Truhlar}},\ }\href {\doibase
  10.1063/1.4939008} {\bibfield  {journal} {\bibinfo  {journal} {J. Chem.
  Phys.}\ }\textbf {\bibinfo {volume} {144}},\ \bibinfo {pages} {024310}
  (\bibinfo {year} {2016})}\BibitemShut {NoStop}%
\bibitem [{\citenamefont {Houston}, \citenamefont {Conte},\ and\ \citenamefont
  {Bowman}(2016)}]{Houston_Bowman_RoamingH2CO_2016}%
  \BibitemOpen
  \bibfield  {author} {\bibinfo {author} {\bibfnamefont {P.~L.}\ \bibnamefont
  {Houston}}, \bibinfo {author} {\bibfnamefont {R.}~\bibnamefont {Conte}}, \
  and\ \bibinfo {author} {\bibfnamefont {J.~M.}\ \bibnamefont {Bowman}},\
  }\href@noop {} {\bibfield  {journal} {\bibinfo  {journal} {J. Phys. Chem. A}\
  } (\bibinfo {year} {2016})}\BibitemShut {NoStop}%
\bibitem [{\citenamefont {Conte}\ \emph {et~al.}(2013)\citenamefont {Conte},
  \citenamefont {Fu}, \citenamefont {Kamarchik},\ and\ \citenamefont
  {Bowman}}]{Conte_Bowman_GaussianBinning_2013}%
  \BibitemOpen
  \bibfield  {author} {\bibinfo {author} {\bibfnamefont {R.}~\bibnamefont
  {Conte}}, \bibinfo {author} {\bibfnamefont {B.}~\bibnamefont {Fu}}, \bibinfo
  {author} {\bibfnamefont {E.}~\bibnamefont {Kamarchik}}, \ and\ \bibinfo
  {author} {\bibfnamefont {J.~M.}\ \bibnamefont {Bowman}},\ }\href@noop {}
  {\bibfield  {journal} {\bibinfo  {journal} {J. Chem. Phys.}\ }\textbf
  {\bibinfo {volume} {139}},\ \bibinfo {pages} {044104} (\bibinfo {year}
  {2013})}\BibitemShut {NoStop}%
\bibitem [{\citenamefont {Ceotto}, \citenamefont {Di~Liberto},\ and\
  \citenamefont {Conte}(2017)}]{ceotto_conte_DCSCIVR_2017}%
  \BibitemOpen
  \bibfield  {author} {\bibinfo {author} {\bibfnamefont {M.}~\bibnamefont
  {Ceotto}}, \bibinfo {author} {\bibfnamefont {G.}~\bibnamefont {Di~Liberto}},
  \ and\ \bibinfo {author} {\bibfnamefont {R.}~\bibnamefont {Conte}},\
  }\href@noop {} {\bibfield  {journal} {\bibinfo  {journal} {Phys. Rev. Lett.}\
  }\textbf {\bibinfo {volume} {119}},\ \bibinfo {pages} {010401} (\bibinfo
  {year} {2017})}\BibitemShut {NoStop}%
\bibitem [{\citenamefont {Kaledin}\ and\ \citenamefont
  {Miller}(2003{\natexlab{a}})}]{Kaledin_Miller_Timeaveraging_2003}%
  \BibitemOpen
  \bibfield  {author} {\bibinfo {author} {\bibfnamefont {A.~L.}\ \bibnamefont
  {Kaledin}}\ and\ \bibinfo {author} {\bibfnamefont {W.~H.}\ \bibnamefont
  {Miller}},\ }\href {\doibase http://dx.doi.org/10.1063/1.1562158} {\bibfield
  {journal} {\bibinfo  {journal} {J. Chem. Phys.}\ }\textbf {\bibinfo {volume}
  {118}},\ \bibinfo {pages} {7174} (\bibinfo {year}
  {2003}{\natexlab{a}})}\BibitemShut {NoStop}%
\bibitem [{\citenamefont {Kaledin}\ and\ \citenamefont
  {Miller}(2003{\natexlab{b}})}]{Kaledin_Miller_TAmolecules_2003}%
  \BibitemOpen
  \bibfield  {author} {\bibinfo {author} {\bibfnamefont {A.~L.}\ \bibnamefont
  {Kaledin}}\ and\ \bibinfo {author} {\bibfnamefont {W.~H.}\ \bibnamefont
  {Miller}},\ }\href {\doibase http://dx.doi.org/10.1063/1.1589477} {\bibfield
  {journal} {\bibinfo  {journal} {J. Chem. Phys.}\ }\textbf {\bibinfo {volume}
  {119}},\ \bibinfo {pages} {3078} (\bibinfo {year}
  {2003}{\natexlab{b}})}\BibitemShut {NoStop}%
\bibitem [{\citenamefont {Halverson}\ and\ \citenamefont
  {Poirier}(2015)}]{Halverson_Poirier_Benzene_2015}%
  \BibitemOpen
  \bibfield  {author} {\bibinfo {author} {\bibfnamefont {T.}~\bibnamefont
  {Halverson}}\ and\ \bibinfo {author} {\bibfnamefont {B.}~\bibnamefont
  {Poirier}},\ }\href@noop {} {\bibfield  {journal} {\bibinfo  {journal} {J.
  Phys. Chem. A}\ }\textbf {\bibinfo {volume} {119}},\ \bibinfo {pages} {12417}
  (\bibinfo {year} {2015})}\BibitemShut {NoStop}%
\bibitem [{\citenamefont {Feynman}\ and\ \citenamefont
  {Hibbs}(1965)}]{feynman_pathintegral_1965}%
  \BibitemOpen
  \bibfield  {author} {\bibinfo {author} {\bibfnamefont {R.~P.}\ \bibnamefont
  {Feynman}}\ and\ \bibinfo {author} {\bibfnamefont {A.~R.}\ \bibnamefont
  {Hibbs}},\ }\href@noop {} {\emph {\bibinfo {title} {Quantum mechanics and
  path integrals [by] RP Feynman [and] AR Hibbs}}}\ (\bibinfo  {publisher}
  {McGraw-Hill},\ \bibinfo {year} {1965})\BibitemShut {NoStop}%
\bibitem [{\citenamefont {Berry}\ and\ \citenamefont
  {Mount}(1972)}]{Berry_Mount_Semiclassical_1972}%
  \BibitemOpen
  \bibfield  {author} {\bibinfo {author} {\bibfnamefont {M.~V.}\ \bibnamefont
  {Berry}}\ and\ \bibinfo {author} {\bibfnamefont {K.}~\bibnamefont {Mount}},\
  }\href@noop {} {\bibfield  {journal} {\bibinfo  {journal} {Rep. on Prog.
  Phys.}\ }\textbf {\bibinfo {volume} {35}},\ \bibinfo {pages} {315} (\bibinfo
  {year} {1972})}\BibitemShut {NoStop}%
\bibitem [{\citenamefont {Liu}\ and\ \citenamefont
  {Miller}(2007)}]{Liu_Miller_linearizedSCIVR_2007}%
  \BibitemOpen
  \bibfield  {author} {\bibinfo {author} {\bibfnamefont {J.}~\bibnamefont
  {Liu}}\ and\ \bibinfo {author} {\bibfnamefont {W.~H.}\ \bibnamefont
  {Miller}},\ }\href@noop {} {\bibfield  {journal} {\bibinfo  {journal} {J.
  Chem. Phys.}\ }\textbf {\bibinfo {volume} {127}},\ \bibinfo {pages} {114506}
  (\bibinfo {year} {2007})}\BibitemShut {NoStop}%
\bibitem [{\citenamefont {Takahashi}\ and\ \citenamefont
  {Takatsuka}(2007)}]{Takahashi_Takatsuka_PhaseQuantization_2007}%
  \BibitemOpen
  \bibfield  {author} {\bibinfo {author} {\bibfnamefont {S.}~\bibnamefont
  {Takahashi}}\ and\ \bibinfo {author} {\bibfnamefont {K.}~\bibnamefont
  {Takatsuka}},\ }\href@noop {} {\bibfield  {journal} {\bibinfo  {journal} {J.
  Chem. Phys.}\ }\textbf {\bibinfo {volume} {127}},\ \bibinfo {pages} {084112}
  (\bibinfo {year} {2007})}\BibitemShut {NoStop}%
\bibitem [{\citenamefont {Tao}\ and\ \citenamefont
  {Miller}(2011)}]{Tao_Miller_Tdepsampling_2011}%
  \BibitemOpen
  \bibfield  {author} {\bibinfo {author} {\bibfnamefont {G.}~\bibnamefont
  {Tao}}\ and\ \bibinfo {author} {\bibfnamefont {W.~H.}\ \bibnamefont
  {Miller}},\ }\href@noop {} {\bibfield  {journal} {\bibinfo  {journal} {J.
  Chem. Phys.}\ }\textbf {\bibinfo {volume} {135}},\ \bibinfo {pages} {024104}
  (\bibinfo {year} {2011})}\BibitemShut {NoStop}%
\bibitem [{\citenamefont {Van~Vleck}(1928)}]{VanVleck_SCpropagator_1928}%
  \BibitemOpen
  \bibfield  {author} {\bibinfo {author} {\bibfnamefont {J.~H.}\ \bibnamefont
  {Van~Vleck}},\ }\href@noop {} {\bibfield  {journal} {\bibinfo  {journal}
  {Proc. Natl. Acad. Sci.}\ }\textbf {\bibinfo {volume} {14}},\ \bibinfo
  {pages} {178} (\bibinfo {year} {1928})}\BibitemShut {NoStop}%
\bibitem [{\citenamefont {Gutzwiller}(1967)}]{Gutzwiller_SCpropagator_1967}%
  \BibitemOpen
  \bibfield  {author} {\bibinfo {author} {\bibfnamefont {M.~C.}\ \bibnamefont
  {Gutzwiller}},\ }\href@noop {} {\bibfield  {journal} {\bibinfo  {journal} {J.
  Math. Phys.}\ }\textbf {\bibinfo {volume} {8}},\ \bibinfo {pages} {1979}
  (\bibinfo {year} {1967})}\BibitemShut {NoStop}%
\bibitem [{\citenamefont {Baranger}\ \emph {et~al.}(2001)\citenamefont
  {Baranger}, \citenamefont {de~Aguiar}, \citenamefont {Keck}, \citenamefont
  {Korsch},\ and\ \citenamefont {Schellhaass}}]{Baranger_Schellhaass_2001}%
  \BibitemOpen
  \bibfield  {author} {\bibinfo {author} {\bibfnamefont {M.}~\bibnamefont
  {Baranger}}, \bibinfo {author} {\bibfnamefont {M.~A.}\ \bibnamefont
  {de~Aguiar}}, \bibinfo {author} {\bibfnamefont {F.}~\bibnamefont {Keck}},
  \bibinfo {author} {\bibfnamefont {H.-J.}\ \bibnamefont {Korsch}}, \ and\
  \bibinfo {author} {\bibfnamefont {B.}~\bibnamefont {Schellhaass}},\
  }\href@noop {} {\bibfield  {journal} {\bibinfo  {journal} {J. Phys. A}\
  }\textbf {\bibinfo {volume} {34}},\ \bibinfo {pages} {7227} (\bibinfo {year}
  {2001})}\BibitemShut {NoStop}%
\bibitem [{\citenamefont {Buchholz}, \citenamefont {Grossmann},\ and\
  \citenamefont {Ceotto}(2016)}]{Buchholz_Ceotto_MixedSC_2016}%
  \BibitemOpen
  \bibfield  {author} {\bibinfo {author} {\bibfnamefont {M.}~\bibnamefont
  {Buchholz}}, \bibinfo {author} {\bibfnamefont {F.}~\bibnamefont {Grossmann}},
  \ and\ \bibinfo {author} {\bibfnamefont {M.}~\bibnamefont {Ceotto}},\
  }\href@noop {} {\bibfield  {journal} {\bibinfo  {journal} {J. Chem. Phys.}\
  }\textbf {\bibinfo {volume} {144}},\ \bibinfo {pages} {094102} (\bibinfo
  {year} {2016})}\BibitemShut {NoStop}%
\bibitem [{\citenamefont {Yamamoto}, \citenamefont {Wang},\ and\ \citenamefont
  {Miller}(2002)}]{Yamamoto_Miller_Fluxcorrelation_2002}%
  \BibitemOpen
  \bibfield  {author} {\bibinfo {author} {\bibfnamefont {T.}~\bibnamefont
  {Yamamoto}}, \bibinfo {author} {\bibfnamefont {H.}~\bibnamefont {Wang}}, \
  and\ \bibinfo {author} {\bibfnamefont {W.~H.}\ \bibnamefont {Miller}},\
  }\href@noop {} {\bibfield  {journal} {\bibinfo  {journal} {J. Chem. Phys.}\
  }\textbf {\bibinfo {volume} {116}},\ \bibinfo {pages} {7335} (\bibinfo {year}
  {2002})}\BibitemShut {NoStop}%
\bibitem [{\citenamefont
  {Heller}(1981{\natexlab{b}})}]{Heller_FrozenGaussian_1981}%
  \BibitemOpen
  \bibfield  {author} {\bibinfo {author} {\bibfnamefont {E.~J.}\ \bibnamefont
  {Heller}},\ }\href {\doibase http://dx.doi.org/10.1063/1.442382} {\bibfield
  {journal} {\bibinfo  {journal} {J. Chem. Phys.}\ }\textbf {\bibinfo {volume}
  {75}},\ \bibinfo {pages} {2923} (\bibinfo {year}
  {1981}{\natexlab{b}})}\BibitemShut {NoStop}%
\bibitem [{\citenamefont {Heller}(1991)}]{Heller_Cellulardynamics_1991}%
  \BibitemOpen
  \bibfield  {author} {\bibinfo {author} {\bibfnamefont {E.~J.}\ \bibnamefont
  {Heller}},\ }\href {\doibase http://dx.doi.org/10.1063/1.459848} {\bibfield
  {journal} {\bibinfo  {journal} {J. Chem. Phys.}\ }\textbf {\bibinfo {volume}
  {94}},\ \bibinfo {pages} {2723} (\bibinfo {year} {1991})}\BibitemShut
  {NoStop}%
\bibitem [{\citenamefont {Shalashilin}\ and\ \citenamefont
  {Child}(2001)}]{Shalashilin_Child_Coherentstates_2001}%
  \BibitemOpen
  \bibfield  {author} {\bibinfo {author} {\bibfnamefont {D.~V.}\ \bibnamefont
  {Shalashilin}}\ and\ \bibinfo {author} {\bibfnamefont {M.~S.}\ \bibnamefont
  {Child}},\ }\href@noop {} {\bibfield  {journal} {\bibinfo  {journal} {J.
  Chem. Phys.}\ }\textbf {\bibinfo {volume} {115}},\ \bibinfo {pages} {5367}
  (\bibinfo {year} {2001})}\BibitemShut {NoStop}%
\bibitem [{\citenamefont {Tamascelli}\ \emph {et~al.}(2014)\citenamefont
  {Tamascelli}, \citenamefont {Dambrosio}, \citenamefont {Conte},\ and\
  \citenamefont {Ceotto}}]{Tamascelli_Ceotto_GPU_2014}%
  \BibitemOpen
  \bibfield  {author} {\bibinfo {author} {\bibfnamefont {D.}~\bibnamefont
  {Tamascelli}}, \bibinfo {author} {\bibfnamefont {F.~S.}\ \bibnamefont
  {Dambrosio}}, \bibinfo {author} {\bibfnamefont {R.}~\bibnamefont {Conte}}, \
  and\ \bibinfo {author} {\bibfnamefont {M.}~\bibnamefont {Ceotto}},\
  }\href@noop {} {\bibfield  {journal} {\bibinfo  {journal} {J. Chem. Phys.}\
  }\textbf {\bibinfo {volume} {140}},\ \bibinfo {pages} {174109} (\bibinfo
  {year} {2014})}\BibitemShut {NoStop}%
\bibitem [{\citenamefont {Di~Liberto}\ and\ \citenamefont
  {Ceotto}(2016)}]{DiLiberto_Ceotto_Prefactors_2016}%
  \BibitemOpen
  \bibfield  {author} {\bibinfo {author} {\bibfnamefont {G.}~\bibnamefont
  {Di~Liberto}}\ and\ \bibinfo {author} {\bibfnamefont {M.}~\bibnamefont
  {Ceotto}},\ }\href@noop {} {\bibfield  {journal} {\bibinfo  {journal} {J.
  Chem. Phys.}\ }\textbf {\bibinfo {volume} {145}},\ \bibinfo {pages} {144107}
  (\bibinfo {year} {2016})}\BibitemShut {NoStop}%
\bibitem [{\citenamefont {Ceotto}\ \emph {et~al.}(2011)\citenamefont {Ceotto},
  \citenamefont {Valleau}, \citenamefont {Tantardini},\ and\ \citenamefont
  {Aspuru-Guzik}}]{Ceotto_AspuruGuzik_Firstprinciples_2011}%
  \BibitemOpen
  \bibfield  {author} {\bibinfo {author} {\bibfnamefont {M.}~\bibnamefont
  {Ceotto}}, \bibinfo {author} {\bibfnamefont {S.}~\bibnamefont {Valleau}},
  \bibinfo {author} {\bibfnamefont {G.~F.}\ \bibnamefont {Tantardini}}, \ and\
  \bibinfo {author} {\bibfnamefont {A.}~\bibnamefont {Aspuru-Guzik}},\ }\href
  {\doibase http://dx.doi.org/10.1063/1.3599469} {\bibfield  {journal}
  {\bibinfo  {journal} {J. Chem. Phys.}\ }\textbf {\bibinfo {volume} {134}},\
  \bibinfo {pages} {234103} (\bibinfo {year} {2011})}\BibitemShut {NoStop}%
\bibitem [{\citenamefont {Hinsen}\ and\ \citenamefont
  {Kneller}(2000)}]{Hinsen_Kneller_SingValueDecomp_2000}%
  \BibitemOpen
  \bibfield  {author} {\bibinfo {author} {\bibfnamefont {K.}~\bibnamefont
  {Hinsen}}\ and\ \bibinfo {author} {\bibfnamefont {G.~R.}\ \bibnamefont
  {Kneller}},\ }\href@noop {} {\bibfield  {journal} {\bibinfo  {journal} {Mol.
  Simul.}\ }\textbf {\bibinfo {volume} {23}},\ \bibinfo {pages} {275} (\bibinfo
  {year} {2000})}\BibitemShut {NoStop}%
\bibitem [{\citenamefont {Harland}\ and\ \citenamefont
  {Roy}(2003)}]{Harland_Roy_SCIVRconstrained_2003}%
  \BibitemOpen
  \bibfield  {author} {\bibinfo {author} {\bibfnamefont {B.~B.}\ \bibnamefont
  {Harland}}\ and\ \bibinfo {author} {\bibfnamefont {P.-N.}\ \bibnamefont
  {Roy}},\ }\href@noop {} {\bibfield  {journal} {\bibinfo  {journal} {J. Chem.
  Phys.}\ }\textbf {\bibinfo {volume} {118}},\ \bibinfo {pages} {4791}
  (\bibinfo {year} {2003})}\BibitemShut {NoStop}%
\bibitem [{\citenamefont {Wehrle}, \citenamefont {Sulc},\ and\ \citenamefont
  {Vanicek}(2014)}]{Wehrle_Vanicek_Oligothiophenes_2014}%
  \BibitemOpen
  \bibfield  {author} {\bibinfo {author} {\bibfnamefont {M.}~\bibnamefont
  {Wehrle}}, \bibinfo {author} {\bibfnamefont {M.}~\bibnamefont {Sulc}}, \ and\
  \bibinfo {author} {\bibfnamefont {J.}~\bibnamefont {Vanicek}},\ }\href
  {\doibase http://dx.doi.org/10.1063/1.4884718} {\bibfield  {journal}
  {\bibinfo  {journal} {J. Chem. Phys.}\ }\textbf {\bibinfo {volume} {140}},\
  \bibinfo {pages} {244114} (\bibinfo {year} {2014})}\BibitemShut {NoStop}%
\bibitem [{\citenamefont {Picconi}\ \emph {et~al.}(2013)\citenamefont
  {Picconi}, \citenamefont {Ferrer}, \citenamefont {Improta}, \citenamefont
  {Lami},\ and\ \citenamefont {Santoro}}]{picconi_santoro_quantumclass_2013}%
  \BibitemOpen
  \bibfield  {author} {\bibinfo {author} {\bibfnamefont {D.}~\bibnamefont
  {Picconi}}, \bibinfo {author} {\bibfnamefont {F.~J.~A.}\ \bibnamefont
  {Ferrer}}, \bibinfo {author} {\bibfnamefont {R.}~\bibnamefont {Improta}},
  \bibinfo {author} {\bibfnamefont {A.}~\bibnamefont {Lami}}, \ and\ \bibinfo
  {author} {\bibfnamefont {F.}~\bibnamefont {Santoro}},\ }\href@noop {}
  {\bibfield  {journal} {\bibinfo  {journal} {Faraday discussions}\ }\textbf
  {\bibinfo {volume} {163}},\ \bibinfo {pages} {223} (\bibinfo {year}
  {2013})}\BibitemShut {NoStop}%
\bibitem [{\citenamefont {Cederbaum}, \citenamefont {Gindensperger},\ and\
  \citenamefont
  {Burghardt}(2005)}]{cederbaum_Burghardt_shorttimeconinters_2005}%
  \BibitemOpen
  \bibfield  {author} {\bibinfo {author} {\bibfnamefont {L.~S.}\ \bibnamefont
  {Cederbaum}}, \bibinfo {author} {\bibfnamefont {E.}~\bibnamefont
  {Gindensperger}}, \ and\ \bibinfo {author} {\bibfnamefont {I.}~\bibnamefont
  {Burghardt}},\ }\href@noop {} {\bibfield  {journal} {\bibinfo  {journal}
  {Phys. Rev. Lett.}\ }\textbf {\bibinfo {volume} {94}},\ \bibinfo {pages}
  {113003} (\bibinfo {year} {2005})}\BibitemShut {NoStop}%
\bibitem [{\citenamefont {Wehrle}(2015)}]{Wehrle_PhDThesis_2015}%
  \BibitemOpen
  \bibfield  {author} {\bibinfo {author} {\bibfnamefont {M.}~\bibnamefont
  {Wehrle}},\ }\href@noop {} {\emph {\bibinfo {title} {Evaluation and analysis
  of vibrationally resolved electronic spectra with ab initio semiclassical
  dynamics, Ph. D. Thesis.}}}\ (\bibinfo  {publisher} {EPFL},\ \bibinfo {year}
  {2015})\BibitemShut {NoStop}%
\bibitem [{\citenamefont {Colbert}\ and\ \citenamefont
  {Miller}(1992)}]{colbert_miller_dvr_1992}%
  \BibitemOpen
  \bibfield  {author} {\bibinfo {author} {\bibfnamefont {D.~T.}\ \bibnamefont
  {Colbert}}\ and\ \bibinfo {author} {\bibfnamefont {W.~H.}\ \bibnamefont
  {Miller}},\ }\href@noop {} {\bibfield  {journal} {\bibinfo  {journal} {J.
  Chem. Phys.}\ }\textbf {\bibinfo {volume} {96}},\ \bibinfo {pages} {1982}
  (\bibinfo {year} {1992})}\BibitemShut {NoStop}%
\bibitem [{\citenamefont {Partridge}\ and\ \citenamefont
  {Schwenke}(1997)}]{partridge_Schwenke_PESH2Omonomer_1997}%
  \BibitemOpen
  \bibfield  {author} {\bibinfo {author} {\bibfnamefont {H.}~\bibnamefont
  {Partridge}}\ and\ \bibinfo {author} {\bibfnamefont {D.~W.}\ \bibnamefont
  {Schwenke}},\ }\href@noop {} {\bibfield  {journal} {\bibinfo  {journal} {J.
  Chem. Phys.}\ }\textbf {\bibinfo {volume} {106}},\ \bibinfo {pages} {4618}
  (\bibinfo {year} {1997})}\BibitemShut {NoStop}%
\bibitem [{\citenamefont {Martin}, \citenamefont {Lee},\ and\ \citenamefont
  {Taylor}(1993)}]{martin_taylo_PESch2o_1993}%
  \BibitemOpen
  \bibfield  {author} {\bibinfo {author} {\bibfnamefont {J.}~\bibnamefont
  {Martin}}, \bibinfo {author} {\bibfnamefont {T.~J.}\ \bibnamefont {Lee}}, \
  and\ \bibinfo {author} {\bibfnamefont {P.}~\bibnamefont {Taylor}},\
  }\href@noop {} {\bibfield  {journal} {\bibinfo  {journal} {J. mol. spectr.}\
  }\textbf {\bibinfo {volume} {160}},\ \bibinfo {pages} {105} (\bibinfo {year}
  {1993})}\BibitemShut {NoStop}%
\bibitem [{\citenamefont {Carter}, \citenamefont {Pinnavaia},\ and\
  \citenamefont {Handy}(1995)}]{carter_handy_exactCH2O_1995}%
  \BibitemOpen
  \bibfield  {author} {\bibinfo {author} {\bibfnamefont {S.}~\bibnamefont
  {Carter}}, \bibinfo {author} {\bibfnamefont {N.}~\bibnamefont {Pinnavaia}}, \
  and\ \bibinfo {author} {\bibfnamefont {N.~C.}\ \bibnamefont {Handy}},\
  }\href@noop {} {\bibfield  {journal} {\bibinfo  {journal} {Chem. phys.
  lett.}\ }\textbf {\bibinfo {volume} {240}},\ \bibinfo {pages} {400} (\bibinfo
  {year} {1995})}\BibitemShut {NoStop}%
\bibitem [{\citenamefont {Lee}, \citenamefont {Martin},\ and\ \citenamefont
  {Taylor}(1995)}]{lee_taylor_PESch4_1995}%
  \BibitemOpen
  \bibfield  {author} {\bibinfo {author} {\bibfnamefont {T.~J.}\ \bibnamefont
  {Lee}}, \bibinfo {author} {\bibfnamefont {J.~M.}\ \bibnamefont {Martin}}, \
  and\ \bibinfo {author} {\bibfnamefont {P.~R.}\ \bibnamefont {Taylor}},\
  }\href@noop {} {\bibfield  {journal} {\bibinfo  {journal} {J. Chem Phys.}\
  }\textbf {\bibinfo {volume} {102}},\ \bibinfo {pages} {254} (\bibinfo {year}
  {1995})}\BibitemShut {NoStop}%
\bibitem [{\citenamefont {Carter}, \citenamefont {Shnider},\ and\ \citenamefont
  {Bowman}(1999)}]{Carter_Bowman_Methane_1999}%
  \BibitemOpen
  \bibfield  {author} {\bibinfo {author} {\bibfnamefont {S.}~\bibnamefont
  {Carter}}, \bibinfo {author} {\bibfnamefont {H.~M.}\ \bibnamefont {Shnider}},
  \ and\ \bibinfo {author} {\bibfnamefont {J.~M.}\ \bibnamefont {Bowman}},\
  }\href@noop {} {\bibfield  {journal} {\bibinfo  {journal} {J. Chem. Phys.}\
  }\textbf {\bibinfo {volume} {110}},\ \bibinfo {pages} {8417} (\bibinfo {year}
  {1999})}\BibitemShut {NoStop}%
\bibitem [{\citenamefont {Vendrell}, \citenamefont {Gatti},\ and\ \citenamefont
  {Meyer}(2007{\natexlab{a}})}]{vendrell_Meyer_zundelspectra_2007}%
  \BibitemOpen
  \bibfield  {author} {\bibinfo {author} {\bibfnamefont {O.}~\bibnamefont
  {Vendrell}}, \bibinfo {author} {\bibfnamefont {F.}~\bibnamefont {Gatti}}, \
  and\ \bibinfo {author} {\bibfnamefont {H.-D.}\ \bibnamefont {Meyer}},\
  }\href@noop {} {\bibfield  {journal} {\bibinfo  {journal} {J. Chem. Phys.}\
  }\textbf {\bibinfo {volume} {127}},\ \bibinfo {pages} {184303} (\bibinfo
  {year} {2007}{\natexlab{a}})}\BibitemShut {NoStop}%
\bibitem [{\citenamefont {Rossi}, \citenamefont {Ceriotti},\ and\ \citenamefont
  {Manolopoulos}(2014)}]{rossi_manolopoulos_TRPDM_2014}%
  \BibitemOpen
  \bibfield  {author} {\bibinfo {author} {\bibfnamefont {M.}~\bibnamefont
  {Rossi}}, \bibinfo {author} {\bibfnamefont {M.}~\bibnamefont {Ceriotti}}, \
  and\ \bibinfo {author} {\bibfnamefont {D.~E.}\ \bibnamefont {Manolopoulos}},\
  }\href@noop {} {\bibfield  {journal} {\bibinfo  {journal} {J. Chem. Phys.}\
  }\textbf {\bibinfo {volume} {140}},\ \bibinfo {pages} {234116} (\bibinfo
  {year} {2014})}\BibitemShut {NoStop}%
\bibitem [{\citenamefont {Hammer}\ \emph {et~al.}(2005)\citenamefont {Hammer},
  \citenamefont {Diken}, \citenamefont {Roscioli}, \citenamefont {Johnson},
  \citenamefont {Myshakin}, \citenamefont {Jordan}, \citenamefont {McCoy},
  \citenamefont {Huang}, \citenamefont {Bowman},\ and\ \citenamefont
  {Carter}}]{hammer_carter_expzundelspectrum_2005}%
  \BibitemOpen
  \bibfield  {author} {\bibinfo {author} {\bibfnamefont {N.~I.}\ \bibnamefont
  {Hammer}}, \bibinfo {author} {\bibfnamefont {E.~G.}\ \bibnamefont {Diken}},
  \bibinfo {author} {\bibfnamefont {J.~R.}\ \bibnamefont {Roscioli}}, \bibinfo
  {author} {\bibfnamefont {M.~A.}\ \bibnamefont {Johnson}}, \bibinfo {author}
  {\bibfnamefont {E.~M.}\ \bibnamefont {Myshakin}}, \bibinfo {author}
  {\bibfnamefont {K.~D.}\ \bibnamefont {Jordan}}, \bibinfo {author}
  {\bibfnamefont {A.~B.}\ \bibnamefont {McCoy}}, \bibinfo {author}
  {\bibfnamefont {X.}~\bibnamefont {Huang}}, \bibinfo {author} {\bibfnamefont
  {J.~M.}\ \bibnamefont {Bowman}}, \ and\ \bibinfo {author} {\bibfnamefont
  {S.}~\bibnamefont {Carter}},\ }\href@noop {} {\bibfield  {journal} {\bibinfo
  {journal} {J. Chem. Phys.}\ }\textbf {\bibinfo {volume} {122}},\ \bibinfo
  {pages} {244301} (\bibinfo {year} {2005})}\BibitemShut {NoStop}%
\bibitem [{\citenamefont {Vendrell}, \citenamefont {Gatti},\ and\ \citenamefont
  {Meyer}(2009{\natexlab{a}})}]{vendrell_Meyer_isotopeeffects_2009}%
  \BibitemOpen
  \bibfield  {author} {\bibinfo {author} {\bibfnamefont {O.}~\bibnamefont
  {Vendrell}}, \bibinfo {author} {\bibfnamefont {F.}~\bibnamefont {Gatti}}, \
  and\ \bibinfo {author} {\bibfnamefont {H.-D.}\ \bibnamefont {Meyer}},\
  }\href@noop {} {\bibfield  {journal} {\bibinfo  {journal} {Angewandte Chemie
  International Edition}\ }\textbf {\bibinfo {volume} {48}},\ \bibinfo {pages}
  {352} (\bibinfo {year} {2009}{\natexlab{a}})}\BibitemShut {NoStop}%
\bibitem [{\citenamefont {Vendrell}, \citenamefont {Gatti},\ and\ \citenamefont
  {Meyer}(2009{\natexlab{b}})}]{vendrell_Meyer_isotopeffects2_2009}%
  \BibitemOpen
  \bibfield  {author} {\bibinfo {author} {\bibfnamefont {O.}~\bibnamefont
  {Vendrell}}, \bibinfo {author} {\bibfnamefont {F.}~\bibnamefont {Gatti}}, \
  and\ \bibinfo {author} {\bibfnamefont {H.-D.}\ \bibnamefont {Meyer}},\
  }\href@noop {} {\bibfield  {journal} {\bibinfo  {journal} {J. Chem. Phys.}\
  }\textbf {\bibinfo {volume} {131}},\ \bibinfo {pages} {034308} (\bibinfo
  {year} {2009}{\natexlab{b}})}\BibitemShut {NoStop}%
\bibitem [{\citenamefont {Vendrell}\ \emph {et~al.}(2009)\citenamefont
  {Vendrell}, \citenamefont {Brill}, \citenamefont {Gatti}, \citenamefont
  {Lauvergnat},\ and\ \citenamefont
  {Meyer}}]{vendrell_Meyer_Jacobianparametriz_2009}%
  \BibitemOpen
  \bibfield  {author} {\bibinfo {author} {\bibfnamefont {O.}~\bibnamefont
  {Vendrell}}, \bibinfo {author} {\bibfnamefont {M.}~\bibnamefont {Brill}},
  \bibinfo {author} {\bibfnamefont {F.}~\bibnamefont {Gatti}}, \bibinfo
  {author} {\bibfnamefont {D.}~\bibnamefont {Lauvergnat}}, \ and\ \bibinfo
  {author} {\bibfnamefont {H.-D.}\ \bibnamefont {Meyer}},\ }\href@noop {}
  {\bibfield  {journal} {\bibinfo  {journal} {J. Chem. Phys.}\ }\textbf
  {\bibinfo {volume} {130}},\ \bibinfo {pages} {234305} (\bibinfo {year}
  {2009})}\BibitemShut {NoStop}%
\bibitem [{\citenamefont {Vendrell}, \citenamefont {Gatti},\ and\ \citenamefont
  {Meyer}(2007{\natexlab{b}})}]{vendrell_Meyer_Zundeldynamics_2007}%
  \BibitemOpen
  \bibfield  {author} {\bibinfo {author} {\bibfnamefont {O.}~\bibnamefont
  {Vendrell}}, \bibinfo {author} {\bibfnamefont {F.}~\bibnamefont {Gatti}}, \
  and\ \bibinfo {author} {\bibfnamefont {H.-D.}\ \bibnamefont {Meyer}},\
  }\href@noop {} {\bibfield  {journal} {\bibinfo  {journal} {Angewandte Chemie
  International Edition}\ }\textbf {\bibinfo {volume} {46}},\ \bibinfo {pages}
  {6918} (\bibinfo {year} {2007}{\natexlab{b}})}\BibitemShut {NoStop}%
\bibitem [{\citenamefont {Vendrell}\ \emph {et~al.}(2007)\citenamefont
  {Vendrell}, \citenamefont {Gatti}, \citenamefont {Lauvergnat},\ and\
  \citenamefont {Meyer}}]{vendrell_Meyer_ZundelHamiltonian_2007}%
  \BibitemOpen
  \bibfield  {author} {\bibinfo {author} {\bibfnamefont {O.}~\bibnamefont
  {Vendrell}}, \bibinfo {author} {\bibfnamefont {F.}~\bibnamefont {Gatti}},
  \bibinfo {author} {\bibfnamefont {D.}~\bibnamefont {Lauvergnat}}, \ and\
  \bibinfo {author} {\bibfnamefont {H.-D.}\ \bibnamefont {Meyer}},\ }\href@noop
  {} {\bibfield  {journal} {\bibinfo  {journal} {J. Chem. Phys.}\ }\textbf
  {\bibinfo {volume} {127}},\ \bibinfo {pages} {184302} (\bibinfo {year}
  {2007})}\BibitemShut {NoStop}%
\bibitem [{\citenamefont {Vendrell}\ and\ \citenamefont
  {Meyer}(2008)}]{vendrell_meyer_zundelquantumdynamics_2008}%
  \BibitemOpen
  \bibfield  {author} {\bibinfo {author} {\bibfnamefont {O.}~\bibnamefont
  {Vendrell}}\ and\ \bibinfo {author} {\bibfnamefont {H.-D.}\ \bibnamefont
  {Meyer}},\ }\href@noop {} {\bibfield  {journal} {\bibinfo  {journal} {Phys.
  Chem. Chem. Phys.}\ }\textbf {\bibinfo {volume} {10}},\ \bibinfo {pages}
  {4692} (\bibinfo {year} {2008})}\BibitemShut {NoStop}%
\bibitem [{\citenamefont {McCoy}\ \emph {et~al.}(2005)\citenamefont {McCoy},
  \citenamefont {Huang}, \citenamefont {Carter}, \citenamefont {Landeweer},\
  and\ \citenamefont {Bowman}}]{mccoy_bowman_VCIzundel_2005}%
  \BibitemOpen
  \bibfield  {author} {\bibinfo {author} {\bibfnamefont {A.~B.}\ \bibnamefont
  {McCoy}}, \bibinfo {author} {\bibfnamefont {X.}~\bibnamefont {Huang}},
  \bibinfo {author} {\bibfnamefont {S.}~\bibnamefont {Carter}}, \bibinfo
  {author} {\bibfnamefont {M.~Y.}\ \bibnamefont {Landeweer}}, \ and\ \bibinfo
  {author} {\bibfnamefont {J.~M.}\ \bibnamefont {Bowman}},\ }\href@noop {}
  {\enquote {\bibinfo {title} {Full-dimensional vibrational calculations for h
  5 o 2+ using an ab initio potential energy surface},}\ } (\bibinfo {year}
  {2005})\BibitemShut {NoStop}%
\bibitem [{\citenamefont {Huang}, \citenamefont {Braams},\ and\ \citenamefont
  {Bowman}(2005)}]{huang_Bowman_ZundelPES_2005}%
  \BibitemOpen
  \bibfield  {author} {\bibinfo {author} {\bibfnamefont {X.}~\bibnamefont
  {Huang}}, \bibinfo {author} {\bibfnamefont {B.~J.}\ \bibnamefont {Braams}}, \
  and\ \bibinfo {author} {\bibfnamefont {J.~M.}\ \bibnamefont {Bowman}},\
  }\href@noop {} {\bibfield  {journal} {\bibinfo  {journal} {J. Chem. Phys.}\
  }\textbf {\bibinfo {volume} {122}},\ \bibinfo {pages} {044308} (\bibinfo
  {year} {2005})}\BibitemShut {NoStop}%
\bibitem [{\citenamefont {Carter}, \citenamefont {Bowman},\ and\ \citenamefont
  {Sharma}(2012)}]{carter_sharma_multimode_2012}%
  \BibitemOpen
  \bibfield  {author} {\bibinfo {author} {\bibfnamefont {S.}~\bibnamefont
  {Carter}}, \bibinfo {author} {\bibfnamefont {J.~M.}\ \bibnamefont {Bowman}},
  \ and\ \bibinfo {author} {\bibfnamefont {A.~R.}\ \bibnamefont {Sharma}},\
  }in\ \href@noop {} {\emph {\bibinfo {booktitle} {American Institute of
  Physics Conference Series}}},\ Vol.\ \bibinfo {volume} {1504}\ (\bibinfo
  {year} {2012})\ pp.\ \bibinfo {pages} {465--466}\BibitemShut {NoStop}%
\bibitem [{\citenamefont {Handy}\ \emph {et~al.}(1992)\citenamefont {Handy},
  \citenamefont {Maslen}, \citenamefont {Amos}, \citenamefont {Andrews},
  \citenamefont {Murray},\ and\ \citenamefont
  {Laming}}]{handy_laming_benzenepes_1992}%
  \BibitemOpen
  \bibfield  {author} {\bibinfo {author} {\bibfnamefont {N.~C.}\ \bibnamefont
  {Handy}}, \bibinfo {author} {\bibfnamefont {P.~E.}\ \bibnamefont {Maslen}},
  \bibinfo {author} {\bibfnamefont {R.~D.}\ \bibnamefont {Amos}}, \bibinfo
  {author} {\bibfnamefont {J.~S.}\ \bibnamefont {Andrews}}, \bibinfo {author}
  {\bibfnamefont {C.~W.}\ \bibnamefont {Murray}}, \ and\ \bibinfo {author}
  {\bibfnamefont {G.~J.}\ \bibnamefont {Laming}},\ }\href@noop {} {\bibfield
  {journal} {\bibinfo  {journal} {Chem. Phys. Lett.}\ }\textbf {\bibinfo
  {volume} {197}},\ \bibinfo {pages} {506} (\bibinfo {year}
  {1992})}\BibitemShut {NoStop}%
\bibitem [{\citenamefont {Grossmann}(2006)}]{Grossmann_SChybrid_2006}%
  \BibitemOpen
  \bibfield  {author} {\bibinfo {author} {\bibfnamefont {F.}~\bibnamefont
  {Grossmann}},\ }\href {\doibase http://dx.doi.org/10.1063/1.2213255}
  {\bibfield  {journal} {\bibinfo  {journal} {J. Chem. Phys.}\ }\textbf
  {\bibinfo {volume} {125}},\ \bibinfo {eid} {014111} (\bibinfo {year}
  {2006}),\ http://dx.doi.org/10.1063/1.2213255}\BibitemShut {NoStop}%
\bibitem [{\citenamefont {Buchholz}, \citenamefont {Grossmann},\ and\
  \citenamefont {Ceotto}(2017)}]{Ceotto_Buchholz_MixedSC_2017}%
  \BibitemOpen
  \bibfield  {author} {\bibinfo {author} {\bibfnamefont {M.}~\bibnamefont
  {Buchholz}}, \bibinfo {author} {\bibfnamefont {F.}~\bibnamefont {Grossmann}},
  \ and\ \bibinfo {author} {\bibfnamefont {M.}~\bibnamefont {Ceotto}},\
  }\href@noop {} {\bibfield  {journal} {\bibinfo  {journal} {J. Chem. Phys.}\
  }\textbf {\bibinfo {volume} {147}},\ \bibinfo {pages} {164110} (\bibinfo
  {year} {2017})}\BibitemShut {NoStop}%
\end{thebibliography}%

\end{document}